\documentclass[pdflatex,sn-basic]{sn-jnl}

\usepackage{graphicx}%
\usepackage{multirow}%
\usepackage{amsmath,amssymb,amsfonts}%
\usepackage{amsthm}%
\usepackage{mathrsfs}%
\usepackage[title]{appendix}%
\usepackage{xcolor}%
\usepackage{textcomp}%
\usepackage{manyfoot}%
\usepackage{booktabs}%
\usepackage{algorithm}%
\usepackage{algorithmicx}%
\usepackage{algpseudocode}%
\usepackage{listings}%
\usepackage{soul}%

\theoremstyle{thmstyleone}%
\theoremstyle{thmstyletwo}%

\theoremstyle{thmstylethree}%

\begin{document}

\title[Modeling prominences and rain]{Modeling multiphase plasma in the corona: prominences and rain}


\author*[1]{\fnm{Rony} \sur{Keppens}}\email{rony.keppens@kuleuven.be}

\affil*[1]{\orgdiv{Centre for mathematical Plasma Astrophysics, Department of Mathematics}, \orgname{KU Leuven}, \orgaddress{\street{Celestijnenlaan 200B}, \city{Leuven}, \postcode{3001}, \country{Belgium}}}
\author[1, 2]{\fnm{Yuhao} \sur{Zhou}}\email{yuhaozhou@nju.edu.cn}
\author[3]{\fnm{Chun} \sur{Xia}}\email{chun.xia@ynu.edu.cn}

\affil[2]{\orgdiv{School of Astronomy and Space Science}, \orgname{Nanjing University},  \city{Nanjing}, \postcode{210023}, \country{People's Republic of China}}
\affil[3]{\orgdiv{School of Physics and Astronomy}, \orgname{Yunnan University},  \city{Kunming}, \postcode{650500}, \country{People's Republic of China}}


\abstract{We review major achievements in our understanding of multiphase coronal plasma, where cool-dense and hot-tenuous matter coexists, brought about by advances in modeling and theory, inspired by observations. We give an overview of models that self-consistently form solar (or stellar) prominences and filaments, or (postflare) coronal rain, and clarify how these different phenomena share a common physical origin, relating radiative losses and coronal heating. While we do not fully understand the coronal heating, multi-dimensional models of solar prominence and rain formation demonstrate how thermal instability triggers condensations, and how their morphology may reveal aspects of the applied heating at play. We emphasize how the many pathways to linear instability due to combined ingredients of heat-loss, gravity, flows, and magnetic topologies are all involved in the resulting nonlinear magnetohydrodynamics. We provide some challenges to future model efforts, especially concerning prominence fine structure, internal dynamics, and their overall lifecycle.}




\maketitle

\setcounter{tocdepth}{3} 
\tableofcontents

\clearpage
\section{Coronal cooling: multiphase galore!}\label{sec-intro}

Although coronal heating, implied by million degree plasma found throughout the solar atmosphere, receives much attention, an equally daunting mystery relates to coronal cooling. Cool material with temperatures down to about 10,000 K happily co-exists with the hot surroundings, as especially evident in mature and long-lived prominences. In energetic extreme-ultraviolet (EUV) or X-ray waveband views, the clear magnetic structuring of the hot corona stands out, with distinct coronal loops and arcades that show variability in all spatial and temporal scales covered by observations. Even though solar physicists agree on a clear role for magnetically aided transport and dissipation of energy into the corona, multiple proxies for chromospheric activity fail to provide a clear correlation between loop brightness and estimated footpoint Poynting fluxes \citep{Judge2024}: only a fraction of all field lines showing significant flux become loaded with hot matter into prominent EUV loops. Furthermore, many of these hot loops show multiphase fine structure: coronal rain blobs down to a few 100 km in width across the assumed magnetic field orientation have been detected and studied statistically \citep{Antolin2015}. Coronal rain is found throughout active region loops \citep{Seray2023}, is linked to EUV-inferred periodicity of a few hours or longer detected in hot loops \citep{Auchere2018}, and may be much more common than our current detection limits allow us to probe. From the thermodynamic viewpoint, prominences and coronal rain blobs show striking similarities in the sense that both imply hundredfold contrasts in density and temperature, surviving whatever heating mechanism is at play in the corona. Perhaps the most tantalizing manifestation of this multiphase corona - meaning that both cold and hot matter are found in close proximity in the same magnetic topology - is the occurrence of postflare coronal rain. Especially more energetic flares of classes M and X demonstrate catastrophically cooled matter raining down loops that only seconds to minutes ago were heated above 10 million degrees \citep{Mason2022}. As pointed out by \cite{Antolin2020}, this omnipresence of multiphase matter in the corona may well be the key to advance our understanding of coronal heating mechanisms. From the modeling perspective, the intricate link between coronal heating and cooling has been explored in idealized settings that gradually progressed from one-dimensional (1D) loops to fully three-dimensional (3D) settings. Especially these model-based findings will be reviewed here, with an emphasis on how mass and energy circulations occur in realistically stratified atmospheres and loops and what governs the formation of cool condensations. Depending on the magnetic topology where modeled condensations form, they may grow and collect into mature prominences, or show purely rain, or prominence-rain hybrids. In fact, \cite{Mason2019} discovered that coronal rain easily develops near null point topologies, as found in a location where an extended ($\sim$ 100 Mm) bipolar region was embedded in more unipolar surroundings. Prominence-rain hybrids are indeed found near nulls, and can show complex-shaped spider prominences resting in sagged field lines above the null, with rain events on widely separated loops \citep{Filippov2023}. These findings challenge traditional views on prominences, such as those reviewed in \cite{Martin1998}. This review deliberately mixes filament and prominence with coronal rain research, as it tries to unify views related to small-scale and large-scale coronal condensations, especially concerning their in-situ formation.

\section{Filaments and prominences: related reviews}\label{sec-reviews}

\begin{figure}[ht]
\centering
\includegraphics[width=0.49\textwidth,height=7cm]{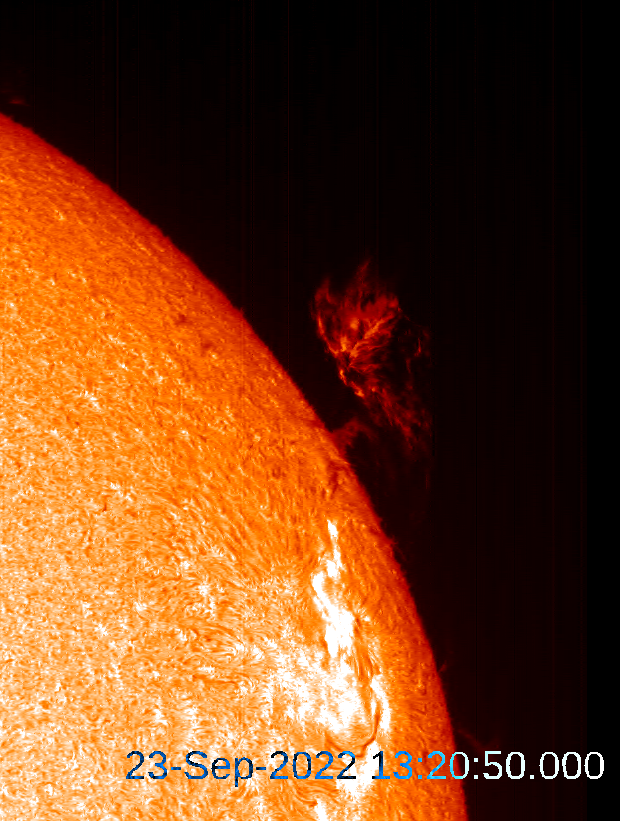}\includegraphics[width=0.49\textwidth, height=7cm]{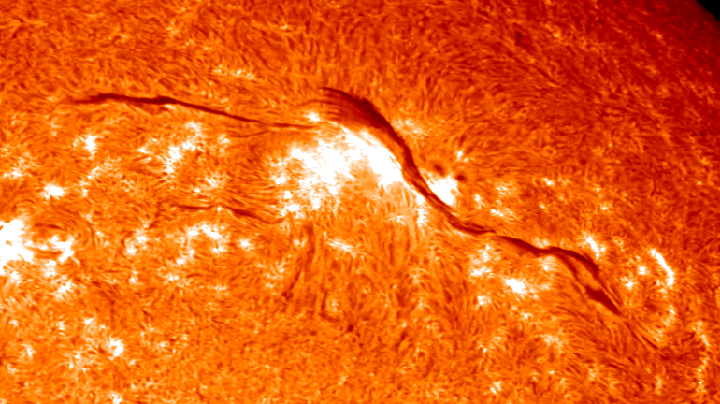}
\caption{H$\alpha$ images made by the CHASE mission \citep{CHASE2022}, where we see a typical prominence (at left, in emission above the limb) versus a clear filament view (at right). Figure credit: Yiwei Ni (Nanjing University).}\label{f-chasefig}
\end{figure}

Large-scale coronal condensations known as filaments (when seen on disk) or prominences (viewed at the limb) have been inferred from ancient descriptions\footnote{\url{https://www2.hao.ucar.edu/education/solar-physics-timeline}}, and spectacular views exist on erupting prominences, like the Grand Daddy eruption from 1946 captured\footnote{\url{https://www2.hao.ucar.edu/education/pictorial/grand-daddy-prominence}} in H$\alpha$. Examples of such wavelength-specific views on prominences and filaments by the CHASE mission \citep{CHASE2022} are shown in Fig.~\ref{f-chasefig}. Their location with respect to active regions introduces the ``active region", ``intermediate" or ``quiescent" prominence terminology, which essentially distinguishes prominences occurring into strong and complex field topology, versus weaker and mostly dipolar magnetic field regions. Obviously, this local magnetic field complexity determines their internal dynamics, as well as their longevity and likelihood to erupt. Quiescent (long-lived) prominences are found above polarity inversion lines (PILs) at higher solar latitudes, and polar crown prominences may even encircle the Sun's pole entirely. The paper by \cite{Hirayama1985} provides an early review, while a living review by \cite{Parenti2014} documents especially the observational characterization of prominence properties. Several - still open - questions relate to their 3D structure in both the magnetic and thermodynamic sense. This 3D structure connects across the different atmospheric layers: from the photosphere, where we know the magnetic fields, to the coronal embedding. A more theory-based view was presented in \cite{Keppens2013}, at a time when multi-dimensional prominence formation models started to gain momentum. Other reviews dedicated to prominence physics highlight \begin{itemize}
    \item 
our partial understanding of their magnetic structure and formation, as inferred from observations and models, the latter mostly without multiphase plasma \citep{Mackay2010}; \item popular cartoon views on three specific formation mechanisms: injection, levitation and  evaporation-condensation \citep{Mackay2010};  
\item an overview of prominence oscillations and their link with magnetohydrodynamic linear wave theory \citep{Mackay2010,Arregui2018};
\item the power of spectral inversion methods that account for departures from Local Thermal Equilibrium called non-LTE, to characterize prominence plasma in conditions that typify their large density-temperature contrasts \citep{Labrosse2010}.
\item some novel insights in filament research collected by \cite{Chen2020}, including the role of helicity, counterstreaming flows, solar tornadoes, and their interrelations.
\end{itemize}
Quite complementary to this review effort is the living review on prominences by \cite{Gibson2018}, with many pointers to relevant theory and model-based insights. \cite{Gibson2018} starts and ends with `fleshing out the magnetic skeleton', implying the progress in magnetohydrodynamic (MHD) simulations that include condensed prominence plasma and energetics. It is especially this progress which we will highlight here. In that spirit, many references to models that predate those with active prominence formation can be found in earlier reviews. Note that many reviews concentrate on the magnetic topologies associated with filaments, mostly divided among flux ropes and sheared dipped arcades. By presenting combined prominence and rain findings in what follows, we emphasize that the underlying mechanisms for condensation formation are indistinguishable and somewhat indifferent to the magnetic topology, as all prevailing models solve the same governing partial differential equations. This was already the case for the coronal loop settings discussed in \cite{Antolin2020}, where coronal rain aspects were emphasized. In books on prominences by \cite{TH1995} and \cite{VialEngvold2015}, the state-of-the-art in the field, respectively, three and one decade ago, can be consulted as well. 

Two independent reviews on prominence modeling at the time of this writing are as follows: \cite{Liakh2025} emphasize especially those numerical models that used the open-source {\tt MPI-AMRVAC} code \citep{Keppens2023}, while \cite{Yuhao2025} extends the discussion to how supporting magnetic structures evolve, and stresses the point that various condensation formation pathways are likely acting together. \cite{Yuhao2025} includes findings from observations, while \cite{Liakh2025} additionally address the role of prominence oscillations, and clarify the various approximations used in synthetic EUV views on modern prominence simulations.

\section{Prelude: force-balance aspects}\label{sec-force}

\subsection{Considerations from Newton's law}\label{s-newton}
\begin{figure}[ht]
\centering
\includegraphics[width=\textwidth]{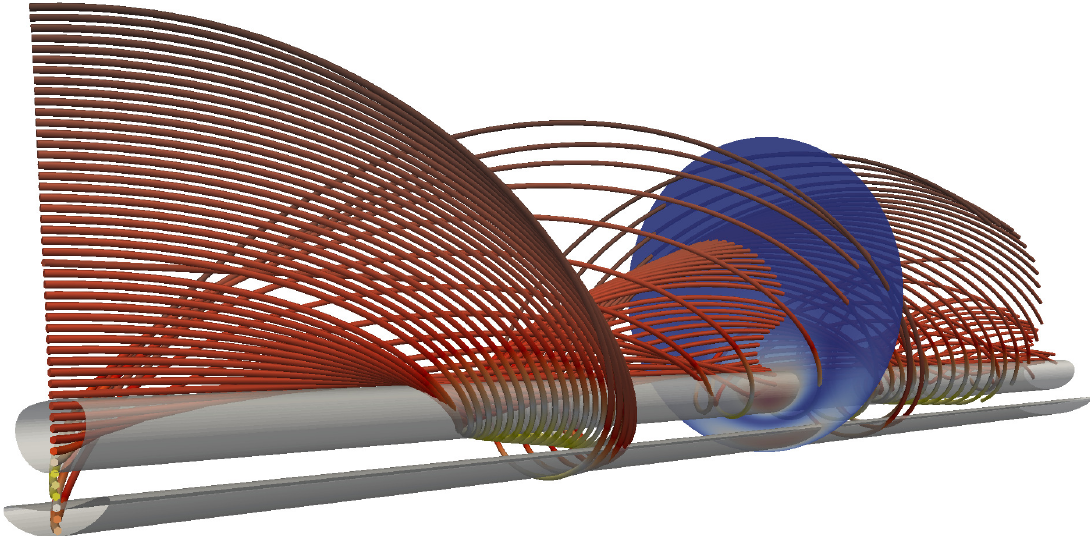}
\caption{A force-balanced flux rope with helical magnetic field lines (red) where higher density matter rests as stacked in the dips (gray isosurfaces). The cross-sectional variation (in blue) quantifies pressure throughout the flux rope, from~\cite{Blokland2011B}.}\label{f-bloklandprom}
\end{figure}

Noting that prominences can be surprisingly long-lived structures (months for quiescent prominences), their large-scale configuration must somehow involve a robust balance of forces. The same can be said for the hot coronal loops that may feature occasional or even cyclic small-scale coronal rain. As an MHD description of the coronal plasma is perfectly appropriate for all observable scales (ion gyroradii in a coronal loop are in the cm range, while observations probe at best 10-100 km), the equation of motion writes generally as
\begin{equation}
    \rho\left(\partial_t \mathbf{v}+\mathbf{v}\cdot\nabla \mathbf{v}\right)=-\nabla p +\rho\mathbf{g}+\frac{1}{\mu_0}\left(\nabla\times\mathbf{B}\right)\times \mathbf{B}+\mathbf{F}_{\mathrm{visc}} \,, \label{momeq}
\end{equation}
where density $\rho$, pressure $p$, velocity vector $\mathbf{v}$, magnetic field vector $\mathbf{B}$ are the relevant physical variables (8 scalars in total for a full 3D scenario), the external (spatially varying) gravitational acceleration is $\mathbf{g}$, the constant $\mu_0$ is the permeability, and a further `viscous' force $\mathbf{F}_{\mathrm{visc}}$ is listed that normally scales with (second order) velocity gradients. Ignoring the fact that all prominences show internal motions and temporal variability, long-lived large-scale structures must be aware of the force equilibrium denoted by
\begin{equation}
    \mathbf{0}=-\nabla p +\rho\mathbf{g}+\mathbf{J}\times \mathbf{B} \,. \label{mombal}
\end{equation}
The Lorentz force uses the current density vector $\mathbf{J}=\left(\nabla\times\mathbf{B}\right)/{\mu_0}$, and a magnetic field with a vanishing current density vector $\mathbf{J}=\mathbf{0}$ is called a potential field.
If we focus attention to a single hot coronal loop as detected in an EUV observation, and assume that the loop is actually tracing the overall magnetic field line shape, the loop obeys the projected hydrostatic balance 
\begin{equation}
0=-\hat{\mathbf{b}}\cdot \nabla p+\rho g_\parallel \,, \label{1Dforce}
\end{equation}
where the unit field line vector $\hat{\mathbf{b}}=\mathbf{B}/|\mathbf{B}|$ also enters the field-aligned gravity component $g_\parallel=\hat{\mathbf{b}}\cdot \mathbf{g}$. 
If we instead focus on the full vectorial expression~(\ref{mombal}), and assess how this force balance can be achieved within a given 3D volume of the solar corona, we realize that the entire magnetic field topology, as well as the possible presence of (volume, surface or line concentrated) currents will play a key role.  Two of the first models relevant for the force equilibrium of prominences simplify the 3D problem posed by Eq.~(\ref{mombal}) by assuming that everything is invariant along the filament axis. Both models then concentrate on 2D force balance in a vertical $(y,z)$ plane across the filament axis (along $x$), and proceed as follows:
\begin{itemize}
\item \cite{Kuperus1974} consider the magnetic topology induced by a line current above (and its mirror line current below) the photosphere: the non-force-free (i.e. $\mathbf{J}\times\mathbf{B}\neq \mathbf{0}$) magnetic field near the line gives an upward force due to an azimuthal field $B_\varphi$ around the filament, which can balance the gravitational pull. 
\item \cite{KS1957} work from the full force-balance equation~(\ref{mombal}), adopt invariance in height $z$, isothermal conditions, and reduce the problem to the notion that upward-oriented, height-self-similar dips in field lines can locally balance gravity with the then upward Lorentz force. This model does account for prominence matter weighing down field lines.
\end{itemize}
Selected examples that go beyond these pioneering models are
\begin{itemize}
    \item \cite{Petrie2007} provides full magnetohydrostatic solutions of Eq.~(\ref{mombal}) for the gravitationally stratified bulk equilibrium of a cool and dense prominence plasma embedded in a near-potential coronal field. The magnetic structure is still adopted as invariant along the filament axis $x$ ($\partial_x\equiv 0$ i.e. 2.5D), but magnetic fields have three components $\mathbf{B}=\left(f(y,z),\partial_z\Psi(y,z),-\partial_y\Psi(y,z)\right)$. This introduces the flux function $\Psi$, whose contours in the $(y,z)$ plane show projected field line shapes.
    \item \cite{Blokland2011B} make the same 2.5D assumption, but exploit the full freedom in the governing MHD equations that dictate which thermodynamic quantity becomes a function of the flux coordinate $\Psi$ only. Computing for the full internal cross-sectional variation of a flux rope topology, nested flux surfaces (contours of constant $\Psi$) can support stacked layers of denser material in a helical 3D field. An example is shown in Fig.~\ref{f-bloklandprom}, where a fully force-balanced state shows field lines (in red) and gray isosurfaces indicate where dense prominence matter levitates in the magnetic dips.
\end{itemize} 

\subsection{Addressing the energy balance}\label{s-energybal}
All models discussed above concentrate on the static force balance Eq.~(\ref{mombal}), and while they provide valuable insights into multidimensional prominence-hosting magnetic topologies, they do not fully address the additional role played by the energy equation, which writes generally in terms of the plasma temperature $T$ as follows.
\begin{eqnarray}
    {\cal{R}}\rho \partial_t T +{\cal{R}}\rho\mathbf{v}\cdot\nabla T + (\gamma-1)p\nabla\cdot \mathbf{v} &=& \left(\gamma-1\right)\left[\rho{\cal{L}}+\nabla\cdot\left(\kappa({\bar{T}}^{5/2})(\mathbf{\hat{b}} \cdot \nabla T)\mathbf{\hat{b}}\right)\right].  \label{energy}
\end{eqnarray}
In this equation, we wrote the ideal gas law as $p={\cal{R}}\rho T$, with gas constant ${\cal{R}}=k_B/\mu$ related to mean particle mass $\mu$ and Boltzmann constant $k_B$, adopted a ratio of specific heats $\gamma$ (typically 5/3) and we introduced two right-hand-side sources that generalize the adiabatic setting where they are assumed to vanish. These two sources involve (1) a net (per unit mass) heat gain-loss term ${\cal{L}}$ that involves any coronal heating mechanism and a radiative-loss term, as well as (2) the anisotropic (purely field-aligned) nature of thermal conduction. The latter has a temperature-dependent parallel conduction coefficient $\kappa({\bar{T}}^{5/2})$ with ${\bar{T}}=T/T_u$ a dimensionless temperature using a temperature unit $T_u$. Again adopting the simplification of static conditions where $\partial_t\equiv 0$ and $\mathbf{v}=\mathbf{0}$, adiabatic settings obey Eq.~(\ref{energy}) trivially, while non-adiabatic effects dictate that
\begin{eqnarray}
    0&=& \rho{\cal{L}}+\nabla\cdot\left(\kappa({\bar{T}}^{5/2})(\mathbf{\hat{b}} \cdot \nabla T)\mathbf{\hat{b}}\right) \,. \label{energybal}
\end{eqnarray}
This thermal balance must be realized throughout most of the (quiet) solar corona, and in essence tells us indirectly that whatever mechanism $h$ (per unit mass) is heating the corona, a steady energy balance requires it to counteract the optically thin radiative losses that contribute as a term $-n_en_H\Lambda(T)$ in $\rho{\cal{L}}$. 
The ionization degree will locally set how the MHD density $\rho$ relates to the number densities for electrons $n_e$ and protons $n_H$, while in fully ionized hydrogen settings we just have optically thin loss prescribed as $-\rho^2\Lambda(T)/m_p^2$ with proton mass $m_p$.
These optically thin losses, combined with conduction, then would give a heating $h$ that perfectly balances both in the form
\begin{eqnarray}
    \rho h &=& n_en_H\Lambda(T) - \nabla\cdot\left(\kappa({\bar{T}}^{5/2})(\mathbf{\hat{b}} \cdot \nabla T)\mathbf{\hat{b}}\right) \,. \label{energyheat}
\end{eqnarray}
We thereby introduced the cooling curve $\Lambda(T)$, which can be precomputed for optically thin radiative loss conditions that depend on temperature and plasma composition.
In macroscopic prominences, the prominence matter itself is at about 10000 K, and interior number densities reach $n\simeq 10^{10} \,{\mathrm{cm}}^{-3}$, while their coronal environment is much hotter and tenuous, making the assumption of purely optically thin radiative losses invalid in their interiors, and requiring consideration of full non-local coupling between radiation fields and plasma, beyond LTE assumptions. This is especially true when one needs to accurately deduce thermodynamic variations, across the transition layer known as the prominence-corona transition region (PCTR). There, thermal conduction as well as details of the varying ionization degree from external to internal prominence conditions must be accounted for.
In cool condensations $n_H$ will be a mixture of protons and neutral hydrogen, depending
on the ionisation degree which strongly depends on photoionisation processes, as detailed in the recent review
by \cite{Heinzel2024}. Recent models \citep{Reep2025} emphasize the need to handle spatio-temporal abundances and account for the more complete dependence of $\Lambda(T_e, n_e, f)$ where $f$ would quantify local, time-varying enhanced or diminished (with respect to photospheric values) $n_X/n_H$ ratios for heavier elements $X$.
Still, a long-lived, quiescent prominence structure (but even a shorter-lived, smaller-scale coronal rain blob in a loop) will try to achieve the approximate energy balance~(\ref{energyheat}). For the 2.5D force-balance models discussed and shown in Fig.~\ref{f-bloklandprom}, realizing this energy balance may select which of the thermodynamic quantities like density $\rho$, temperature $T$, or entropy $S$ becomes a flux function $f(\Psi)$, which may well evolve during its lifetime. 

A thorough consideration of both force Eq.~(\ref{mombal}) and steady energy balance Eq.~(\ref{energyheat}) was conducted in \cite{Low2012A}, by adopting a Kippenhahn-Schl\"uter model augmented with energy considerations. \cite{Low2012A} concluded that insisting on a static model, a collapse to a singularly thin mass sheet is inevitable which in turn invokes the need for resistive decay. This brings in the resistive induction equation as
\begin{equation}
    \partial_t \mathbf{B} = \nabla \times \left(\mathbf{v}\times\mathbf{B}-\eta\mathbf{J}\right)\,, \label{induction}
\end{equation}
where the resistivity $\eta$ will act to destroy the perfect frozen-in nature of ideal MHD at current concentrations.
For the `static' prominence models in \cite{Low2012A}, the PCTR coincides with tangential discontinuities in the magnetic fields and hence in localized discrete current layers, which must dissipate resistively. This causes slippage across the field and continuous cross-field mass transport \citep{Low2012}, while the interior (and exterior) of the prominence continuously restores both force and energy balance. In that sense, internal dynamics is inevitable in `static' settings.  Interestingly, \cite{Heasley1976} identified a similar `collapse' of a magnetohydrostatic equilibrium to a geometrically thin sheet, in their efforts to combine both radiative and ideal magnetohydrostatic equilibrium requirements in a Kippenhahn-Schl\"uter setup. While \cite{Heasley1976} did not invoke resistive decay, but rather focused on the intricate problem of (non-LTE) radiative, force and statistical equilibrium properties for illuminated slab models for quiescent prominences, it is clear that PCTR conditions combine both resistive and non-LTE physics aspects that must be improved in realistic MHD simulations. Combining ideal magnetohydrostatic Kippenhahn-Schl\"uter-type force considerations with 2D non-LTE radiative transfer was done in \cite{Heinzel2001}, constructing magnetically confined, but vertically infinite threads with internal, horizontal 2D structure. 
We will point out in the following sections that modern multi-dimensional prominence simulations which solve for the nonlinear evolutions dictated by both Eq.~(\ref{momeq}) and Eq.~(\ref{energy}) in various magnetic topologies indeed confirm the analytic predictions on resistive slippage, and make quiescent prominences fascinating long-lived macroscopic structures which show rich smaller-scale dynamics, trying to ensure an evolution towards the combined conditions posed by Eq.~(\ref{mombal}) and Eq.~(\ref{energyheat}). In that respect, coronal cooling and heating research is intricately linked, as argued in Section~\ref{sec-intro}. 

This suspected cooling-heating link is in line with earlier findings which assumed an ad-hoc fixed shape for a prominence-hosting field line, on which the problem reduces to the combined solution of the projected force balance Eq.~(\ref{1Dforce}) along with the static energy Eq.~(\ref{energyheat}). In \cite{Dahlburg1998} such 1D static equilibrium models for prominences were constructed, where it was argued that a footpoint-concentrated heating term $h$ would be required, as well as a centrally dipped field line portion where the prominence matter resides. \cite{Dahlburg1998} considered essentially the coronal part of the dipped loop only, adopting their boundary conditions to the upper chromosphere. 
The fact that 1D static solutions, from low chromosphere to corona, are too idealized for hot loops was pointed out by \cite{Kuin1982}, noting that the rapid temperature variation through the thin transition region effectively couples the near-isothermal coronal loop part with the chromosphere, where locally the conduction is much less efficient. A static solution that incorporates this transition region between chromosphere and corona has a temperature structure along the loop fully determined by loop length and adopted heating rate. \cite{Kuin1982} deduced 1D, time-dependent equations governing the entire loop-averaged density and temperature, approximating the chromosphere-to-corona coupling in terms of a deviation from such a static solution. This showed the possibility of limit cycle behavior for loop-averaged thermodynamic quantities, essentially bringing in time-dependency in the problem. When modeling low chromosphere to coronal regions, where also (rain or prominence) condensations reside in the coronal region, there are multiple transition regions in the domain, not only the low-lying chromosphere-to-corona one, but also one surrounding each condensation fragment. Together with the findings that the PCTR invokes localized current concentrations, realistic models must handle all three nonlinear, time-dependent laws posed by Eqs.~(\ref{momeq}), (\ref{energy}) and (\ref{induction}). 

\subsection{Dip-only model considerations}\label{s-diponly}
\begin{figure}[ht]
\centering
\includegraphics[width=\textwidth]{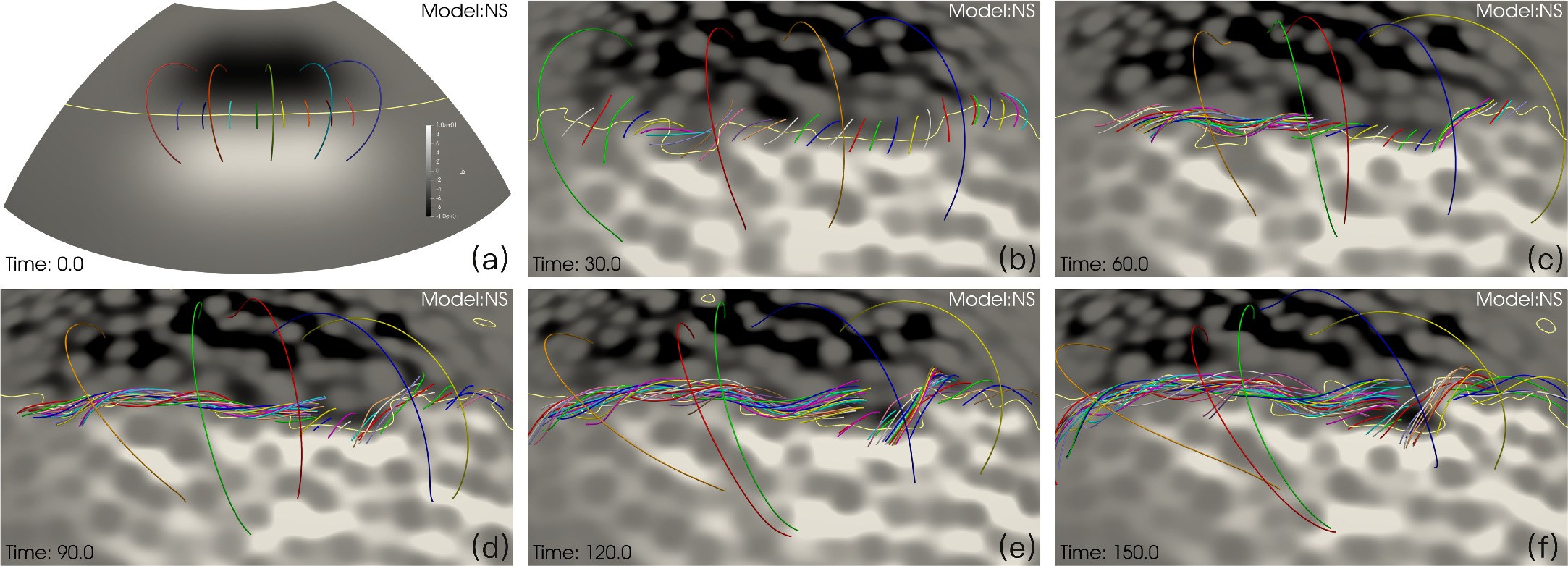}
\caption{The magnetofrictional evolution of a large spherical dipole region (panel (a)), when subjected to synthetic supergranular motions that impact the magnetogram (panels (b) through (f)). Eventually, a large-scale nonlinear force-free field that has a magnetic flux rope topology emerges (panel (f)). From \cite{CHUN2024}.}\label{f-chunfig}
\end{figure}

Since all prominence models invoke locally concave-upward, or dipped, field lines in the tenuous coronal environment, it has become customary to solve for purely force-free 3D magnetic fields, obeying $\mathbf{J}\times\mathbf{B}= \mathbf{0}$. One then labels the concave upward field line segments as ``prominence" sites, even though there is no (hot or cold) plasma in these models at all. This approach boils down to adopting a uniform, zero beta approximation, but the variation of plasma beta $\beta=2 \mu_0 p/B^2$ when a prominence or any kind of condensation is incorporated can reach unit values \citep{Heinzel1999}. 
An example of such a nonlinear force-free field configuration is shown in Fig.~\ref{f-chunfig}, taken from \cite{CHUN2024}, where the self-consistent formation of a polar crown filament configuration was demonstrated. The authors use a magnetofrictional approach to solve for the gradual evolution of a bipolar arcade in a large spherical domain, with an initial (top left panel in Fig.~\ref{f-chunfig}) east-west PIL at $55^\circ$ latitude. Magnetofrictional modeling uses the induction Eq.~(\ref{induction}) with a driving velocity that ultimately settles the magnetic field $\mathbf{B}$ on a force-free configuration. At each stage, a nonlinear force-free field configuration is established numerically, while the bottom (photospheric) boundary is driven by a time-varying realistic supergranular velocity field \citep{Liu2022}. \cite{CHUN2024} investigate the role of latitude-dependent Coriolis forces, of changes in the original PIL orientation, and a dependence on hemisphere location. The particular model shown in Fig.~\ref{f-chunfig} (panel (f)) convincingly reproduces a large magnetic flux rope structure with extended dipped regions where prominence matter may be hosted, as needed for polar crown filaments.

By coloring the dips in magnetic nonlinear force-free fields as prominence sites, we are also assuming that all visual thermodynamic structuring is always perfectly field-aligned. Theoretically, this is not required by Eq.~(\ref{mombal}). The only thing that follows directly is the stratification along field lines from Eq.~(\ref{1Dforce}). Further specifying to 2.5D (say, $x$-invariant as in Fig.~\ref{f-bloklandprom}) conditions, this becomes a relation that links pressure and density variations along constant magnetic flux $\Psi(y,z)$ contours. The full 3D whole-prominence fine structure models presented by \cite{Gunnar2015,Gunnar2015B} instead start from a force-free magnetic topology and augment it with purely field-aligned plasma that obeys hydrostatic equilibrium in selected dipped sections. By adopting an observationally inspired cross-sectional extent (1000 km) of each plasma-loaded field line, and invoking a related criterion to label field lines as independent, a heuristic fine-structured model is constructed that can be used to confront emission and absorption properties of prominence and filaments, respectively. This model was adjusted \citep{Gunnar2015B} to mimic time-evolution, by changing the underlying nonlinear force-free field in a boundary-driven magnetofrictional setting similar to Fig.~\ref{f-chunfig}. Synthetic H$\alpha$ images could reproduce key features in observations, such as barb creation, or dark voids. Despite these successes, the force-free backbone field adopted ignores how condensed matter may change the magnetic topology locally, and does not address finer-scale dynamics.

In what follows, we highlight efforts that go beyond purely force-balanced models, as we concentrate on time-dependent, multi-dimensional momentum and energy evolutions dictated by Eqs.~(\ref{momeq}) and (\ref{energy}). This shifts our attention to those physical processes that actually control prominence (or rain) formation itself, and brings in prominence internal dynamics. This calls for a thorough understanding of linear stability theory.

\section{Linear MHD theory and the instability zoo}\label{s-zoo}

\subsection{Ideal MHD spectroscopy}\label{ss-ideal}
In the previous section, we noted that many model efforts reduce the 3D, fully nonlinear, time dependent problem posed by momentum and energy Eqs.~(\ref{momeq}) and (\ref{energy}) to a balance of forces and the overall energy budget, as expressed in Eqs.~(\ref{mombal}) and (\ref{energyheat}). These force and thermal balance equations describe the background or ``equilibrium" configuration that must hold on average in long-lived structures like prominences. However, to understand how they {\em form}, or to study what drives {\em small-scale internal dynamics}, time-dependence must be considered. Linear MHD theory, which quantifies this time-dependent aspect by computing all normal modes with a time-dependence given by $\exp(-i\omega t)$, provides us with direct information on the overall stability of the equilibrium against small perturbations. Computing all the normal modes is termed `MHD spectroscopy' \citep{goed2019}, where eigenvalue-eigenfunction pairs obey the full set of linearized MHD equations. The spread of the eigenvalues throughout the complex-valued $\omega$-plane dictates how an initial value problem will evolve, as shown analytically in ideal MHD by \cite{Hans1998}. A static configuration with one-dimensional variation (such as a stratified atmosphere or radially varying loop) has Alfv\'en and slow frequencies that form (stable wave, i.e. with real-valued $\omega=\omega_R$) continuous ranges which play a decisive role in the time evolution, apart from all discrete normal modes. A discrete mode is an isolated $\omega$ value that corresponds to more globally varying eigenfunctions (although they can occur in sequences where fine-scale eigenmode structure develops at specific locations), while a continuum mode is like a truly singular local resonance. In a static, 1D MHD equilibrium setting, discrete normal modes may be purely growing (or damped), with imaginary values for their eigenfrequencies $\omega=i\nu$, providing us with growth rates $\nu$ for instabilities. Countless studies of specific equilibrium configurations have meanwhile identified many routes to instability that all relate to discrete modes, including all well-known hydrodynamic instabilities such as effective-gravity-driven Rayleigh-Taylor instability (RTI), shear-flow-driven Kelvin-Helmholtz instability (KHI), or the zoo of MHD instabilities that can act as triggers for violent flux rope eruption processes in the solar atmosphere \citep[see, e.g.][]{Rev2019}. This ideal MHD spectrum of waves and instabilities is governed mathematically by (up to two) self-adjoint operators, which return in a field-theoretical treatment of the time-reversible ideal MHD equations \citep{Keppens2016}. This insight makes the eigenmode spectrum physically relevant at any time instance of a fully nonlinearly evolving configuration, so extends its validity beyond analyzing modes for static and thermally balanced settings. 

\subsection{From thermal instability to thermal continuum}\label{ss-TI}
Deviations from ideal, time-reversible MHD are in the right-hand-side terms in the nonlinear energy Eq.~(\ref{energy}), which quantify instantaneous non-adiabatic effects at play such as optically thin radiative loss and thermal conduction. \cite{Parker1953} was the first to realize that this introduces a rather specific linear instability route: one fully dictated by heat-loss processes encoded in ${\cal{L}}$. In his analysis, \cite{Parker1953} considered a partial linearization of Eq.~(\ref{energy}), just focusing on the terms ${\cal{R}}\partial_tT=(\gamma-1){\cal{L}}$, and pointed out how the temperature dependence of ${\cal{L}}$ can cause thermal runaway. This thermal instability route was further analysed in depth by \cite{Field1965}, who already argued for its relevance in explaining prominence formation, but also in astrophysical settings beyond the solar realm. \cite{Field1965} presented dispersion relations for uniform hydro and MHD settings, included the effects of anisotropic thermal conduction, and analysed condensation eigenmodes for fully stratified atmospheres. Also in laboratory plasmas, thermal instability has been realized as being the cause of radiative condensations (so-called ``marfes") that form in tokamak edge plasmas \citep{Drake1987}. In modern tokamaks equipped with divertors, these condensations also occur near the X-point topologies, where they are termed X-point radiators, and they are affected by impurity-driven radiative losses \citep{Stroth2022}.
The theory of thermal instability has been revisited in a number of papers, such as \cite{Balbus1986,Waters2019,Falle2020}. 
However, the emphasis in these works was on dispersion relations for uniform media and on discrete condensation eigenmodes, which would demonstrate rather global eigenfunction variations. Observations of filament fine-structure as well as coronal rain itself rather confront us with truly localized variation, so any formation mechanism should preferentially explain this substructure during formation. 

Such a more local response can be linked to the findings of 
\citet{vdl1991}, who discovered that a static MHD, infinite straight cylinder (flux rope) model allows for a thermal continuum, in addition to the slow and Alfv\'en continuum wave ranges discussed earlier. This thermal continuum can be purely unstable and couples to the slow modes, while all discrete modes can be influenced by non-adiabatic processes \cite[e.g. discrete Alfv\'en modes can turn overstable and form prominences as argued in][]{Keppens1993}. This thermal continuum generalizes the thermal instability analysed by \citet{Parker1953} and \citet{Field1965} and likely explains the coronal rain phenomenon in loops \citep{KeppensTC2025}, where again in-situ localized runaway is witnessed. Typical solar coronal (atmosphere or loop) conditions with optically thin radiative losses can hence show fragmentary condensations, due to an exponential growth of the thermal continuum or associated thermal discrete eigenfunctions. In \citet{vdl1991B} (for cylinders) and \citet{vdlslab1991} (for slabs), the linear MHD analysis accounted for anisotropic thermal conduction, with both parallel as well as finite perpendicular conduction $\kappa_\perp$ (i.e. across the magnetic surfaces). Accounting for small, but non-zero $\kappa_\perp$ implied that highly fine-structured eigenfunctions were found for thermal modes, as a natural explanation for the fine-structure in prominences. 

\citet{ireland1992} generalized the MHD normal mode analysis from \cite{vdl1991B} for the cylindrical case to include finite resistivity $\eta$, and confirmed that (1) $\kappa_\perp=0=\eta$ has a stable Alfv\'en continuum next to a coupled (third order in $\omega$) slow-thermal continuum; (2) when $\kappa_\perp\ne 0$ while $\eta=0$, we still have an Alfv\'en continuum, but now an isothermal slow continuum and a dense set of discrete (quasi-continuum) thermal modes; and (3) $\kappa_\perp=0$ and $\eta\ne 0$ removes both the Alfv\'en and the slow continuum and shifts the thermal continuum, while replacing the original thermal continuum range by densely packed discrete modes (a quasi-continuum). 

The fact that linear MHD theory that incorporates all relevant conduction and resistive effects always produces nearly singular, fine-structured unstable thermal eigenmodes is consistent with the omnipresence of multithermal fine structure throughout the chromosphere to the corona. 
For the laboratory counterpart of radiative condensations called ``marfes" in tokamaks \citep{Drake1987}, a clear link with thermal continua in axisymmetric multi-dimensional tokamak equilibria has been demonstrated by \citet{DePloey2000}. Note that \cite{DePloey2000} quantified these continua for an axisymmetric 2.5D equilibrium governed by Eq.~(\ref{mombal}) with gravity ignored, where the three orthogonal directions $\nabla p$, $\mathbf{J}$ and $\mathbf{B}$ define the nested, donut-shaped flux surfaces $\Psi$. Most recently, the thermal continuum for coronal flux ropes (as straight cylinder structures) with both helical magnetic and flow fields in the configuration was analysed by \cite{Hermans2024}, with an excellent agreement between analytic predictions on (Doppler shifted) continua, discrete thermal modes and numerical eigenmode computations.

\subsection{Linking thermal instability with in-situ condensations}\label{ss-link}

\begin{figure}[ht]
\centering
\includegraphics[width=\textwidth]{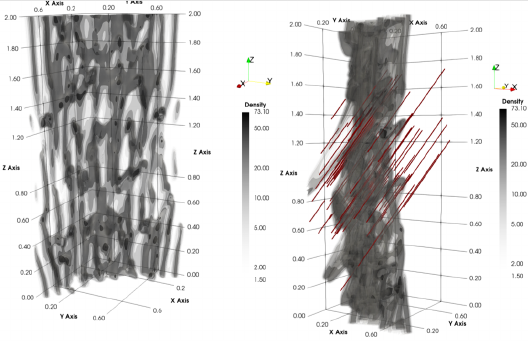}
\caption{The 3D density structure resulting from TI in a local coronal volume. The view at right also shows (in red) the background (nearly) homogeneous magnetic field, clearly not aligning with the condensation fine structure. From~\cite{Claes2020}.}\label{f-claesfig}
\end{figure}

The most convincing means to show that thermal instability is the natural explanation for any in-situ forming condensations, is to simulate this process ab-initio in a uniform magnetized medium that is subject to radiative losses. This was done for solar coronal settings in a series of papers \citep{Claes2019,Claes2020,Hermans2021}, where \cite{Claes2019} revisited the thermal instability (TI) theory from \cite{Field1965} and its dispersion relation by quantifying all eigenfunctions for homogeneous MHD media. Using this information, one can simulate a local box of coronal plasma of which we know the exact (complex-valued) eigenmodes and growthrates for instabilities, and clearly link the linear growthrate findings with any runaway condensation. This was done in 2D settings by \cite{Claes2019} and \cite{Hermans2021}, where the latter showed the influence on the condensation onset of the cooling curve $\Lambda(T)$ in Eq.~(\ref{energyheat}). One recurring finding from these local-box nonlinear evolutions was that condensations originally orient perpendicular to the local magnetic field, to then become subject to thin-shell instabilities that cause their disruption. By also extending the simulations to full 3D, \cite{Claes2020} pointed out that the thin-shell instabilities introduce fine-structure and cause condensation fragments to `follow' the field lines, although the strands that form do not align with the uniform magnetic field adopted. This is shown in Fig.~\ref{f-claesfig}, where the density structure due to TI and its consecutive thin-shell disruption is illustrated. This finding may well have major implications, as filament observations usually infer the (unmeasurable) magnetic field topology from the filament strands. 

The works discussed thus far always adopted a single-fluid MHD viewpoint on the corona, with local radiative losses as quantified in a cooling table $\Lambda(T)$ appropriate for optically thin settings. As soon as an actual in-situ condensation is established, its associated density-temperature variation may well violate some of the inherent assumptions in this approach. For instance, the degree of ionization may vary drastically across the condensation \citep{Heinzel2024}. The linear theory for TI in plasma-neutral mixtures, as well as the link with nonlinear runaway condensation formation in local coronal boxes, was presented in \cite{Beatrice2024}. The (temperature dependent) ionization degree and coupling between neutrals and charges influence the thermal mode growth. When ionization-recombination is incorporated, the neutral fraction internal to the condensation can increase substantially from its negligible value in the hot corona prior to condensation onset.

\subsection{Re-inventing linear thermal instability theory}\label{ss-wheel}

Although the theoretical description of the thermal instability (TI), along with its physical consequences, were introduced and discussed at length (37 pages) in the seminal work by \cite{Field1965}, there are many works that in essence merely revisit its findings. This is also true for \cite{Claes2019}, which emphasized the polarizations  - or eigenfunctions - of the perturbations, needed to selectively initiate a nonlinear MHD simulation and follow non-adiabatic MHD modes (slow, Alfv\'en, fast or entropy) into the nonlinear regime.  Worth pointing out here is the unfortunate confusion that can arise when revisiting the established theory on TI for uniform plasmas. This is particularly evident for the much simpler case of a uniform, radiating gas, where a pure hydrodynamic description suffices. This case is fully derived in Section II of \cite{Field1965} and it is well-known that a uniform, non-adiabatic gasdynamic medium, where mass conservation, momentum equation~(\ref{momeq}), and energy equation~(\ref{energy}) are linearized assuming plane-wave perturbations of the form $\exp(-i\omega t+i\mathbf{k}\cdot\mathbf{x})$, yields a third order polynomial for the dispersion relation $\omega(k)$ as
\begin{equation}
    \omega^3-i\omega^2\frac{(\gamma-1){\cal{L}}_T}{{\cal{R}}}-\omega c_0^2 k^2-i(\gamma-1)\left(\rho_0{\cal{L}}_\rho-T_0{\cal{L}}_T\right)k^2=0 \,. \label{druni}
\end{equation}
Linearization is here done about a static and uniform medium (constant density $\rho_0$ and temperature $T_0$, setting the uniform sound speed $c_0$) and for simplicity, we omitted the added effect of thermal conduction (this is present in the equivalent eq.~(15) from \cite{Field1965}). The medium is also assumed to achieve a thermal equilibrium, i.e. ${\cal{L}}_0=0$ for the equilibrium, implying that heating/cooling effects perfectly balance. The coefficients of this third order polynomial contain partial derivatives of the heat-loss function ${\cal{L}}_T=\frac{\partial{\cal{L}}}{\partial T}$ and ${\cal{L}}_\rho=\frac{\partial{\cal{L}}}{\partial \rho}$, which again need to be evaluated for the equilibrium. We note in passing that all these restrictive assumptions (uniform medium, or adopting a perfect thermally balanced state) have been relaxed and generalized in \cite{KeppensTC2025}. As fully described in \cite{Field1965}, where it uses complex growthrates $n=-i\omega$, this governing third order polynomial (even with conduction effects incorporated) allows a complete mathematical categorization of all possibilities for its three roots: depending on its coefficients, we can visualize in an appropriate state-space (Fig.~1 of \cite{Field1965}) where we find three real roots, or where we instead have one real root and a complex conjugate pair of roots. More importantly, we can predict when we have a purely growing instability ($n$ real and positive, such as the TI) or when we encounter overstable (i.e. growing amplitude, moving) eigenmode pairs: these are then overstable acoustic modes, which come in a forward-backward pair. The latter may also coalesce and turn into two real roots, so we may encounter up to three real $n$ solutions. In \cite{Waters2019}, this is revisited and instead of the complete discussion in the statespace as used by \cite{Field1965} (his Fig.~1), an equivalent discussion emphasizing the wavelength dependence and the ratio of the coefficient of the second order ($\sim\omega^2$) and constant term from Eq.~(\ref{druni}) is provided. It is thereby unfortunate that the cases where up to three modes can be unstable are termed `entropy', `slow isochoric', and `fast isochoric', at least from the viewpoint that slow and fast modes have a clear meaning in MHD, while the hydro case only allows for entropy and (potentially coalesced) acoustic modes.  A similar cautionary comment applies to the work by \cite{Kolotkov2020}, who derive exactly the same dispersion relation as above (i.e. a revisit of eq.~(15) from \cite{Field1965}), but this time calling it representative for a thermal mode and a pair of slow MHD waves on an `infinite strength magnetic field'. This `infinite magnetic field' also renders the plasma beta parameter zero, and hence simplifies wave dynamics to a purely field-aligned, pressure and energy mediated description, so a hydro setting is sufficient. \cite{Kolotkov2020} then assumes there is freedom in the heat-loss function $\rho{\cal{L}}= \rho h - n_en_H\Lambda(T)$ from equation~(\ref{energyheat}), by taking a heating function $h \propto \rho^a T^b$ with $a$ and $b$ left as free parameters. The consequences of varying $a,b$ are then illustrated, arguing for an `active medium' scenario \citep{Kolotkov2021} where `slow MHD' waves are influenced by thermal misbalance and vice-versa. One may question the adopted density-temperature dependence of the unknown heating $h(\rho,T)$, especially when non-uniformity of the background is incorporated, although the observation stands that wave damping or heating encodes information on its variation.

Even more caution is due when following up on the recent claim by \cite{Waters2025} that `a seperate much simpler linear instability' exists, which is the isochoric mode initially discovered by \cite{Parker1953}.  Only linearizing the temperature evolution equation, \cite{Parker1953} argued that any setup where a positive ${\cal{L}}_T$ is present (note our difference in sign convention for ${\cal{L}}$ with earlier works), may lead to runaway with exponential growth 
$\exp((\gamma-1){\cal{L}}_T \,t)/{\cal{R}}$. \cite{Field1965} already discussed that this isochoric criterion can be obtained from the general dispersion relation~(\ref{druni}), in a fairly specific limit which is `rarely met in practice'. Mathematically we could set $k=0$ and obtain this mode from the dispersion relation~(\ref{druni}), but this implies considering an infinitely homogeneous mode that everywhere modifies temperature (up to infinity) in a uniform medium. It was clarified in \cite{KeppensTC2025} that in any non-homogeneous generalization, this is a spurious mode (mathematically corresponding to a range of apparent singularities of the governing ordinary differential equation or ODE), while a true continuum is found that generalizes the TI to its more relevant stratified settings. That thermal continuum (TC) is a genuine singularity for the ODE, and has physically relevant, ultra-localized eigenfunctions, a fact familiar from MHD spectroscopy where Alfv\'en and slow continua play similar roles in charting and organizing the natural eigenoscillations of a plasma. Exact eigenmode computations for 1D semi-circular loops (the topic of section~\ref{s-1d}) also confirmed the absence of this spurious mode, making its discussion irrelevant when interpreting nonlinear evolutions. In that context, opinion papers like \cite{Klimchuk2019}, arguing for a clear distinction between thermal non-equilibrium (TNE) and thermal instability, are missing the point that any instant of an actual nonlinear, time-evolving state can be diagnosed spectrally, where eigenmodes (stable and unstable) are fully accounting for spatial variation and eigenfrequencies quantify growthrates that may rival those of the evolving background. Situations may arise where it is impossible to reach a perfect stationary state (despite having e.g. constant-in-time heating applied to a loop) and this is stated as the characterizing property of a TNE state. This limit cycle behavior (irrespective of condensation formation) is well studied and understood from 1D loop models, as discussed in section~\ref{s-1d}, but every condensation that actually forms and becomes a rain blob or prominence is well associated with a manifestation of the TI. 

\subsection{Further MHD spectroscopic findings}\label{ss-further}

As summarized in Section~\ref{ss-TI}, the thermal instability (TI) of a radiative uniform plasma gets replaced by an entire thermal continuum (TC) when we analyse the MHD spectrum of inhomogeneous media, like plane-parallel layers \citep{vdlslab1991} or cylindrical flux tubes \citep{vdl1991B,Hermans2024} with internal equilibrium variations. This TC is also present in hydrodynamic settings accounting for gravitational stratification, besides the radiative losses, as demonstrated in \cite{KeppensTC2025}. This includes the fixed-magnetic-field-line view on a coronal loop, where the basic state is one governed by projected, purely field-aligned motion and (mass and energy) conservation laws that may be aware of the area-variation along the loop. TC can even be quantified for thermally imbalanced states, i.e. states that deviate from the equality~(\ref{energyheat}). Since we already found that in-situ condensations (such as those forming in local coronal volumes as shown in Fig.~\ref{f-claesfig}) directly relate to the TI, we can expect that the TC modes play a role in actual time-evolutions of stratified, magnetized atmospheres as well as along coronal loops. How exactly the TC impacts linear as well as nonlinear time-dependent evolutions is yet to be appreciated fully, and is a topic of active research. An example of such study is the work by \cite{Jordi2025}, which considers the intricate link between resistive tearing modes and the TC for a force-free Harris sheet equilibrium, using a combined linear MHD spectral and nonlinear simulation approach. Such studies are inspired by findings from multi-dimensional nonlinear MHD simulations that include resistivity and optically thin radiative losses, where intricate interplay between resistive tearing and TI/TC condensations was demonstrated: 2D island chains could form trapped condensations within plasmoids \citep{Sen2022} and 3D reconnecting current sheets were shown to drive condensation formation in their vicinity \citep{Sen2023}. While these studies ignored gravity, they did represent typical coronal (temperature, density and plasma beta) conditions, hinting that these mode interactions must be at play in fully realistic settings as well.

\begin{figure}[ht]
\centering
\includegraphics[width=\textwidth]{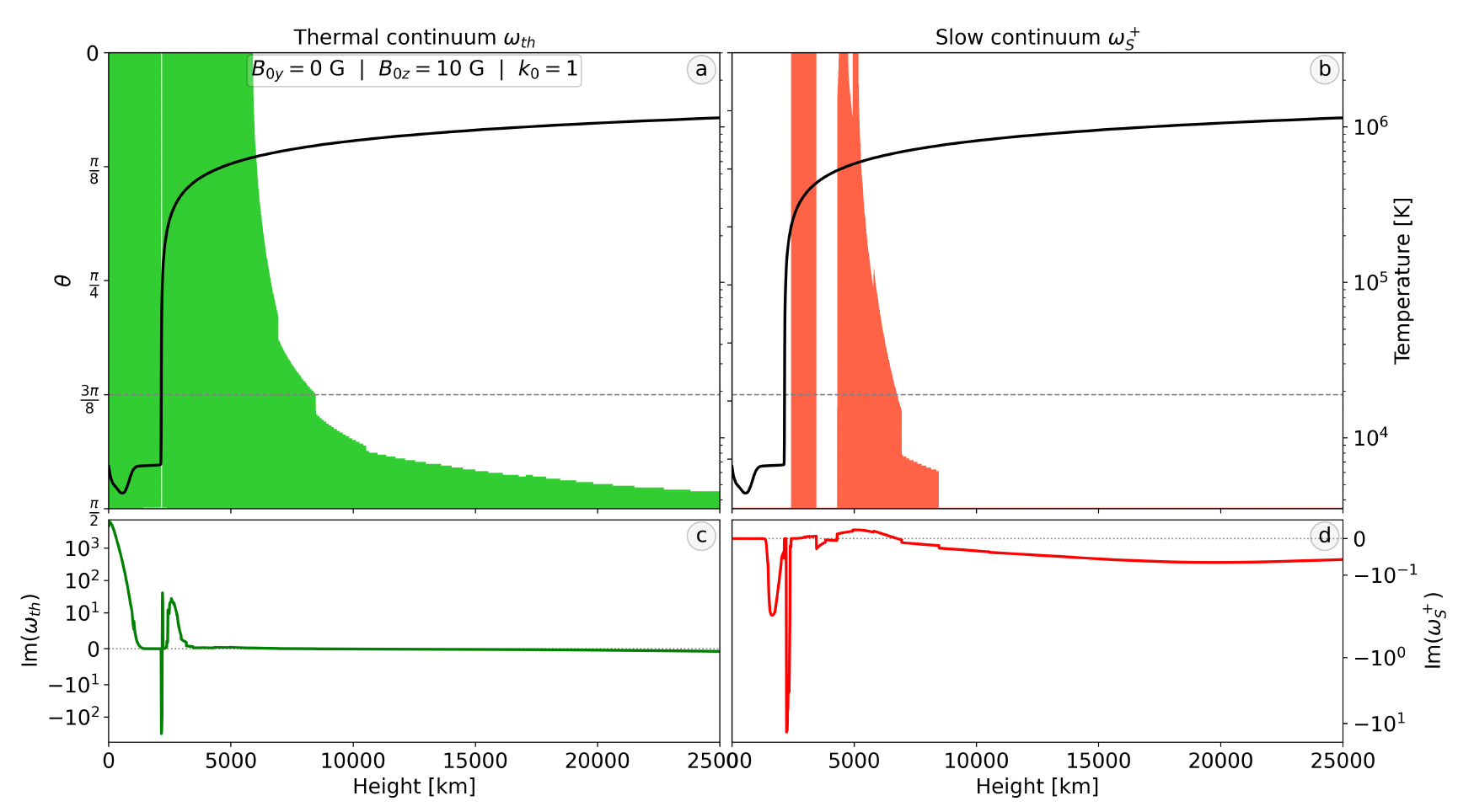}
\caption{For a realistic solar atmosphere (with temperature-density as function of height) from photosphere to corona, augmented with a horizontal, uniform 10 G field, a full spectroscopic analysis identifies all shaded regions as containing unstable thermal (top left panel a) or overstable slow (top right panel b) eigenmodes. All panels have height varying horizontally, and the top panels also vary the angle $\theta$ between the horizontal wavevector and the magnetic field. At the specific angle indicated by a dashed horizontal line in panels a and b, the lower panels show the growthrates of thermal (bottom left panel c) or slow (bottom right panel d) modes. From~\cite{Claes2021}.}\label{f-claesfig8}
\end{figure}

In that context, an important insight from modern MHD spectroscopy was presented by \cite{Claes2021}, who computed all linear eigenmodes for gravitationally stratified, plane-parallel atmospheres with possibly height-varying, sheared horizontal magnetic field. After showing how the spectrum changes from adiabatic to non-adiabatic cases, this study performed MHD spectroscopy in a fully realistic solar atmosphere. While the adiabatic case is analytically tractable when taking a uniform magnetic field and constant density scale height, allowing to recover and extend work on stable magneto-atmospheric waves~\citep{Nye1976}, the solar atmosphere case had a realistic photosphere to coronal variation from 0 to 25 Mm height, including the transition region variation. Computing all MHD eigenmodes in thermally balanced setting (obeying Eq.~(\ref{energyheat})), it was found that unstable thermal modes prevail through the chromosphere and into the corona, while overstable slow modes manifest themselves mostly in the low corona. Figure~\ref{f-claesfig8} demonstrates this by showing the height regions identified as unstable, for both thermal and slow modes. The full non-adiabatic MHD eigenmode spectra can also be computed for purely chromospheric, or purely coronal regions, and it showed how thermal instability is nearly unavoidable. This aligns with the observational fact that the chromosphere is highly structured and filled with chromospheric fibrils, and in a sense `explains' the multiphase nature of the solar corona (showing rain and prominences). 

That the linear MHD spectrum not only explains in-situ condensation formation, but also signals actual internal dynamics, was pointed out in \cite{Blokland2011}, who computed the entire continuous spectrum for the prominences in flux ropes as shown in Fig.~\ref{f-bloklandprom}. In such 2.5D force-balanced settings governed by Eq.~(\ref{mombal}), the nested flux contours that form the flux rope may have unstable continua, which now quantify modes that may grow near an isolated flux contour. If, during any time in the flux rope evolution, the density ends up being constant along the flux contours, it turns out that instability may arise when the flux-surface projected Brunt-V\"ais\"al\"a frequency $N^2_{\mathrm{BV,pol}}$ becomes negative, as given by
\begin{equation}
    N^2_{\mathrm{BV,pol}}=-\left[\frac{\mathbf{B}_{\vartheta}\cdot\nabla p}{\rho B}\right]\left[\frac{\mathbf{B}_{\vartheta}\cdot\nabla S}{\gamma S B}\right] \,, \label{cci}
\end{equation}
where $S$ denotes the entropy.
The $\mathbf{B}_{\vartheta}$ in this equation is the magnetic field (component) as quantified in the non-orthogonal, straight field line coordinate system that is fully known once the solution to the force balance Eq.~(\ref{mombal}) is available. This Convective Continuum Instability (CCI) is similar to those identified earlier for axisymmetric accretion tori or in toroidally rotating tokamak scenarios, and has meanwhile been identified in fully nonlinear, multi-dimensional prominence forming flux rope scenarios~\citep{Jack2021}. It leads to a rapid matter redistribution occurring along the flux contour, and acts as a seed for consecutive thermal instability. Similar links between coronal rain formation and the linear stability criteria expressed in Eq.~(\ref{cci}) were demonstrated for 3D coronal rain formation in \cite{Moschou2015}. 

These findings on the possibility for unstable continuum modes (both TC and CCI modes) are to be contrasted with stability quantification for discrete Rayleigh-Taylor or interchange modes, which usually use growthrate estimates that adopt simple two-layer equilibrium models. In reality, the richness of the (evolving) MHD spectrum for a magnetic flux rope or arcade system will impact its nonlinear evolution in ways which we do not fully understand just yet. We now turn our attention to insights gathered from purely nonlinear simulations based on the time-evolving governing equations.

\section{Findings from 1D nonlinear hydro evolutions}\label{s-1d}

If we ignore all the intricacies of multi-dimensional MHD, we can reduce our problem to a 1D hydrodynamic view of what happens along a pre-fixed coronal loop or arcade field bundle. Then, the given field line shape dictates the flow along it, where we solve for a velocity $v_\parallel(s,t)=\hat{\mathbf{b}}\cdot\mathbf{v}$ and $s$ is the coordinate along the loop. This may incorporate an area-variation $A(s)$, related to a field strength variation $B(s)$ that ensures flux conservation as in $B(s)A(s)=\mathrm{constant}$. In any case, we then have mass conservation
\begin{equation}
    \partial_t\rho+\nabla\cdot( \rho\mathbf{v})=\partial_t\rho+\frac{1}{A}\partial_s(A\rho v_{\parallel})=0 \,, \label{masscons1D}
\end{equation}
along with the field-line projected variant of Eq.~(\ref{momeq}) as well as the energy equation~(\ref{energy}). Although this eliminates all MHD processes (as it projects away the Lorentz force), the field line shape can be taken arbitrary and may include a dipped section such as those considered in the dip-only models discussed in Section~\ref{s-diponly}. The linear waves retained in such models are only non-adiabatically modified pressure-driven p-modes, while the TC remains a robust ingredient of the (hydrodynamic) normal mode spectrum \citep{KeppensTC2025}. This, in combination with steep gradients in temperature and density at transition region heights, makes the reduction to 1D still relevant and challenging to compute, especially given the limit cycle possibilities for loop-averaged quantities, as pointed out by \cite{Kuin1982}.

\subsection{Evaporation-condensation models}\label{ss-evap}
That such a 1D loop model can indeed form a stable prominence near the loop top was shown convincingly by \cite{Mok1990}, where an initially semicircular field line was heated near its legs with an exponentially decaying heating function. For this specific configuration, the decay distance must be shorter than 12 \% of the total length of the loop to trigger condensation. This ultimately triggered the formation of a TI-induced condensation, although the link with the stability criteria of \cite{Field1965} was not explicitly made. The loop connected the deep chromosphere to the corona, and the shape of the field line was altered to have an imposed central dip, once the density of the condensation exceeded some threshold. This allowed the condensation to persist stably in the corresponding gravitational well.

Analogous simulations of a loop with a prefixed central dipped region were done by \cite{Antiochos1991}, where the authors emphasized how a uniform weak heating plus an additional strong heating localized near loop footpoints with an exponentially decaying heating function can trigger condensation to form a chromospheric-density prominence. The process of chromospheric evaporation from footpoint-localized heating was shown to be essential for prominence formation. This evaporation-condensation scenario was later revisited with fully grid-adaptive 1D simulations in \cite{Antiochos1999}, allowing to reach quasi-steady-state conditions with a central condensation of size 5000 km.

\begin{figure}[th]
\centering
\includegraphics[width=\textwidth,height=12cm]{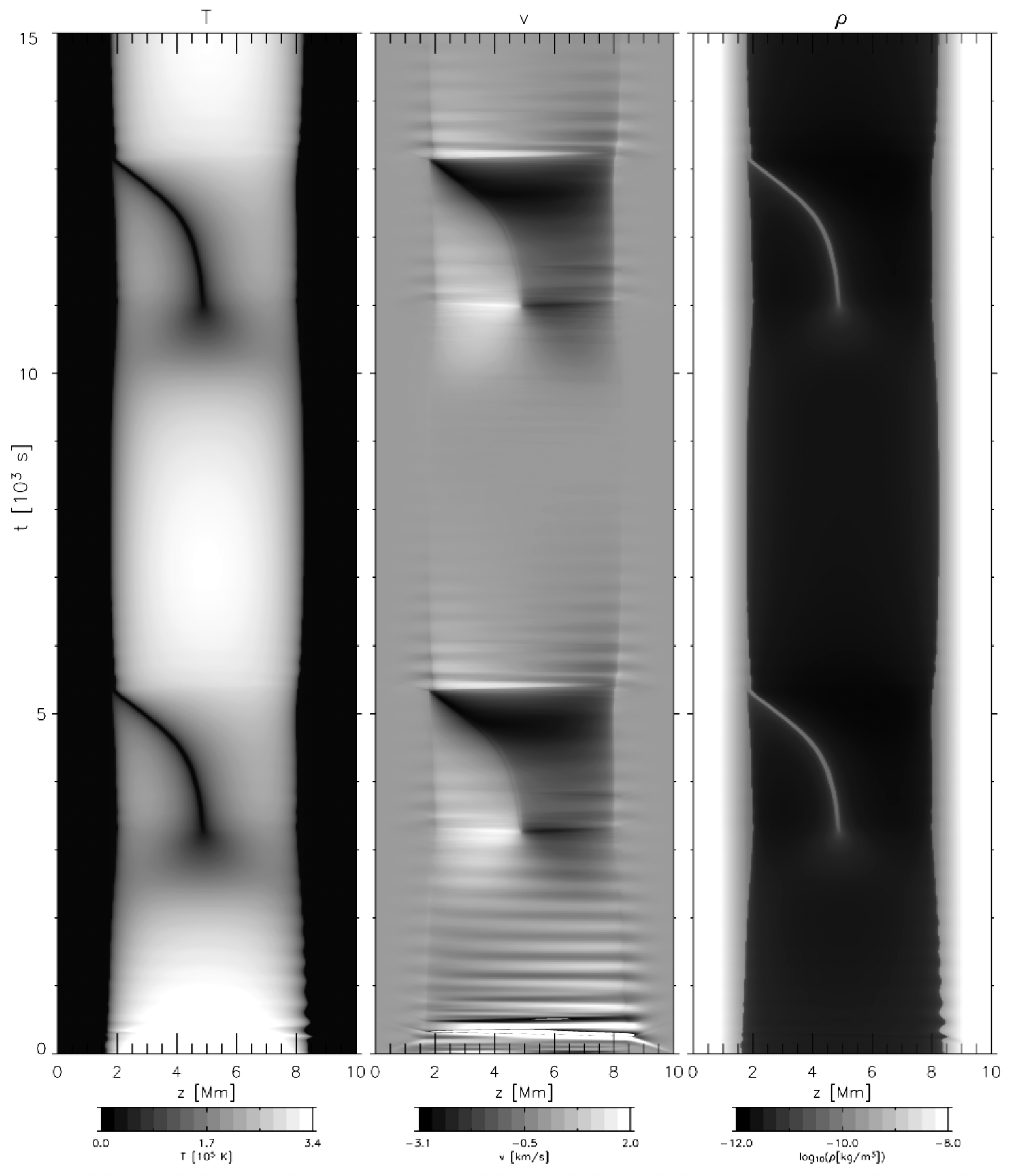}
\caption{Temperature, velocity and density evolution in a 10 Mm semi-circular loop, showing cyclic condensations that rain down. Each variable is shown as a function of the loop-aligned horizontal coordinate, versus time (vertically). From~\cite{Muller2003}.}\label{f-mullerfig}
\end{figure}

This evaporation-condensation mechanism was subsequently investigated under asymmetric heating conditions, in similar loop settings \citep{Antiochos2000}. The loop itself extended 320 Mm, showing a long, shallow dipped section. A cyclic behavior with condensation formation, movement, and destruction was witnessed, and the authors coined this time-dependent behavior `Thermal Non-Equilibrium' or TNE. This inherent time-dependent evolution - despite having no time-dependent heating imposed on the loop - was backed up with order-of-magnitude estimates of the pressure imbalance across a condensation, causing the condensation to move. Condensations formed due to a $\Lambda(T)\propto \bar{T}^{-b}$ power law for the cooling function. This power index $b$ plays a determining role, fully consistent with the linear TI and TC theory, where the partial derivatives of the heat-loss function ${\cal{L}}$ with respect to density ${\cal{L}}_\rho$ and temperature ${\cal{L}}_T$ enter the dispersion relation (our Eq.~(\ref{druni})) and set the entire TC range. Based on their 1D model, the authors argued that the TNE state of a coronal loop forming condensations is always nearly in hydrostatic equilibrium (i.e. obeying Eq.~(\ref{1Dforce})), but fails to achieve thermal balance as expressed in our Eq.~(\ref{energyheat}). As recently shown in \cite{KeppensTC2025}, this thermal imbalance can be fully incorporated in linear spectroscopy and then clarifies the role of TC, as well as of non-adiabatically modified p-modes, in the observed evolution. Curiously, \cite{Antiochos2000} even argued that magnetic dips hosting prominence matter must be preexisting, since a condensation would tend to move away from the loop apex, before it can weigh it down. Of course, prominence-weight-induced dips can only be realized in multi-dimensional MHD settings \citep{Xia2012}, and are at the very heart of the \cite{KS1957} model to ensure internal force balance in prominences. 

The same model ingredients were used in semi-circular, short (10 Mm) loop settings by \cite{Muller2003} and the cyclic formation and falling down of condensations were evident, as illustrated in Fig.~\ref{f-mullerfig}. This work already made it clear that the processes behind coronal rain are identical to those that dictate prominence formation, and loop length and shape are hence relevant parameters to vary, besides those controlling the heating and cooling prescriptions. Follow-up modeling \citep{Muller2005} confronted the obtained rain blob speeds with observations, finding a very satisfactory agreement with speeds of the order of 100 km/s, decelerating when approaching the upper chromosphere. This places their speeds below free fall, and pressure gradients that build up across the falling blobs nicely explain their slowing velocity patterns.

\cite{Mendoza2005} again focused on semi-circular loops, but varied the heating to mimic randomly appearing Gaussian energy pulses near the footpoints. Both the loop lengths, as well as the (controlled) time between energy impulses, affect the outcome, with condensations only found beyond a critical elapsed time between pulse injections. It should be noted that the semi-circular loop lengths considered only ranged from 5 to 30 Mm, so they would be rather low-lying loops.

\cite{Karpen2005} updated the evaporation-condensation model by taking into account area variation $A(s)$, updating the cooling curve $\Lambda(T)$, and adopting a field line shape extracted from an actual 3D sheared-arcade ideal MHD simulation. By adopting asymmetric localized heating between left and right footpoint, new dynamic features were found, including the appearance of paired condensations that eventually merged. A similar strategy was followed in \cite{Luna2012}, where again individual field lines from a 3D MHD configuration of a sheared arcade were simulated independently, but the visualizations performed translated the 1D hydro findings into their fixed 3D magnetic geometry. Both condensation threads and blobs were ubiquitous, with threads residing in dipped field line sections, and blobs falling to the chromosphere.

\cite{Susino2010} used randomly selected snapshots of 1D hydro loop models to assimilate a synthetic differential emission measure (DEM) of a multi-stranded coronal loop. Condensations formed when impulsive heating was localized at footpoints, but they did not affect the synthesized DEMs.

That each condensation that forms is ultimately caused by TI was demonstrated in \cite{Xia2011}, by showing direct agreement with the criteria of TI. This work investigated filament formation in a loop with a shallow dip, using grid-adaptive computations with a yet more realistic cooling curve prescription. Even when the chromospheric heating is turned off after some finite time, filament growth can continue due to siphon flows established towards the filament thread. Extending these 1D simulations to cover multiple days after turning off the heating, such that the filament length saturates, \cite{Yuhao2014} established how this length depends on the dipped field line parametrizations, like its half-width, depth and altitude.

\cite{KlimchukLuna2019} presented approximate analytical formulae for predicting when limit cycle behavior (or TNE) is almost guaranteed. In a simple symmetric (say, semi-circular) loop, it will be impossible to get a real stationary state achieved when the following condition holds
\begin{equation}
1+\frac{A_{\mathrm{tr}}R_{\mathrm{tr}}}{A_{\mathrm{c}}R_{\mathrm{c}}} < \frac{H_{\mathrm{foot}}}{H_{\mathrm{apex}}} \,. \label{tne}
\end{equation}
This formula contains (averaged) volumetric heating rates $H$ for loop footpoint regions versus its apex, and the cross-section of the loop is $A_{\mathrm{tr}}$ at transition region heights versus $A_{\mathrm{c}}$ for its coronal section. The radiative losses per unit area (in $\mathrm{erg}\,\mathrm{cm}^{-2}\,\mathrm{s}^{-1}$) for transition region versus corona appear as $R_{\mathrm{tr,c}}$. This relates directly to our equation~(\ref{energyheat}), since $H=\tilde{\rho}\tilde{h}$ (where $\tilde{f}$ is an averaged $f$) and an energy balance in a loop of length $L$ becomes $A_{\mathrm{c}}L H = A_{\mathrm{c}}L \tilde{n}^2 \tilde{\Lambda}(\tilde{T}_{c})+A_{\mathrm{tr}} R_{\mathrm{tr}}$, hence requiring the total heating of the coronal loop to balance the summed coronal radiative losses with losses across the transition region. The transition region losses are then inserted from an order-of-magnitude estimate based on the downwards conductive heat flux from the corona, i.e. setting $R_{\mathrm{tr}}=\tilde{\kappa}(\tilde{T}_c) \tilde{T}_c/L$. This leads to estimates of the temperature $\tilde{T}_c$ throughout the coronal section, and using the ideal gas law, the balance fixes the entire coronal loop thermodynamics. The TNE condition from Eq.~(\ref{tne}) then simply results from distinguishing the heating in a loop-footpoint-concentrated, versus a kind of overall (weaker) background heating near the apex, which causes a conflict in the temperature as estimated from $H_{\mathrm{foot}}$ versus $H_{\mathrm{apex}}$. Note that the footpoint heating must hence always exceed the background one, in order for cyclic behavior to be at play. We note in particular that there is no mention of any partial derivatives of the heat/loss function ${\cal{L}}$ which feature prominently in all TI criteria, such as in the dispersion relation~(\ref{druni}) or its generalization to the thermal continuum \citep{KeppensTC2025}. \cite{KlimchukLuna2019} provide a second condition relevant for asymmetries between footpoints (in either heating or geometry or both) and verify their predictions with 1D hydro simulations. Other aspects, like the required numerical resolution as well as heating timescales, are discussed in \cite{Johnston2019}.

\subsection{Model extensions and parameter surveys}\label{ss-survey}

By extending the model to a 1.5D setup, where also Alfv\'en waves can be incorporated, \cite{Antolin2010} showed that coronal rain formation can encode information on the coronal heating mechanism. To model Alfv\'en wave heating, the model is extended with a velocity $v_\phi(s,t)$ and magnetic field component $B_\phi(s,t)$ quantifying rotational flow and twist about the flux tube axis. By contrasting 1D hydro scenarios with parametrized `nanoflare heating', versus 1.5D setups with Alfv\'en wave heating, the former was more prevalent to lead to coronal rain events. This links with the finding from formula~(\ref{tne}), in the sense that Alfv\'en heating becomes more uniform over the loop, while the nanoflare setup was concentrated at footpoints. These rain-oriented simulations adopted a semi-circular loop shape. This 1.5D approach was also adopted in \cite{Yoshihisa2025}, where instead of a quasi-steady, localized heating, condensations could be triggered by a single heating event in a dipped coronal loop. Alfv\'en waves converted to longitudinal compressive waves (p-modes) that steepened into shocks, and ultimately triggered condensation formation. A key finding from \cite{Yoshihisa2025} is that the total amount of heating integrated over time is a discriminating factor for TI onset: even short pulses of sufficient strength can trigger condensation formation and a modification to formula~(\ref{tne}) was suggested.

Another interesting extension in 1.5D was presented by \cite{vanHoven1992}, where the authors added equations for the perpendicular velocity $v_\perp(s,t)$, and the self-consistent weight-induced field deformation $B_\perp(s,t)$. In a two-stage simulation they followed the thermal-instability driven condensation, and the resulting bending of the field, to provide a stably supported prominence. The role of a siphon flow towards the loop apex, set up by a pressure drop as radiative losses increase, was pointed out as well.

Triggered by observationally established, long-period EUV intensity variations in active region loops \citep{Auchere2014}, a thorough parameter survey of 1D hydro loop models was performed by 
\cite{Froment2018}. The authors linked the cyclic condensation formation and inherent time-dependent variation of coronal loops that undergo specific heating and cooling conditions to the detected periodicities. Using a suite of 1020 different simulations, \cite{Froment2018} varied loop geometry and the adopted heating configurations. All loop geometries could realize repeated heating-cooling cycles, again termed as TNE manifestations, if the heating prescription was favorable: especially stratified heating with footpoint concentration was found necessary. Cyclic behavior in active region loop models clearly prefers a specific combination of parameters, while actual condensations (coronal rain) and long period intensity variations can occur together. An even more extended parameter survey by \cite{Pelouze2022} included up to 9000 individual 1D hydro runs, augmenting earlier studies with especially asymmetric loop and heating conditions. To understand why some loops show time-dependent quasi-periodic variations (TNE cycles) without rain, versus others where actual condensations appear, the loops were taken to mimic an actual rain-producing event \citep{Auchere2018}, and systematic scans concluded that asymmetric loops are less likely to produce rain, unless their imposed heating compensates for the asymmetry. \cite{Pelouze2022} highlighted that the same magnetic loop could lead to prominence formation, or to periodic time variation with or without rain condensations, depending on the imposed (parametrized) heating. \cite{Kucera2024} changed the heating prescription to one consisting of many impulsive, discrete energy pulses (so-called nanoflares) along the loop, varying their location and frequency. Fully randomized nanoflares were less favorable to condensation formation, although the added freedom in frequency and location on the imposed pulses could occasionally trigger rain events, while equivalent steady heating showed no thermal instability. Repeated pulses with short time separations, located near footpoints, do produce condensations, consistent with steady parametrized heating findings.

The evaporation-condensation scenario to form coronal condensations feeds on the fact that excess (coronal) heating leads to excess conductive flux from corona to chromosphere, inducing evaporation, and as the density in the corona increases, more radiative loss results. Heating may then no longer balance the losses, in other words, a thermal misbalance results. As clarified in \cite{KeppensTC2025}, such a thermal misbalance impacts the entire linear spectrum of non-adiabatic p-modes and TC modes, and since the spectrum of normal modes essentially dictates the early temporal evolution, it is perfectly possible that the same loop structure subjected to varying heating $H(s,t)\equiv \rho(s,t)h(s,t)$ as appearing in the net-heat-loss function ${\cal{L}}$ will cause the loop to evolve differently. For both parametric studies by \cite{Froment2018} and \cite{Pelouze2022}, it would be of interest to quantify how the entire linear spectrum of non-adiabatic p- and TC eigenmodes changes by the thermal imbalance due to the adopted heating prescriptions, and to see whether and how the linear spectrum relates to the various nonlinear evolutions (with and without rain formation). In the same spirit of addressing the role of the imposed heating, \cite{Huang2021} clarified that the 1D hydro model can even unify two different pathways to prominence formation: one based on evaporation-condensation, versus one in which chromospheric matter is injected into the corona. This unified model in essence varied the $H(s,t)$ prescription to heat either the lower chromosphere or the upper one. Taking a Gaussian pulse for the localized heating in both space and time, a lower heating deposition led to pressure-induced injection of matter, while heating the upper chromosphere fitted the traditional evaporation-condensation scenario. These findings did not account for important non-LTE effects, especially relevant in (lower) chromospheric regions, and were obtained for a loop with a centrally dipped portion.

Most 1D results discussed thus far considered either semi-circular loops, or adopted field lines extracted from a sheared arcade setup, with the latter looking at rather long field lines with extended dipped sections. For actual large-scale prominence settings, dipped field line sections exist in both sheared arcades and magnetic flux rope setups, and the flux rope topology is believed to be dominant \citep{Ouyang2017}. Therefore, repeating the 1D hydro approach in flux rope settings was realized by \cite{Jinhan2022}, using an analytical flux rope model where the twist can be controlled. This twist turned out to dictate the detailed distribution of threads throughout the flux rope, as well as their dynamic evolution. It was also discovered that the filament spine does not really align with the flux rope axis, or the underlying PIL, so observational deductions on magnetic field topology from the filament morphology must be treated with caution. 

Another recent extension of the 1D hydro models on evaporation-condensation (and TNE or TI/TC condensations) is the work by \cite{Scott2024}, who studied how a transonic 1D solar wind on an open, radially expanding field line can give rise to very similar cyclic behavior. When condensations formed, their formation height related to the scale of the imposed footpoint heating. To interpret these findings with linear theory and the role of TI and TC, an extension of the work by \cite{KeppensTC2025} to diagnose all non-adiabatic, normal modes for moving and transonic configurations is called for.

Other modern extensions of the 1D hydrodynamic viewpoint on prominence and coronal condensation formation include the pioneering study of \cite{Veronika2025}, where a two-fluid model is adopted that accounts for both plasma and neutral species. Indeed, the ionization fraction within a prominence structure differs markedly from the coronal environment, where full ionization prevails. The evaporation-condensation prominence formation scenario, involving TI to trigger the runaway, fully carries over to a two-fluid setting. Pronounced two-fluid effects appear in shocks that accompany the first complete condensation. Decoupling effects also appear in the PCTR, with velocity differences between both species on the order of 100 m $\mathrm{s}^{-1}$. Although these are challenging to detect, transient decouplings have been inferred in prominence observations of precisely these magnitudes \citep{Khomenko2016,Zapior2022}.

A study by \cite{David2022} analyzes a similar plasma-neutral scenario, but in a 2D setting, and finds that ion-neutral decouplings of order of 1 km $\mathrm{s}^{-1}$ may develop on falling coronal rain blobs. The rain blob was initially inserted as a localized density enhancement in both charges and neutrals, and then followed when falling through an isothermal corona with purely vertical, initially uniform magnetic field. It would be of interest to include the actual TI-based condensation formation as well in future plasma-neutral studies for gravitationally stratified, multi-dimensional settings. Note that the study of ad-hoc inserted, falling rainblobs, as influenced by gravity, density ratio and pressure variations has been studied in idealized settings in various studies, such as \cite{Oliver2014,Hillier2025}.

\subsection{Frozen-Field Hydrodynamics: towards 3D models}\label{s-ffhd}
All findings discussed thus far adopt rigid magnetic field backgrounds, where the magnetic field variation at most enters through an area-expansion $A(s)$ along the fixed field line, as in the mass conservation law from Eq.~(\ref{masscons1D}). Analogous to the dip-only models by \cite{Gunnar2015}, visualizations can adopt some finite cross-sectional size and a heuristic inter-thread spacing, to create the illusion of 3D structure as showcased in \cite{Luna2012} for sheared arcades, or by \cite{Jinhan2022} for flux ropes. 

In contrast, truly 3D volume-filling hydrodynamic evolutions in any (frozen) magnetic topology can be studied as well, following the technique introduced by \cite{Mok2005}. Using our earlier notation of the local unit vector $\hat{\mathbf{b}}$ along the fixed magnetic field, the governing equations read as
\begin{eqnarray}
    \partial_t\rho+\nabla\cdot\left(\rho v_\parallel \hat{\mathbf{b}}\right) & = & 0 \,, \label{ffhdrho}\\
    \partial_t\left(\rho v_\parallel\right)+\nabla\cdot\left[\left(\rho v^2_\parallel + p \right)\hat{\mathbf{b}}\right] &=& \rho g_\parallel +p\left(\nabla\cdot\hat{\mathbf{b}}\right) \,, \label{ffhdmom}\\
    \partial_t e +\nabla\cdot\left[(e+p)v_\parallel\hat{\mathbf{b}}\right] & = & \rho g_\parallel v_\parallel +\nabla\cdot\left[\kappa(T)(\hat{\mathbf{b}}\cdot\nabla T)\hat{\mathbf{b}}\right]+\rho{\cal{L}} \,, \label{ffhde}   
\end{eqnarray}
where we write the equations in terms of the momentum density $\rho v_\parallel$ and total energy density $e=\rho v_\parallel^2/2+p/(\gamma-1)$. The magnetic field is still a fully externally fixed, 3D topology, forcing only field-aligned flow $\mathbf{v}=v_\parallel \hat{\mathbf{b}}$. This frozen-field hydro model still can not address weight-induced dipping of field lines, or any of the relevant MHD instabilities (Rayleigh-Taylor, interchange, CCI), but it provides a computationally affordable way to study parametrized heating prescriptions contained in $\rho{\cal{L}}$. 

This frozen-field hydro model was adopted to study filament formation in arcades by \cite{ffHDI2024}, followed by filament formation in a 3D twisted flux rope in \cite{ffHDII2025}. This approach allows for exactly the same freedom in parametric surveys that relate to the evaporation-condensation scenario: one can impose any 3D field topology, vary the spatio-temporal behavior of the heating function and switch the cooling table $\Lambda(T)$. It has the advantage that one can prescribe heating dependencies aware of local magnetic field $h(\mathbf{B}(\mathbf{x}))$, or any of the observationally established heating dependencies $H(B,\rho,T)=\rho h(B,\rho,T)$. Moreover, many heating models, as suggested by early scaling laws \citep{RTV1978}, depend not only on local values like $B$, $\rho$, and $T$, but also on the local magnetic field line length $L$. In full 3D MHD models, computing $L$ at every point and time step is often too expensive, since field line integration is costly. However, in the frozen-field HD approach, the magnetic field is static. This means the field line lengths only need to be computed once, at the start of the simulation. After that, they can be used as fixed parameters to guide spatial heating profiles. This makes it practical to include $L$-dependent heating models, which would be difficult to implement in dynamic MHD setups. Its volume-filling nature also ensures that one can directly translate the setup to synthetic observables by performing line-of-sight integrations, and the area-variation we had to introduce artificially in a single field line setting is now incorporated automatically, by means of divergence evaluations of physical fluxes.

\begin{figure}[ht]
\centering
\includegraphics[width=\textwidth]{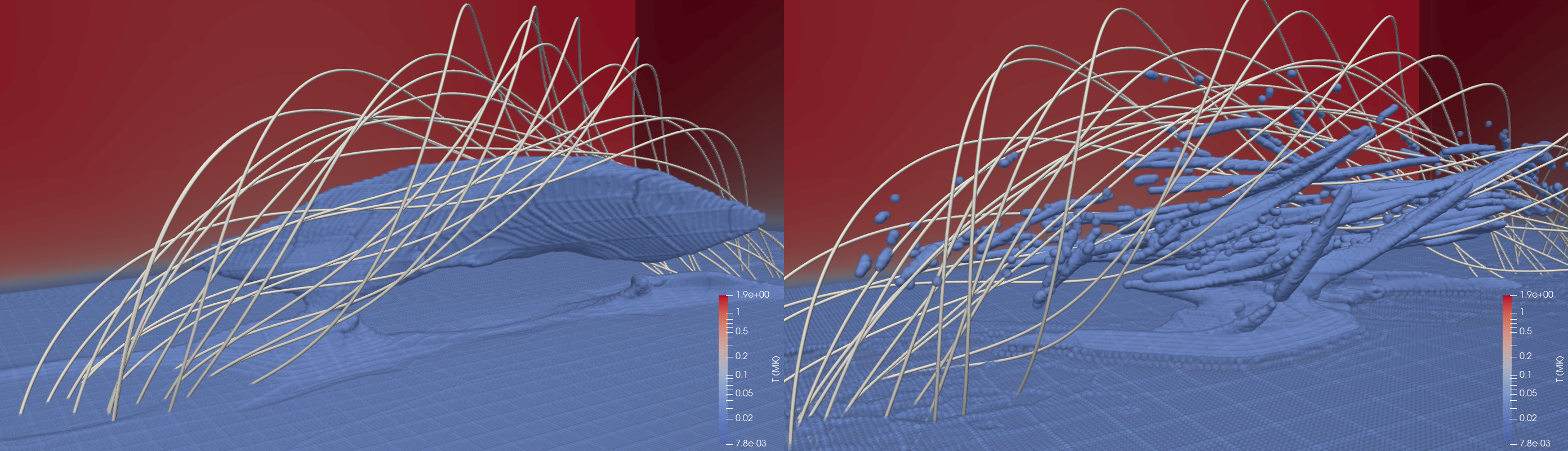}
\caption{
Prominence structures formed in a twisted flux rope under different heating prescriptions. 
Left panel: A relatively stable and coherent prominence forms under steady, footpoint-concentrated heating. 
Right panel: When the heating is randomized in both space and time, the resulting prominence develops fine structure and exhibits more dynamic behavior. 
In both panels, the background color denotes atmospheric temperature, the blue contour traces the 15{,}000~K isocontour (highlighting the cool prominence), and the gray lines show selected magnetic field lines.
}\label{f-yuhaoffhd}
\end{figure}

As an example, Fig.~\ref{f-yuhaoffhd} shows a filament formed in a twisted flux rope (much like the one adopted in \cite{Jinhan2022}), where we contrast the obtained prominence structure when the heating is varied. In the study at left, taken from \cite{ffHDII2025}, the flux rope was subjected to a foot-point concentrated heating, which was applied steadily for several hours before it was turned off, and a relatively monolithic, steady prominence (indicated by the blue contour outlining the 15{,}000~K isocontour region in the figure) is formed that occupies the entire bottom of the flux rope. If we instead change the heating to a more randomized (in both space and time) prescription, the prominence forms pronounced fine-structure, and behaves more dynamically, as indicated by the fragmented and evolving structure in the right panel. This is again a clear indication that the evaporation-condensation scenario, along with the linear hydrodynamic spectrum enriched by TI/TC eigenmodes, may well allow us to deduce knowledge on the unknown local coronal heating, through the detailed morphology and dynamic behavior of prominences. 

\section{Condensations forming and evolving in MHD}\label{s-mhd}

Even though the essentially 1D hydro models discussed in the previous section~\ref{s-1d} provide insight into the mass and energy circulation in idealized loop or arcade flux bundles, all observed condensations are best described in a multi-dimensional magnetohydrodynamic (MHD) setting, where we have to face the full nonlinear set of equations~(\ref{momeq})-(\ref{energy})-(\ref{induction})-(\ref{masscons1D}). These nonlinear PDEs describe the macroscopic plasma dynamics fully, and are aware of the entire zoo of linear waves and instabilities mentioned in Section~\ref{s-zoo}. Especially in such multi-dimensional MHD settings, many breakthroughs in modeling prominence and coronal rain formation occurred in the past decade. We organize these according to magnetic topology in this section, discussing models with 2D arcades, models targeting flux rope cross-sectional evolutions, up to full 3D topologies. A complementary means to classify all these models is provided in Table~\ref{t-clas}, which serves to emphasize the actual formation pathway of the condensations. In that respect, multi-dimensional MHD has gone significantly beyond the evaporation-condensation model, which is the one that is almost primarily researched within 1D models. Next to injection, levitation and evaporation-condensation, the table lists novel formation mechanisms like plasmoid-fed-prominence-formation (PF$^2$), emergence-driven prominence formation, or reconnection-levitation mechanisms, which all invoke the ubiquitous reconnection that occurs throughout the dynamic solar corona. In every formation pathway, the ultimate local trigger for actual condensation is invariably linked to thermal instability (TI), as discussed in Sections~\ref{ss-TI}-\ref{ss-link}.

\begin{sidewaystable}
\caption{Multi-dimensional MHD simulations self-consistently forming coronal condensations.}\label{t-clas}
\begin{tabular*}{\textwidth}{@{\extracolsep\fill}lccccc}
\toprule%
Formation & D & Configuration & Atmosphere &  Remarks & Reference \\
\midrule
in-situ TI  &  2.5D & Arcade & Corona & Footpoint shear & \cite{Choe1992} \\
&  2.5D & Arcade/FR & Corona & KS and KR\footnotemark[1]  & \cite{Choe1998} \\
   &  2.5D & Flux Rope & Corona & levitate-condense & \cite{Kaneko2015B} \\
       &  2.5D & Flux Rope & Corona & CCI-TI verified & \cite{Jack2021} \\
       &  2.5D & Flux Rope & Corona & varied heating & \cite{Brughmans2022} \\
              &  2.5D & Flux Rope & Corona & rotation & \cite{Valeriia2023} \\
                            &  3D & Flux Rope & Corona & Rayleigh-Taylor & \cite{Jack2022} \\
     &  3D & Flux Rope & Corona & Reconnect-Condense & \cite{Kaneko2017} \\
          &  3D & Flux Rope & Corona & Rayleigh-Taylor & \cite{Kaneko2018} \\
          &  3D & Flux Rope & Corona & RT and rain & \cite{Donne2024} \\
          & 3D & fan-spine & Corona & null-reconnection & \cite{Beatrice2025} \\
\midrule
evaporate   &  2D & Fixed arc & Chrom.+cor. & counterstreaming & \cite{Zhou2020} \\
       
         &  2D & Fixed arc & Chrom.+cor. & multi-threaded & \cite{Jercic2023} \\
          &  2.5D & Streamer & Chrom.+cor. & TNE in helmet streamer & \cite{Schlenker2021} \\
  &  2.5D & Arcade & Chrom.+cor. & TI verified & \cite{Xia2012} \\
 &  2.5D & Arcade & Chrom.+cor. & coronal rain & \cite{Fang2013} \\
  &  2.5D & Arcade & Chrom.+cor. & coronal rain & \cite{Fang2015} \\
    &  2.5D & Arcade & Chrom.+cor. & random heating & \cite{Li2022} \\
        &  2.5D & Arcade & Chrom.+cor. & flux emergence & \cite{Li2023} \\
 &  2.5D & Arcade & Chrom.+cor. & flux rope forms & \cite{Keppens2014} \\
   &  2D & Arcade & Chrom.+cor. & cyclic heating & \cite{Zhou2023} \\
  &  2.5D & Arcade & Chrom.+cor. & varied heating & \cite{Jercic2024} \\
  &  2.5D & Arcade & Chrom.+cor. & X-point reconnect & \cite{Craig2025} \\
       &  2.5D & Arcade & Conv.+Chr.+cor. & TI induced strands & \cite{Antolin2022} \\
  &  3D & Flux Rope & Chrom.+cor. & barb formation & \cite{Xia2014} \\
    &  3D & Flux Rope & Chrom.+cor. & Fine structured & \cite{Xia2016} \\
        &  3D & Bipole Arc & Chrom.+cor. & coronal rain & \cite{Xia2017} \\
           &  3D & Arcade & Chrom.+cor. & coronal rain & \cite{Moschou2015} \\
                                 &  3D & Bipolar AR & Conv.+Chr.+cor. & coronal rain & \cite{Kohutova2020} \\
                      &  3D & Bipolar AR & Conv.+Chr.+cor. & coronal rain & \cite{Zekun2024} \\
\midrule
levitate & 2.5D & streamer & Chrom.+cor. & FR erupts & \cite{Linker2001JGR...10625165L} \\
 & 2.5D & Flux Rope & Chrom.+cor. & erupts  & \cite{Zhao2017} \\
& 2.5D & Flux Rope & Chrom.+cor. & erupts  & \cite{Zhao2020} \\
& 3D & Flux Rope & Chrom.+cor. & erupts and drains  & \cite{Xing2025} \\
\midrule
plasmoid-fed
& 2.5D & Flux Rope & Chrom.+cor. & PF$^2$ & \cite{Zhao2022} \\
\midrule
emergence
& 2.5D & Flux Rope & Chrom.+cor. & reconnection-fed & \cite{Li2025} \\
& 3D & Flux Rope & Corona & erupts & \cite{Fan2017,Fan2018} \\
& 3D & Flux Rope & Corona & erupts & \cite{Fan2019} \\
& 3D & Flux Rope & Corona & erupts and drains & \cite{Fan2020} \\
injection & 2D & Arcade & Chrom.+cor. & unified model & \cite{Huang2025}\\
\midrule
postflare rain 
& 2.5D & Arcade & Corona & non-force-free & \cite{Samrat2024} \\
& 2.5D & standard flare & Chrom.+cor. & dark flare loops & \cite{Ruan2021} \\
& 3D & standard flare & Chrom.+cor. & synthetic views & \cite{Ruan2024} \\
\botrule
\end{tabular*}
\footnotetext[1]{Both Kippenhahn-Schl\"uter and Kuperus-Raadu prominence types were obtained}
\end{sidewaystable}

\subsection{2D magnetic arcade evolutions}\label{s-2darcade}

We first discuss magnetic arcade setups, which may involve bipolar or quadrupolar magnetic field configurations, and either look at a multi-dimensional MHD evolution in a vertical plane (perpendicular to the solar surface, containing the direction of gravity) or in the actual curved 2D surface that corresponds to the magnetic flux surfaces. The latter will be termed `fixed arcade' models, since there we do not allow field line bending in the direction of gravity, but only within the magnetic surface itself.

\subsubsection{Deforming arcade models}
\begin{figure}[ht]
\centering
\includegraphics[width=0.48\textwidth]{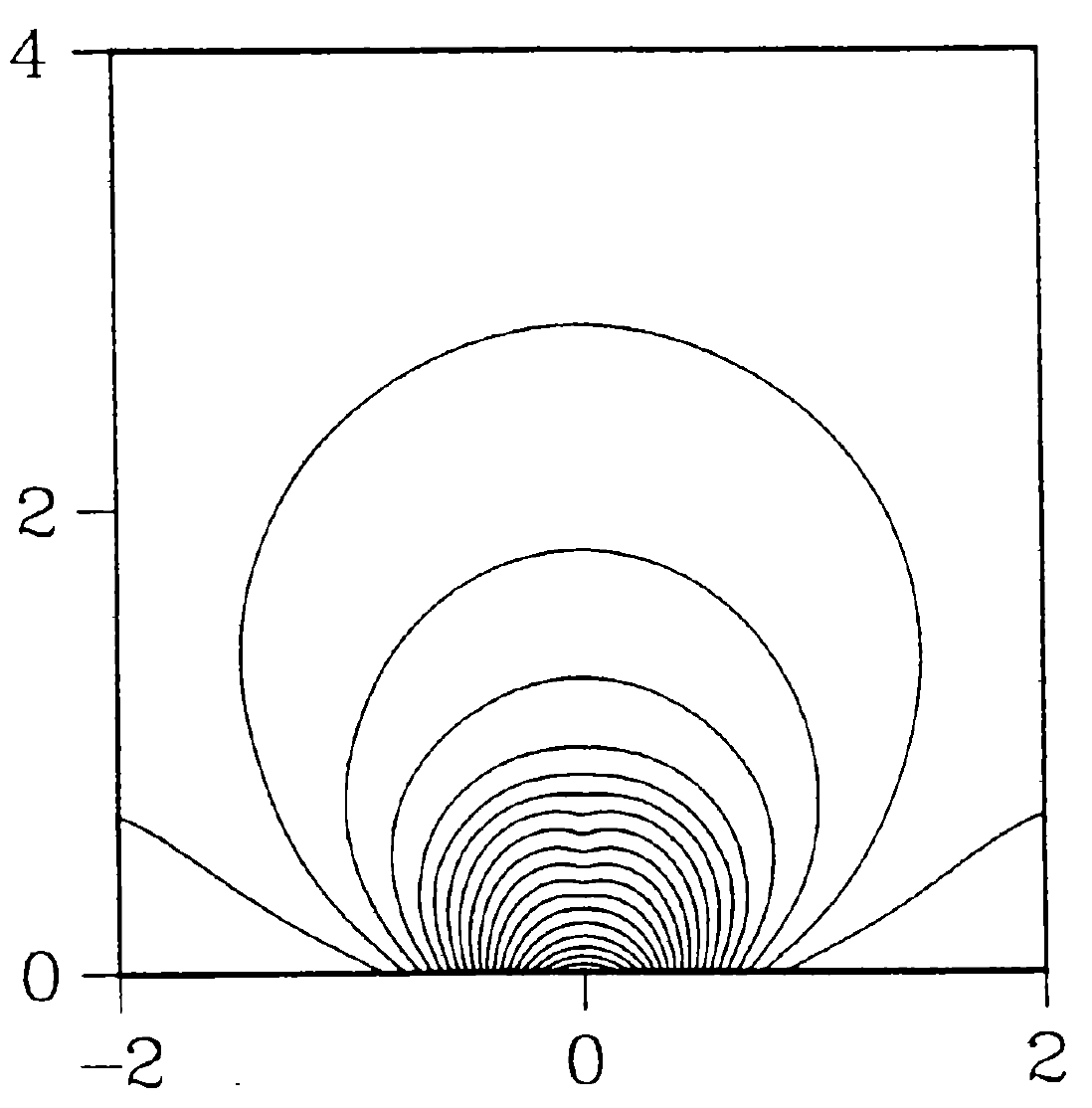}\includegraphics[width=0.48\textwidth]{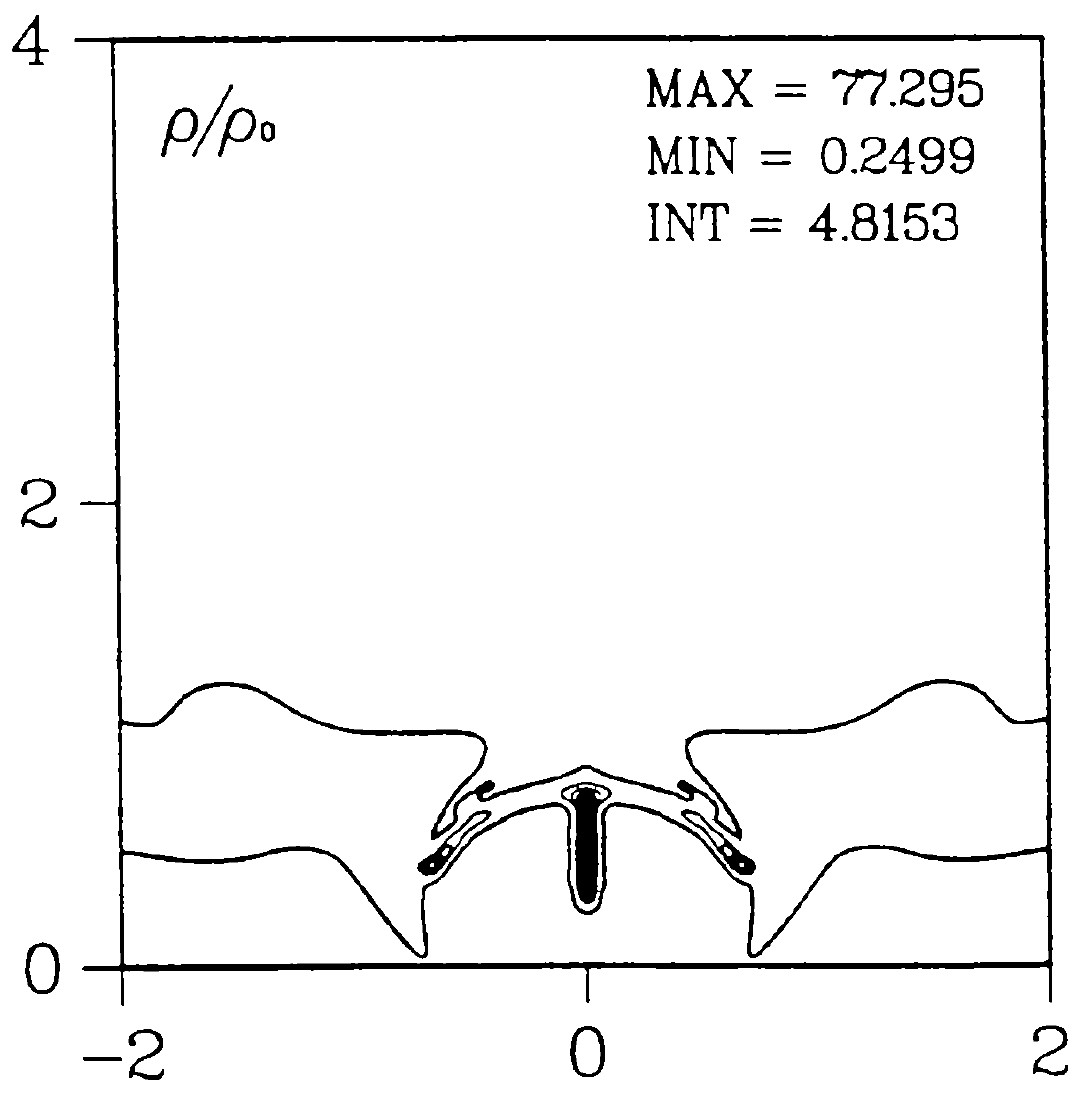}
\caption{A snapshot of the early MHD simulation by \cite{Choe1992}, after the prominence has formed by in-situ TI. The arcade magnetic field develops a locally dipped section centrally (left panel), weighed down by the in-situ condensation as seen in density (right).}\label{f-choelee}
\end{figure}

\cite{Choe1992} demonstrated the clear possibility for in-situ condensations formed by TI, with a pioneering model of prominence formation in a coronal arcade disturbed by shearing the arcade field lines. The prominence weight suffices to dip the central arcade field lines, and a Kippenhahn-Schl\"uter type topology develops naturally, as shown in Fig~\ref{f-choelee}. The authors considered only the corona, so there is no other formation pathway than TI/TC mediated runaway cooling. In a later conference proceedings, 
\cite{Choe1998} reported that the same in-situ TI pathway can be used to form both Kippenhahn-Schl\"uter, as well flux-rope embedded Kuperus-Raadu prominence types, and even found a prominence condensing in between two bipolar arcades. 

By including the chromosphere and transition region, the first evaporation-condensation based model in a footpoint-heated, bipolar 2.5D arcade by \cite{Xia2012} quantified the Kippenhahn-Schl\"uter force balance throughout the prominence, while the initial in-situ condensation was shown to agree with linear TI instability criteria. By varying the magnetic topology to a quadrupolar arcade with a centrally dipped portion from the beginning, \cite{Keppens2014} could follow the evaporation-condensation scenario into the hours-long evolution of a prominence. In their study, the filament weight ultimately triggered reconnection and the formation of a flux rope filled with prominence matter, while also coronal-rain-like drainage occurred during the evolution. 

Staying in a quadrupolar setting, considering only 2D in-plane vector fields, \cite{Zhou2023} made a clear link between the observationally established `winking filament' behavior and the evaporation-condensation scenario. By making the localized footpoint heating cyclic, the resulting stretching and up-down movement of the prominence can indeed result in a periodic appearing and disappearing from H$\alpha$ line center and line wings. Using synthetic H$\alpha$ filament views integrating along a vertical line-of-sight, clear `winking' was demonstrated with a periodicity that differs from the imposed cyclic heating, as understood from a forced oscillator analogy. The H$\alpha$ synthesis was based on the approximate method as introduced by \cite{Heinzel2015}, but \cite{Jack2023} shows excellent agreement with more advanced non-LTE computations for a similar 3D MHD prominence model. 

\cite{Huang2025} presented 2D MHD simulations of a dipped arcade, where a flux emergence event (an ephemeral bipole) is mimicked by a time-evolving bottom boundary prescription. This leads to (anomalously controlled) reconnection when the bipole emerges near a footpoint of the pre-existing dipped field. It was shown how the height of the resulting reconnection (lower to upper chromosphere) indeed recovers the 1D proof-of-principle study by \cite{Huang2021} which unified injection and evaporation prominence formation pathways: prominences could form in both ways, but this time in 2D setups, as a result of reconnection-induced heating.

\begin{figure}[ht]
\centering
\includegraphics[width=0.5\textwidth,height=5cm]{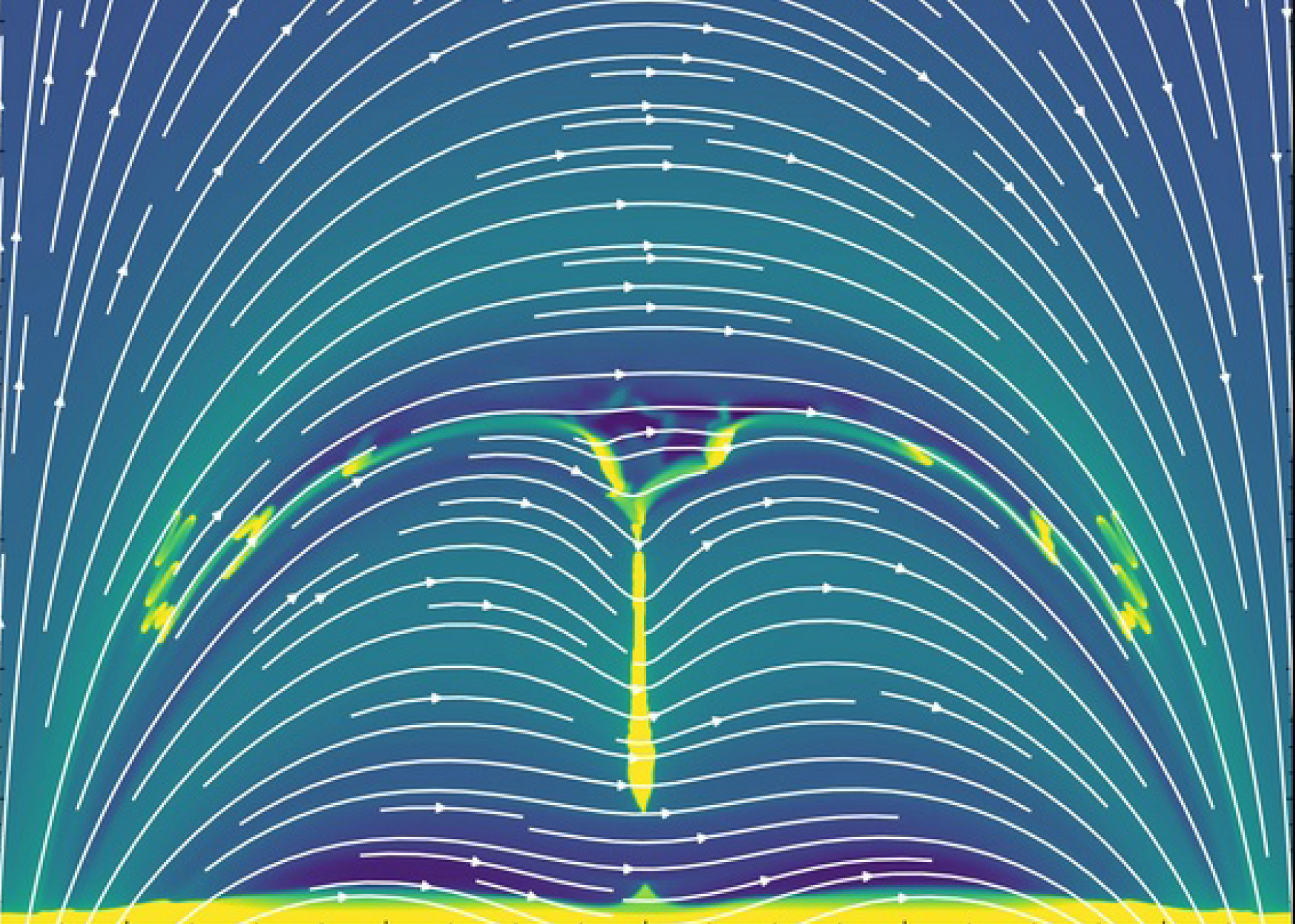}\includegraphics[width=0.5\textwidth,height=5cm]{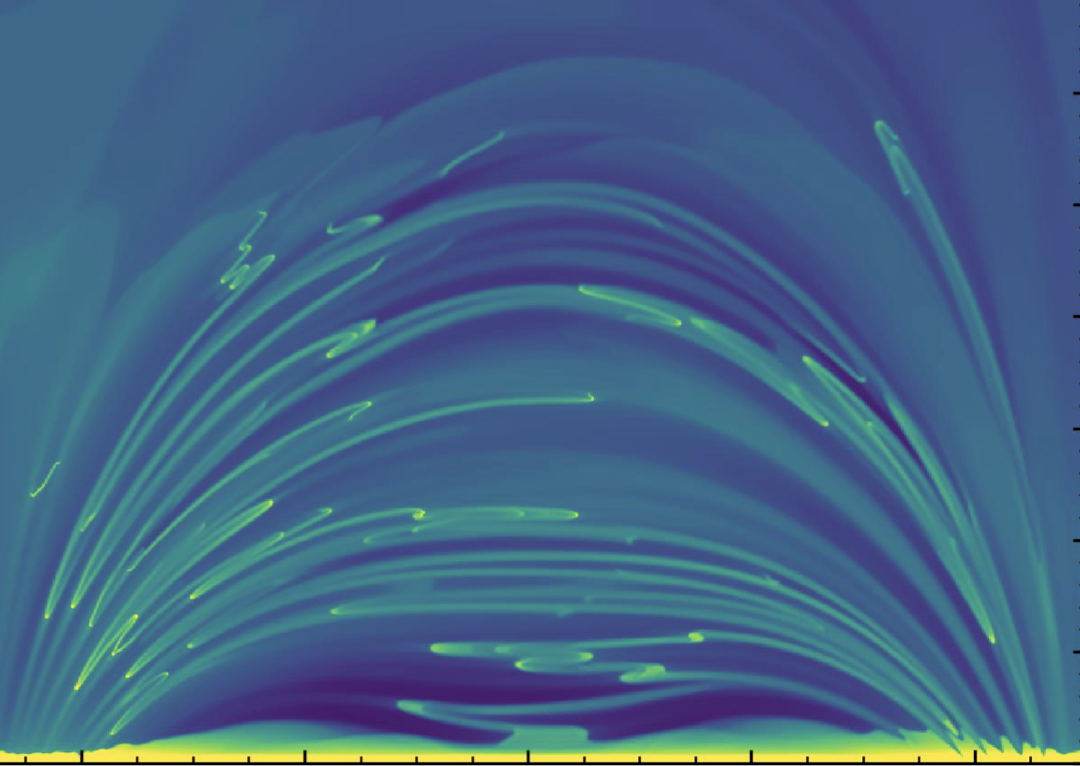}
\caption{Prominence topologies differ greatly for steady (left) versus stochastic (right) heating scenarios, as demonstrated by \cite{Jercic2024}. Shown is the density variation from low chromosphere to corona, with the magnetic topology indicated at left. The domain shown is 100 Mm $\times$ 80 Mm. Steady heating forms a vertical prominence, while stochastic heating favors dynamically evolving, horizontal threads.}\label{f-vero}
\end{figure}

\cite{Jercic2024} performed 2.5D simulations of a chromosphere to corona quadrupolar arcade, where the obtained prominence topologies differed markedly depending on the imposed heating. As illustrated in Fig.~\ref{f-vero}, the same topological setup can form a more vertical prominence structure (at left), or show a more horizontal, fragmented prominence (at right), when steady (left) versus impulsive heating (at right) is acting near the upper chromosphere. Stochastic heating leads to threadlike structures that are highly dynamic, while the steady heating shows a typical vertical slab-like prominence, where the weight-induced field line deformation can even trigger localized reconnection high up in the corona. Since the two scenarios were not exactly tailored to deliver the same amount of energy injected into the corona, and both scenarios invoked several parameters (like pulse durations and amplitudes for the stochastic case), follow-up studies could parametrically explore which of the two topologies prevails, depending on total injected energies as spread over background and localized (steady to stochastic) regimes.
Finally, by changing the 2.5D quadrupolar arcade such that its central X-point is embedded in the corona, \cite{Craig2025} formed dynamic prominences, that demonstrated reconnection-based drainage through the coronal X-point. 

In summary, 2D deforming arcade models have already shown rich morphologies in the obtained prominences, and the weight-induced effects combined with the possibility for field line reconnection, can cause rather dynamic prominence-rain hybrid evolutions. In line with the findings of simplified 1D fixed-loop models, the spatio-temporal details of the imposed heating have a direct impact on the prominence appearance and evolution.

\subsubsection{Fixed arcade models}\label{s-fixedarc}

\begin{figure}[ht]
\centering
\includegraphics[width=\textwidth,height=5cm]{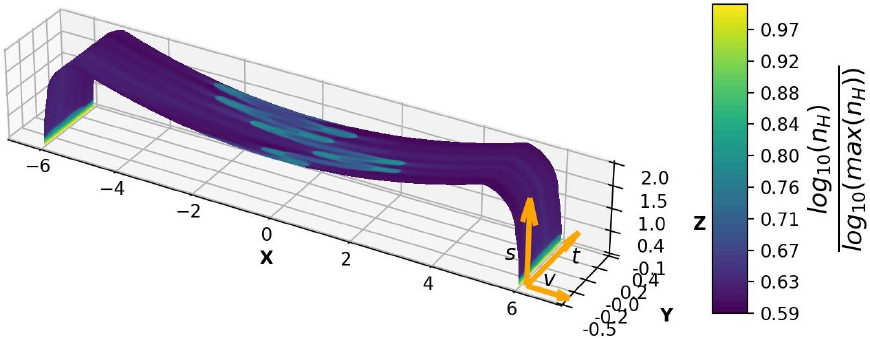}
\caption{Stochastically heated fixed arcade, forming threaded prominences, as seen in this density view, from \cite{Jercic2023}. The 2D MHD simulation maintains the arcade shape shown, but allows for cross-field coupled evolutions with field line bending possible in the curved surface. The threaded prominence structure evolves dynamically due to randomized heating at both feet.}\label{f-vero2}
\end{figure}

As a meaningful 2D MHD generalization of the 1D fixed-field line approach, \cite{Zhou2020} introduced a 2D model of a fixed arcade where we allow for field line bending in the given magnetic surface. The idea is to follow the thermodynamic as well as magnetic evolution in the curved surface that connects the left to right footpoint regions of an arcade. By pioneering an imposed randomized heating near both footpoint regions (random in space and time, but exponentially stratified along the arcade), \cite{Zhou2020} demonstrated the formation of threadlike structures, accompanied with counterstreaming flows. Indeed, random evaporation can drive longitudinal oscillations of threads, as well as inter-thread unidirectional flow patterns, which alternate. A follow-up study by \cite{Jercic2023} quantified the effect of the randomized heating pulse heights and amplitudes on the self-consistent multi-threaded prominence formation. An example is shown in Fig.~\ref{f-vero2}, illustrating the fixed arcade concept. Condensation rates were found proportional to the heating pulse amplitudes, and in full agreement with rates inferred from observations.

\subsubsection{To rain or not to rain}\label{ss-rain2d}

Several of the 2D arcade simulations discussed showed that while forming a large scale prominence, smaller coronal rain condensations may occur as well, such as seen in the left panel of Fig.~\ref{f-vero}. The heating prescription as well as the magnetic topology and strength combine to create conditions that may prevent the collection of condensed matter into a macroscopic prominence. The first multi-dimensional demonstration of actual coronal rain was presented by 
\cite{Fang2013}, where a bipolar arcade with footpoint localized, but steady, heating showed blob characteristics that directly matched with observations. The model demonstrated the possibility of blobs which condensed in-situ in the corona and then evaporated while falling prior to reaching the chromosphere, or blobs that get siphoned over the apex of the loops. Blob widths and lengths that averaged 400 to 800 km (in adaptive simulations that resolved cells of 78 km at best) must be compared with current insights on coronal rain fine structure. Higher-resolution (up to 20 km detail), longer term studies by \cite{Fang2015} followed the cyclic occurrence of coronal rain showers in arcades. They identified how each coronal rain blob can be accompanied by fully 2D-shaped rebound shocks due to pressure-mediated siphon flows, while counterstreaming flows may break up blob strands into several segments. Blob deformations, and blobs impacting the transition region to merge with chromospheric matter, were discussed in detail, along with the PCTR structure that forms around each individual rain feature.
We note in passing that the basic model ingredients (topology of the field, steady left-right symmetric heating) from \cite{Xia2012}, where a monolithic prominence was formed, versus those in \cite{Fang2013}, where the first multi-dimensional rain occurred, are rather similar. A key difference is the enforced symmetry at the midplane adopted in \cite{Xia2012}, which was relaxed (and broken by numerical discretization errors) in \cite{Fang2013}. It is still relevant to investigate which key factors (such as field strength, effective resolutions, adopted numerical discretization) allow rain versus prominence formation in such 2D and 3D settings. Another important finding from \cite{Fang2013,Fang2015} concerns the sympathetic cooling identified in these 2D setups, where the first blob causes local (Lorentz force) perturbations on neighbouring field lines that are close to TI, initiating condensation formation across the entire arcade. This effect should be studied in more detail in full 3D settings, as it impacts the obtained rain blob morphologies in as yet incompletely understood fashion.

More recently, \cite{Li2022} revisited the same 2D setup with randomly heated arcades, as opposed to the steady heating applied in \cite{Fang2013,Fang2015}. The improved statistics on blob widths and lengths showed that blobs with areas less than 0.5 Mm$^2$ dominated the population, obtained by counting more than 6000 individual blobs over the simulated 10 hour period. By translating the simulations to synthetic EUV observations, and quantifying periodicities in the lightcurves, the study showed periods of several tens of minutes to hours, in agreement with the observations by \cite{Auchere2014}.

All studies mentioned so far kept the magnetic topology at the (low chromospheric) base fixed, and concentrated on TI induced condensations within evaporation-condensation cycles. In reality, the chromosphere-to-corona magnetic topology may change drastically due to e.g. flux emergence events. Starting with a coronal-rain-supporting arcade, \cite{Li2023} investigated how flux emergence impacts the multiphase thermodynamics of the arcade. All rain blobs were forced to the footpoint opposite to the emergence location, and the reconnection associated with the emerging flux showed how a multi-thermal jet forms, where plasmoids containing cool dense matter interfere with a hot jet component.

In a self-consistent radiative MHD simulation that spanned from sub-photosphere to corona, \cite{Antolin2022} reported the in-situ formation of rain blobs due to TI, influenced by topological changes in the coronal field as an originally open field line becomes closed and connected across the periodic sides. Although the heating in such models is stated to be self-consistent, it is mostly due to Joule heating and hence numerically dominated as the actual physical resistivity regime in the corona is outside the scope of any numerical simulation. Nevertheless, it was shown how the clump formation leads to a slightly enhanced magnetic field concentration in the blob, leading to a fundamental field strand of transverse size (width) of around 400 km. This is fully consistent with the findings from the chromosphere-to-corona-only studies discussed above. The clear analysis of the basic process of TI-induced field strand formation and how the rain blob deforms and impacts the transition region, leading to detectable UV brightenings, augmented insights gained from the pioneering study by \cite{Fang2015}. 

\subsection{Flux rope models in 2.5D}\label{s-2dfr}

\begin{figure}[ht]
\centering
\includegraphics[width=0.32\textwidth,height=5cm]{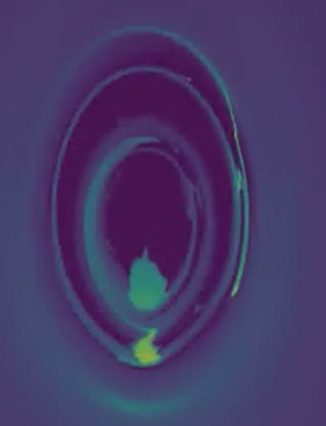}
\includegraphics[width=0.32\textwidth,height=5cm]{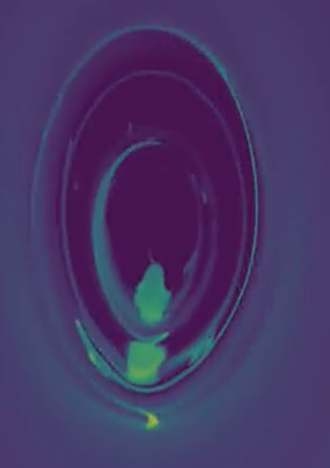}
\includegraphics[width=0.32\textwidth,height=5cm]{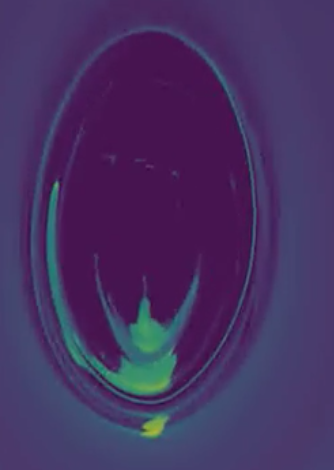}
\caption{Density snapshots illustrating various stages in the formation of a prominence in a flux rope, from \cite{Jack2021}. Note how the nested flux rope surfaces clearly dictate the dynamic redistribution of TI-induced condensations.}\label{f-jack}
\end{figure}

While Section~\ref{s-2darcade} focused on condensation formation in 2D arcade-like simulations, we already mentioned that especially for realizing long-lived prominence structures, the observations seem to indicate a preference for flux rope embedded material \citep{Ouyang2017}. In that context, various recent models follow a formation scenario that does not depend on evaporation-condensation, but rather on levitation-condensation as introduced by \cite{Kaneko2015B}. Indeed, considering only a coronal volume where there is no chromosphere and transition region to evaporate from, the idea is that magnetic arcades can be deformed by systematic footpoint motions to form a flux rope (with an underlying reconnection X-point in 2.5D settings). In that process, part of the lower coronal material, which is stratified and hence slightly denser, gets lifted, and ultimately the thermodynamic conditions within the formed flux rope become liable to TI/TC condensation. This leads to inverse polarity prominence types, and creating the flux rope from an initial arcade involves converging motions, and possibly shearing of the arcade field as well. This levitation-condensation process was revisited at an extremely high resolution (down to 6 km in cell sizes) by \cite{Jack2021}, where the first convincing link with linear MHD spectroscopy findings was made. Indeed, it was found that in typical coronal settings, the flux rope initially realizes an internal density variation that is almost constant on the nested flux surfaces, implying that the Convective Continuum Instability from Eq.~(\ref{cci}) will induce thermodynamic changes on all flux surfaces where the projected Brunt-V\"ais\"al\"a frequency turns negative. Distinct condensations then form as a direct result of TI, and these condensations can form throughout the flux rope. Pressure imbalances (and gravity) ultimately cause the denser blobs to collect towards the lower dipped flux rope part, where baroclinic considerations (unaligned density and pressure gradients) cause further finer-scale motions. In the later stages where a monolithic prominence resides in the dipped region, the simulation showed a clear resistive slippage across the field, as argued for theoretically by \cite{Low2012}. Density views on the forming prominence, during the very dynamic TI-regulated condensation and redistribution stages are provided in Fig.~\ref{f-jack}. The same simulation was also used to demonstrate the revolutionary multi-dimensional non-LTE radiative transfer capabilities of the radiance cascades technique, implemented in the DexRT code \citep{Osborne2025}. The fine-structured prominences as shown in Fig.~\ref{f-jack} pose severe challenges to multi-dimensional non-LTE post-processing, as the traditional short characteristics approach suffers from artificial rays due to their finite angle coverage. At the same time, multi-dimensional spectroscopy is needed to properly handle the shadowing effects due to the multi-layered prominence structure.

In many multi-dimensional MHD simulations on prominence and coronal rain formation, the net heat-loss function ${\cal{L}}$ used in equation~(\ref{energy}) may have a heating contribution at its initial time that does not exactly ensure the balance expressed in Eq.~(\ref{energyheat}). Inspired by the findings from 1D hydro loop models as discussed in Section~\ref{s-1d}, where besides the loop shape, the imposed spatio-temporal heating is varied to trigger different loop behavior, multi-dimensional MHD models frequently adopt a combination of a weak, vertically stratified background heating that is meant to balance the losses, with localized (steady or stochastic) footpoint heating. In the corona-only models from \cite{Kaneko2015B,Jack2021}, there is no such footpoint heating, but only the weaker background heating to balance the density-temperature dependent radiative losses. Starting in isothermal settings, the heating function can just adopt the same exponential density variation we expect in a hydrostatic setting, thereby ensuring the balance from Eq.~(\ref{energyheat}) at startup. However, observations of coronal loops have led to `scaling laws', quantifying the heating that loops experience on the basis of observable quantities like loop length, typical magnetic field strength, and typical loop density. Therefore, a thorough 2.5D parametric survey of the levitation-condensation scenario was performed by \cite{Brughmans2022}, to determine especially the role of different background heating prescriptions on the prominence formation. Besides the typical exponential decay, the authors used locally mixed prescriptions where $H\equiv \rho h \propto B^\alpha \rho^\beta$ for varying $(\alpha,\beta)$ combinations, and introduced a new dynamic way to automatically bring in the ignored third dimension (i.e. the `length' of the loop). This latter approach implies that the field line length, which varies from flux rope center to edge, really imposes a reduced heating within the flux rope interior, so that an automated tracking of the flux rope shape and an accompanying reduction in the heating inside the flux rope is appropriate. This reduction proved to allow the formation of more massive prominences, as it can trigger TI in many more locations of the flux rope interior, which then again collect dynamically in the bottom region of the flux rope. A visualization of the 2.5D simulation as it evolves in the phase space of $(1/\rho,T)$ further allowed to detect both isobaric and non-isobaric evolutionary stages in the TI-mediated condensation formation.

By making the background heating stochastic (as done first for fixed arcades in \cite{Zhou2020}), \cite{Valeriia2023} demonstrated that the levitation-condensation route to prominence formation in a flux rope almost necessarily induces prominence rotation. Indeed, such more randomized heating leads to slight left-right asymmetries when the original arcade deforms (by converging and shearing) to the flux rope topology. This implies that after the reconnection, net rotational motion is present within the (coronal) flux rope interior. When then the TI manifests, we should see an expected spin-up due to contraction during the condensation formation, in accord with angular momentum conservation. Since asymmetric conditions and stochastic heating are surely active throughout the solar corona, we would expect to find observational evidence in favor of strong (order 60 km/s) rotation at early stages of prominence formation in flux ropes. The study by \cite{Valeriia2023} already provided synthetic EUV impressions of the rotating prominence, while a follow-up study by 
\cite{Alex2024} performed full non-LTE spectroscopy on the model. It was found that the rotation should be clearly detectable in Mg II and Ly $\alpha$ or Ly $\beta$ spectral lines, while H $\alpha$ and Ca II lines show only faintly detectable rotation signatures. 

\subsection{3D MHD evolutions}\label{ss-3Dmhd}

Fully 3D MHD models of coronal condensation formations emerged in the last decade, and they can again be classified according to the prevailing magnetic topology, or whether coronal-only, chromosphere-to-corona, or even sub-photosphere to corona regions are included (see Table~\ref{t-clas}). Naturally, extensive parametric surveys of 3D models have not yet been realized. Still, many findings of the 1D and 2D surveys carry over, especially concerning the role of parametrized (or numerically realized) heating in evaporation-condensation settings. New with respect to the 2D or 2.5D models, is that even from the linear MHD spectroscopy view, 3D models open up the pathway to magnetic Rayleigh-Taylor and interchange dynamics, which is most effective when the associated wave vector $\mathbf{k}$ of the perturbation can orient perpendicular to the local magnetic field $\mathbf{B}$. This is essentially prohibited in the 2D or 2.5D settings of all models discussed thus far, but can always be achieved in 3D. The fact that Rayleigh-Taylor dynamics and magneto-thermal convection is observed very clearly in quiescent prominences \citep{Berger2011}, has inspired many modelers to study such buoyancy-driven dynamics for density contrasts representative for prominence-corona transitions. Examples of such studies, that go all the way to the magneto-convective phase, are \cite{Keppens2015,XiaRK2016,Madhurjya2023,Thomas2024}, while a review of Rayleigh-Taylor related insights for prominence physics is provided in \cite{Hillier2018}. Since most of those models do not include the actual in-situ condensation phase, they are not further discussed here. Instead, we here review those 3D MHD works where non-eruptive prominence structures, or sustained coronal rain events, are realized self-consistently. 

\subsubsection{Prominences}\label{ss-prom}

\begin{figure}[ht]
\centering
\includegraphics[width=\textwidth,height=8cm]{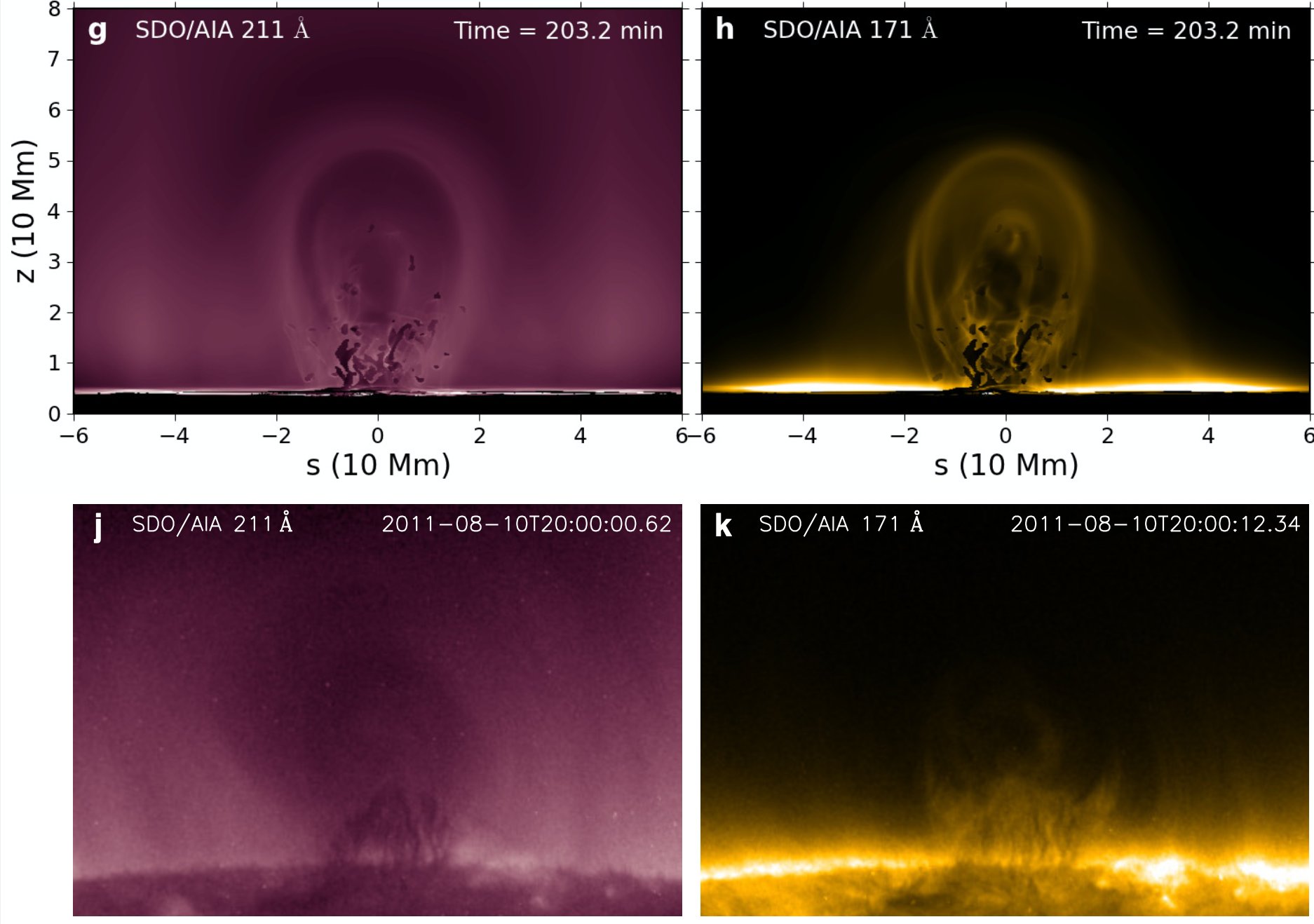}
\caption{A synthetic EUV view oriented along the prominence spine of the model of \cite{Xia2016} (top row), compared to actual observations in the same EUV channels (bottom row).}\label{f-xia}
\end{figure}

The first successful in-situ condensation of a coherent prominence structure was demonstrated by \cite{Xia2014}. The approach taken was to initiate a full 3D MHD simulation from chromosphere to corona, including all the relevant processes listed in the energy equation~(\ref{energy}), from a prior 3D isothermal MHD study where a coronal arcade was deformed into a flux rope. The effective resolution of 460 km per cell on the domain of 240 $\times$ 180 $\times$ 120 Mm$^3$ was modest (using three levels of automated grid refinement), but sufficient to follow the initial mass loss when adjusting thermodynamically to the chromosphere-to-corona settings, causing a slight uplifting of the flux rope. This in essence realizes the coronal levitation-condensation pathway discussed earlier (introduced by this name in the later 2.5D study of \cite{Kaneko2015B}) and triggered in-situ TI condensation. The prominence eventually settled in the dipped portion of the flux rope, and even had a barb feature corresponding to an earlier mass drainage at one of the prominence ends. Despite the modest resolution, the synthetic EUV views on the prominence nicely demonstrated the appearance of a dark coronal cavity, where the horn structures corresponded to prominence-loaded helical field lines that separate from overlying prominence-free fields lines through the cavity. The cavity itself is the clear signature of in-situ plasma redistribution, happening as a result of the TI process.

A further leap forward was realized by \cite{Xia2016}, where a higher-resolution simulation realized the first fully dynamic, 3D fragmented prominence structure. This work also succeeded in quantifying the mass circulation from chromosphere to corona, where (1) individual dense prominence fragments drain into the chromospheric reservoir; (2) a continuous plasma evaporation at the flux rope feet was enforced; and (3) the in-situ TI process replenished the prominence mass during several hours. Technically, the model again used a prior isothermal MHD arcade-to-flux rope deformation, but realized a more thermodynamically relaxed full chromosphere-to-corona model that was then in a third stage subjected to footpoint-localized heating at both ends of the flux rope. Hence, a true evaporation-condensation scenario resulted, where the increased effective resolution (now using four mesh levels, down to 250 km cell sizes) allowed to form a more fine-structured prominence. Representative EUV synthetic views are shown in the top row of Fig.~\ref{f-xia}, and can be contrasted with actual observational counterparts shown in the bottom row. The dynamics of individual prominence fragments can be directly compared with blob dynamics of coronal rain as studied in bipolar arcades \citep{Fang2013,Fang2015}.

Further 3D MHD progress was also made on coronal-only simulations, where a higher resolution is affordable from the very start: as long as no in-situ condensations form, there is no transition region (which challenges modern numerical discretizations) anywhere in the volume. Modern shock-capturing techniques, as well as advanced algorithmic treatments of the parabolic thermal conduction and local optically thin loss terms, can fully cope with the sudden appearance of PCTR structures as coronal condensations form by in-situ TI. \cite{Kaneko2018} performed a 3D simulation of a levitation-condensation setup, in essence repeating the 2.5D setup from \cite{Kaneko2015B} in a periodically treated third dimension oriented along the PIL underlying the flux rope axis. By randomizing the converging motions along this added direction, they were able to trigger Rayleigh-Taylor type deformations on the prominence body, which still formed a fairly monolithic vertical structure in the bottom of the flux rope. 
Revisiting the same levitation-condensation scenario of \cite{Kaneko2015B} in a 3D coronal volume, \cite{Jack2022} achieved an unprecedented 21 km effective resolution (using four AMR levels on a 24 $\times$ 30 $\times$ 24 Mm$^3$ domain). This allowed to finally resolve the scales needed to follow the magnetic Rayleigh-Taylor process during the prominence formation phase. Producing both filament H$\alpha$ views and prominence EUV observations, this simulation could reconcile the more thread-like on-disk views with vertically structured prominence views. Baroclinic effects on the falling fingers in the Rayleigh-Taylor process were analysed in detail, and the supplementary material of this publication provided quantitative links with linear MHD spectroscopic findings, comparing all terms that contribute to (magnetically modified) Brunt-V\"ais\"al\"a frequencies. The simulation in essence reproduced a purely coronal in-situ condensation contained in a periodic section of a flux rope, but the obtained widths of vertical fingers and plumes compared favorably to observations. A follow-up non-LTE spectroscopic synthesis in \cite{Jack2023} showed that the approximate H$\alpha$ synthesis used in the original publication agreed well with a true non-LTE treatment (albeit performed in a 1.5D sense, where individual vertical columns are analysed separately). 

Another coronal-only prominence formation model was put forth in \cite{Kaneko2017}, involving reconnection between adjacent arcade fields that effectively doubled the length of coronal field lines in the resulting flux rope. This implied that an in-situ TI condensation could set in, where the authors argued that the newly formed longer field lines now exceeded the Field length. This length scale can be deduced from a dimensional analysis on the energy balance expressed in Eq.~(\ref{energyheat}), setting  $\lambda_F^2=\kappa(\bar{T}^{5/2}) T/n^2\Lambda(T)$, and fieldlines with lengths exceeding this $\lambda_F$-limit would be more favorable to TI-driven condensations. However, it is noted in \cite{KeppensTC2025} that this length scale does not directly feature in a full eigenmode spectrum of a (expanding) 1D loop structure. Therefore, it can only serve as a qualitative measure in comparing the stabilizing role of thermal conduction along the field lines with the optically thin driving term of the TI. \cite{Kaneko2017} showed convincing synthetic EUV views on their 3D prominence structure and its evolution, and termed this formation process reconnection-condensation.

\begin{figure}[ht]
\centering
\includegraphics[width=0.68\textwidth,height=6cm]{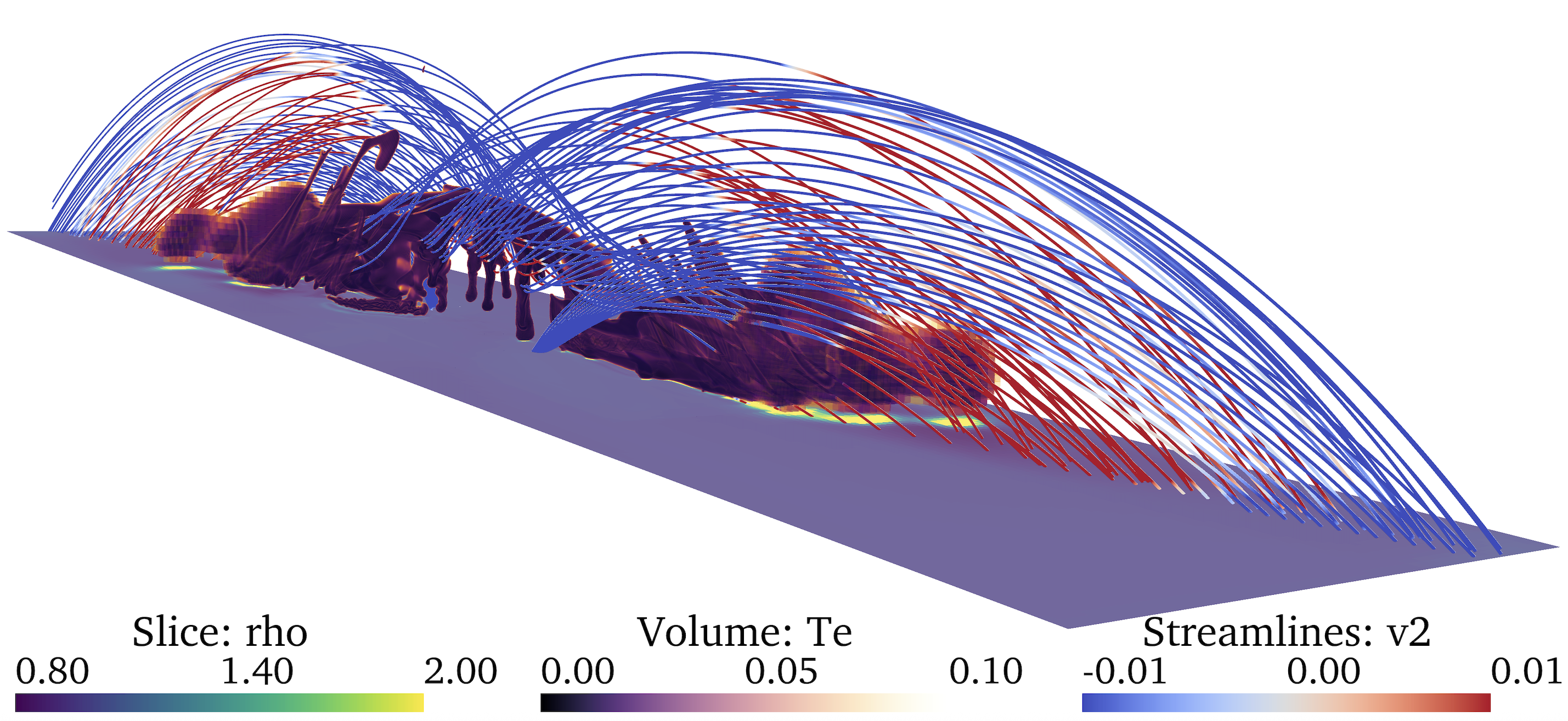}
\includegraphics[width=0.3\textwidth,height=6cm]{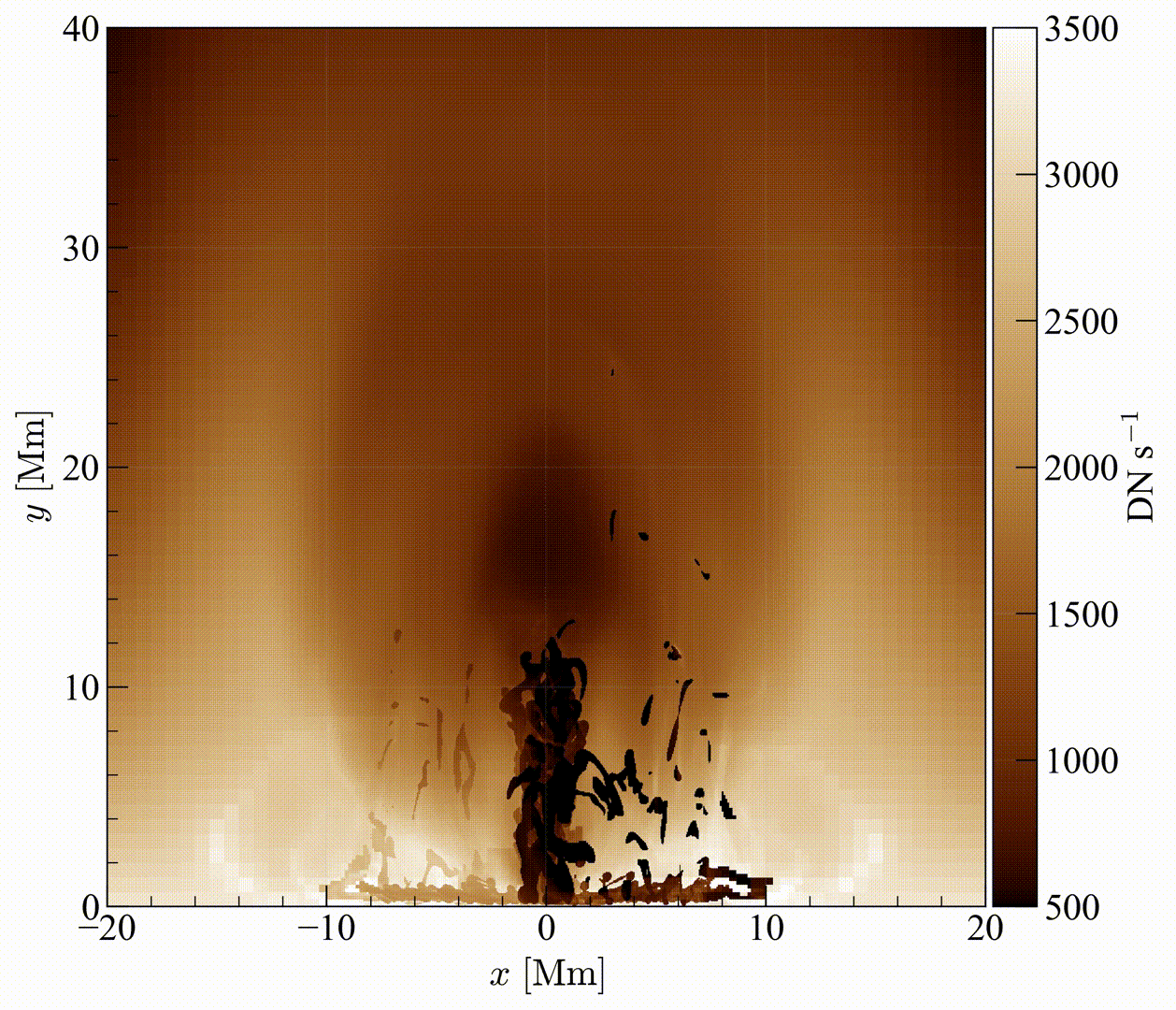}
\caption{The magnetic field and density structure (left panel) of a 3D prominence formed by reconnection-condensation, as modeled by \cite{Donne2024}. Note the Rayleigh-Taylor fingering occurring in the central, reconnected flux rope region. At right, a synthetic EUV view along the prominence axis is shown.}\label{f-donne}
\end{figure}

This reconnection-condensation, coronal-only prominence formation was revisited recently by \cite{Donne2024}, targeting to resolve further details by improving their 120 km cell-size to a 41 km resolution. It is to be noted that the actual formation process is very similar to the levitation-condensation discussed earlier (and showcased in \cite{Kaneko2015B,Kaneko2018,Jack2022}), as the double arcade system is forced to reconnect into an elongated flux rope structure by centrally converging motions, and hence the appearance of the flux rope and the underlying reconnection are present in all these models. The condensation trigger is due to TI/TC processes, and \cite{Donne2024} showed that the in-situ catastrophic condensations lead to siphoning flows that in turn feed the prominence with matter from the low corona. This implies that mature prominences may also form without any active evaporation process near the upper chromosphere-transition region heights, with TI being the main mass-collecting agent. An impression of the model is shown in Fig.~\ref{f-donne}, and it can be seen that the central portion of this prominence also develops clear Rayleigh-Taylor fingering, while in the early stages coronal rain develops above the formed flux rope, as it compresses the material in higher layers during its early rise. The formed prominence is highly fine-structured, as can be seen in the virtual prominence view along its axis displayed at right in Fig.~\ref{f-donne}. Note again that all these corona-only, 3D models by construction exclude any role for condensation triggering by the chromosphere-corona coupling at the underlying transition region, which is (over)emphasized in `thermal non-equilibrium' scenarios from simplified 1D hydro loop models. It is still unclear which prominence or rain properties (mass distribution, morphologies, counter-streaming flows) will change when we combine these setups with the added possibility of the evaporation-condensation pathway.

\subsubsection{Coronal Rain}\label{ss-rain}

As discussed already for the lower-dimensional (1D, 2D or 2.5D) models, 3D scenarios can lead to more coronal-rain-like evolutions, depending on the magnetic topology and field strength, on the imposed heating model, or both. The first truly 3D coronal rain simulation study by \cite{Moschou2015} used a potential quadrupolar arcade (a 3D variant of the topology shown in cross-sectional view in Fig.~\ref{f-vero}) and combined a weak steady exponentially varying background heating (to maintain the corona) with a footpoint localized heating to evaporate upper chromospheric matter. Especially the central, dipped region showed runaway cooling due to TI, and blobs were found to form and develop Rayleigh-Taylor and interchange dynamics. Cool blobs were seen to gradually move downwards until they end up being guided by the lower-lying fields. The modest effective resolution (roughly 200 km per grid cell) still allowed to capture about 20-30 individual blobs for a duration of over half an hour physical time, with most blobs at temperatures of 20000 K. The nonlinear evolution was physically understood by quantifying several linear stability criteria (e.g. those related to the CCI from Eq.~(\ref{cci}) or to TI) at consecutive times during the simulation. In that sense, linear spectroscopic findings could make direct links with the obtained nonlinear behavior.

By increasing resolution (to about 80 km) and changing the magnetic topology to a weak bipolar setup, \cite{Xia2017} demonstrated that rain can indeed show clear Rayleigh-Taylor driven deformations that divert rain downwards from weaker to stronger magnetic field regions, where blobs eventually follow field lines. The synthetic EUV views also differed from the typical purely-field aligned behavior seen in strong active region loops, but may well represent multiphase dynamics in less-studied, lower field regions. The simulation could provide good statistics on blob properties and their tendency to align their velocity with the local field, with evidence for a coronal rain shower in the later phase of the simulation.  The field strengths adopted were of order 3 Gauss at altitudes of 30 Mm, and the bottom dipolar field was only 120 Gauss, hence clearly outside parameters for active regions. One may contrast these magnetic field values with an observationally inferred value of 350 Gauss at 25 Mm above the solar limb, in a spectropolarimetric observation of a limb flare loop \citep{Kuridze2019}.

\begin{figure}[ht]
\centering
\includegraphics[width=\textwidth]{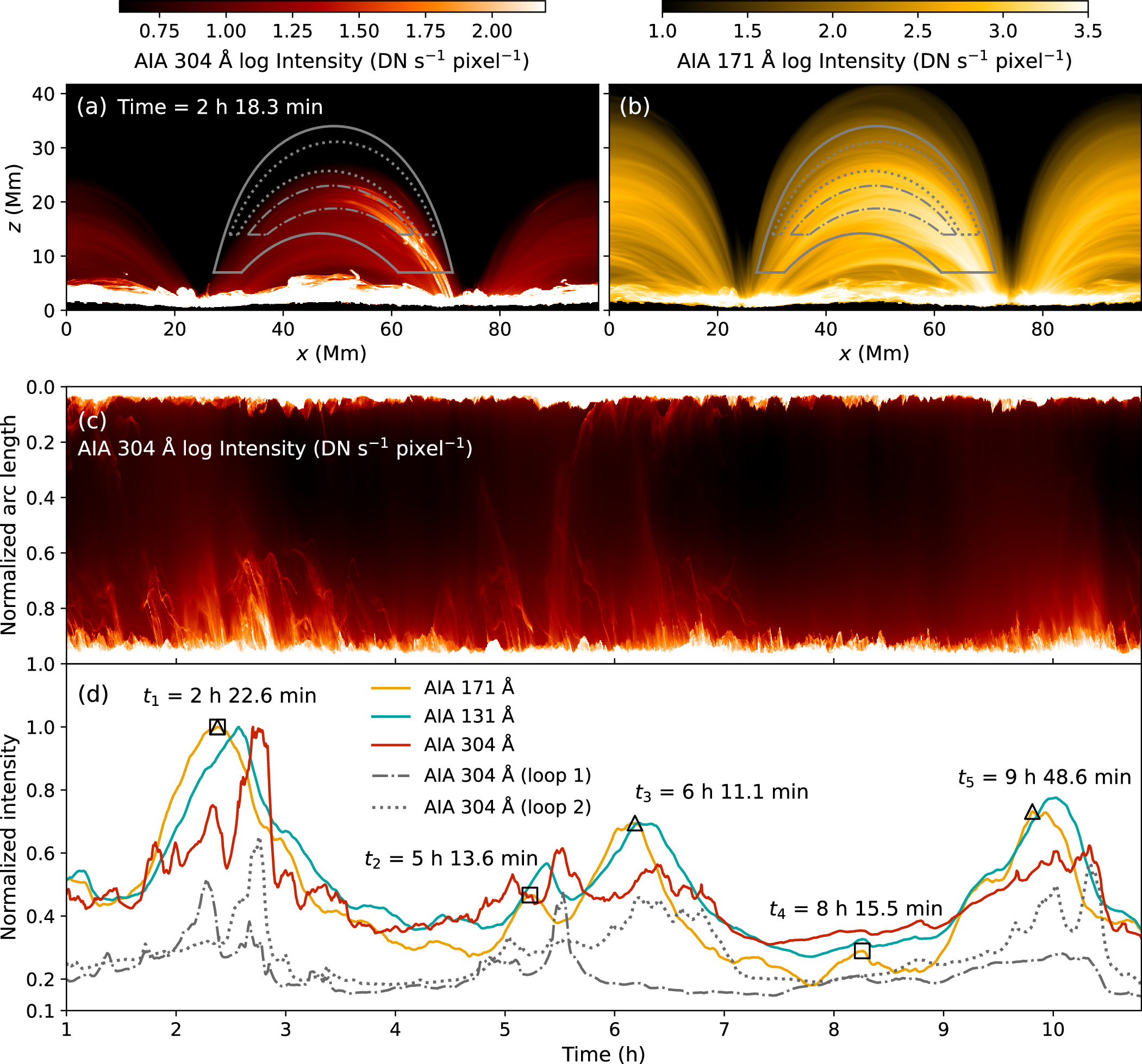}
\caption{Synthetic view in two EUV AIA channels of the 3D coronal rain simulation by \cite{Zekun2024} (top panels). A space-time plot along arched field lines is shown in panel (c), while synthetic AIA lightcurves (panel (d)) show characteristic periodicities and relative phase shifts that resemble observations.}\label{f-zekun}
\end{figure}

While \cite{Moschou2015,Xia2017} did not include the (sub)photosphere, meanwhile more self-consistent sub-photosphere to corona models emerged as well. This involves radiative MHD treatments that handle the corresponding optically thick-to-thin radiative transitions properly. They realize the coronal heating by (mostly numerically enhanced) Ohmic dissipation within convectively shuffled coronal loops. In \cite{Kohutova2020}, a dipolar region was simulated where the resulting impulsive heating events allowed for thermal instability to set in on several field line bundles. Compared to the studies mentioned thus far, cell resolution was about 50 km, the field strength at coronal heights reached a higher 10 G, while the domain extended only to 14 Mm above the photosphere, restricting attention to low-lying loops. Analyzing a selection of field lines which displayed condensations, clear parallels with 1D studies could be made in so far that asymmetric heating conditions are known to influence TNE and TI, although also open field lines displayed condensation formation.

Finally, \cite{Zekun2024} performed 3D coronal rain simulations of very extended duration, accounting for the realistic 2000 G photospheric fields accompanying a (bipolar) pair of sunspots, again from subphotosphere to corona (now up to 41 Mm above the solar surface). This simulation recovered the full cyclic behavior of rain (as realized earlier in 2D settings in \cite{Fang2015,Li2022}) by covering almost half a day of physical time. Synthetic EUV views in multiple channels recovered periodicities and also relative temporal shifts, fully consistent with the observations \citep{Auchere2018}, as demonstrated in Fig.~\ref{f-zekun}. Using the linear TI criteria, the spatio-temporally varying location of all rain blobs matched well with predicted instability sites.

It will be of interest to conduct more full 3D, extended duration simulations, for conditions from more quiescent to active region magnetic fields and simple to complex topologies, allowing to make a clearer link with observational findings on the shape and lifetime of coronal rain showers \citep{Seray2022,Seray2023}.

\subsection{Erupting prominences}\label{ss-erupt}

While all model efforts discussed thus far greatly advanced our understanding of how prominences and rain blobs form, they all focused on magnetic topologies that stayed quasi-steady while condensations evolved. At the same time, the most spectacular views on prominences are obtained during eruptions. While there are countless efforts to model the initiation and evolution of Coronal Mass Ejections (CMEs) and how they in turn relate to linear MHD criteria \citep{Rev2019}, prominence matter (which makes up most of the ejected material) is rarely included. Here, we briefly summarize models where self-consistently formed prominences appear within actual eruptions. 

In \cite{Linker2001JGR...10625165L}, a 2.5D axisymmetric simulation was presented where prominence formation by levitation occurred within a helmet streamer. In a spherical domain that starts at the top of the chromosphere, a boundary-controlled shear flow is acting within the helmet streamer, which subsequently changes to a flux reduction by controlling the boundary tangential electric field. In this part of the evolution, a flux rope forms that lifts chromospheric matter to coronal heights. Depending on the amount of flux reduction, the flux rope with filament can erupt or remain stably suspended.

In \cite{Zhao2017}, a Cartesian 2.5D simulation from chromosphere to high corona (up to 250 Mm) follows what happens when an initial linear force-free arcade gets deformed into an (erupting) flux rope by imposing converging motions at its base. Using up to 7 adaptive grid levels, details down to 24 km could be resolved, while the resulting magnetic reconnection was mitigated by the included resistive MHD effects. As the flux rope forms, it levitates chromospheric material, and this originally chromospheric matter became trapped in the erupting flux rope as a mature prominence. Height-time traces of the path followed by the flux rope and the embedded prominence matched closely, while dynamical aspects (such as plasmoid formation) in the current sheet forming between the erupting flux rope and the lower-lying flare loop system could be identified. Indeed, such mesoscale details arise when islands merge into the flux rope, disturbing the erupting prominence with internal fast shocks \citep{Zhao2020}.

A related study by \cite{Zhao2022} introduced a novel `plasmoid-fed prominence formation' (or PF$^2$) scenario that actively relies on plasmoid-mediated dynamics to make a prominence appear and erupt. Instead of actively deforming an arcade as done in \cite{Zhao2017}, the initial condition starts from a flux rope that is in a catastrophe state, where the possibility for a steady equilibrium is lost. Catastrophe pathways to CMEs are studied frequently in MHD modeling, but \cite{Zhao2022} showed that the current sheet forming underneath the erupting flux rope can collect chromospheric matter into it, and when the current sheet thins and becomes plasmoid unstable, the cool and dense chromospheric matter can get trapped within individual plasmoids. The upwards traveling ones then act to feed the bottom of the eruption with seed prominence matter, further augmented with sympathetic in-situ TI condensations. The main aspect of this PF$^2$ model is its intrinsically threaded and fine-structured nature of the erupting prominence, where details of reconnection, tearing and thermal instabilities all interplay. Future work should look at full 3D realizations of this erupting prominence pathway, and observations may well identify specific signatures of this form-while-erupt scenario, as opposed to prominences that live quietly for extended periods of time, and suddenly get ejected.

\begin{figure}[ht]
\centering
\includegraphics[width=\textwidth]{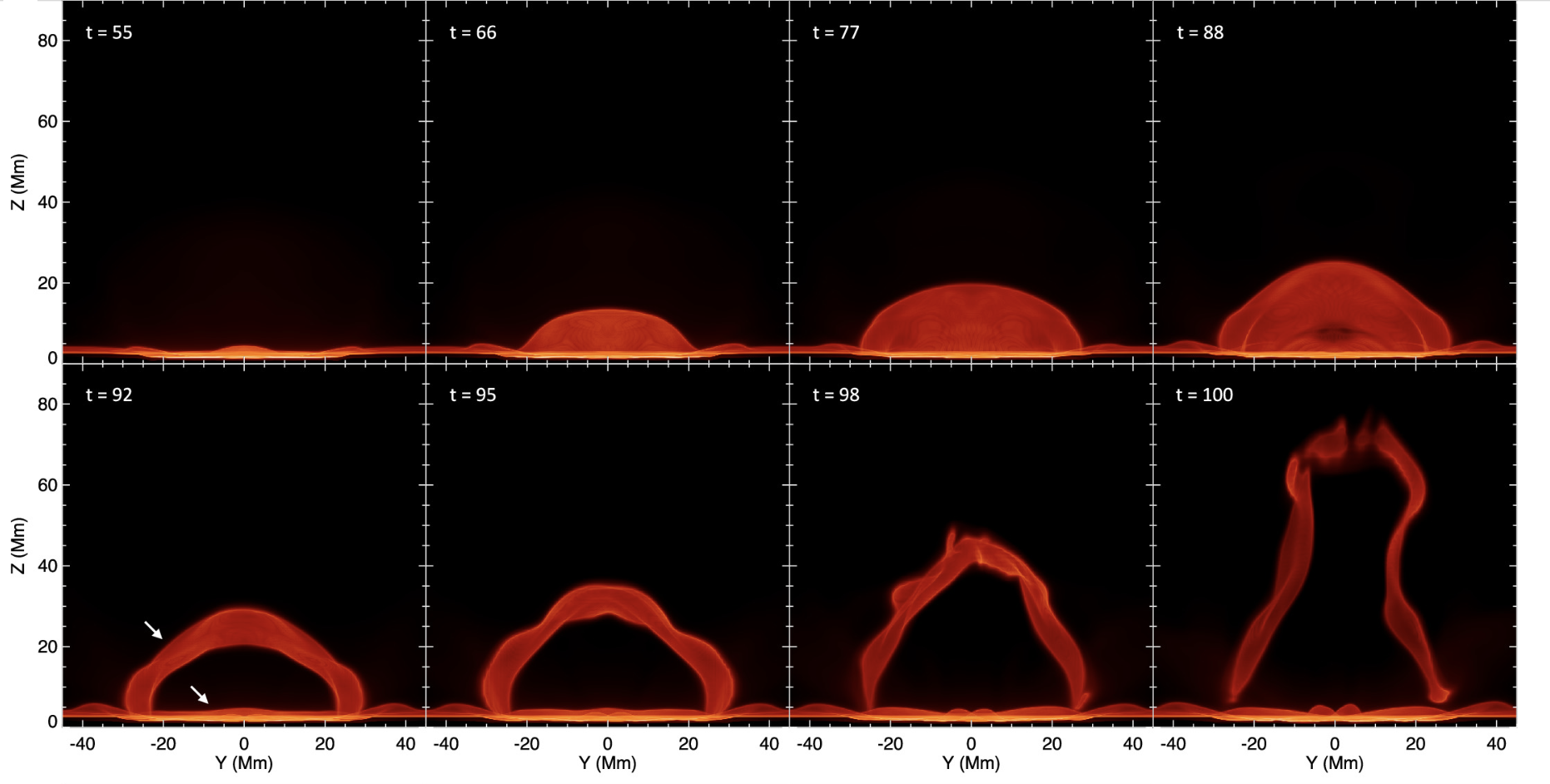}
\caption{Synthetic EUV image at 304 \AA\, from the 3D CME simulation of \cite{Xing2025}, where a filament gets levitated, splits (bottom left panel) and the top filament drains while erupting.}\label{f-xing}
\end{figure}

Full 3D, but corona-only, simulations of erupting prominences were pioneered in \cite{Fan2017}, where a spherical domain extended up to 11 solar radii. The initial state focuses on a helmet streamer located between radially stretched field lines that contain ambient solar wind, and the lower boundary emerges a twisted flux rope structure within the helmet streamer. The lower boundary treatment mimicked the behavior of a transition region where evaporation could happen, by driving the base pressure in a prescribed e-folding time to the one consistent with the conductive heat flux. This prescription introduced some heuristic parameters, which could be tuned to result in runaway cooling and prominence formation (TI induced condensations), while the twisted rope deforms and erupts (due to linear MHD kink instability). \cite{Fan2018} made further detailed comparisons between cases with and without prominence formation, highlighting how non-force-free the magnetic field is when cool material is present, and how the drainage of filament matter can aid in speeding up the final eruption. Synthetic EUV SDO/AIA views for prominences \citep{Fan2019} along the flux rope axis clearly resembled observations (with prominence horns enclosing the cavity, much like those discussed in \cite{Xia2014}). Further follow-up simulations by \cite{Fan2020} combined the form-and-erupt scenario to one where large-amplitude longitudinal oscillations of the 3D prominence structure were initiated as well. They found a negligible effect of those (damped) oscillatory motions on the 3D eruption and drainage. 

Most recently, \cite{Xing2025} presented a 3D chromosphere-to-corona simulation of a CME event where shearing and converging footpoint driving forms a flux rope in a 3D Cartesian box, where a mature filament forms from levitating chromospheric matter. A splitting is observed in both the magnetic and filament topology shortly before the eruption, while the drainage of filament mass drives the slow rise phase of the CME event. A synthetic view on the eruption is shown in Fig.~\ref{f-xing}, where the bottom left row ($t=92$ view) indicates the split in an erupting and a low-lying filament.

Efforts to model erupting prominences are very timely, since they are now extensively observed by the Metis coronagraph onboard of Solar Orbiter mission, as e.g. demonstrated in \cite{Russano2024}. Metis can follow eruptions at distances up to 10 solar radii, and combines first time imaging in the ultraviolet H-I Ly-$\alpha$ line with polarized white light, yielding unique plasma diagnostics. Noteworthy in that respect is the work by \cite{Dion2025}, presenting a parametric survey of 2.5D CME events as triggered by a combination of converging-shearing footpoint motions in an initial arcade setup. The study covers cases without and with eventual eruptions, but all have self-consistently formed prominences due to thermal instability. \cite{Dion2025} find that erupting prominences may evaporate in situ due to Ohmic heating occuring at localized current concentrations, and that tearing and plasmoids on the current sheet underneath the erupting, prominence-carrying flux rope can contribute to the internal fine structure of observed CMEs.

\subsection{Other topologies and condensation formation pathways}\label{ss-topol}

Most models discussed thus far consider traditional arcade or flux rope evolutions, which represent closed field topologies with both ends anchored to the photosphere. The main physical process behind the thermal instability (a delicate interplay between heating and cooling under a coronal-specific radiative-loss prescription) is somewhat indifferent to the magnetic topology: local runaway condensation formation may happen also in open field regions. While recent 1D hydro models \citep{Scott2024} indeed confirm this possibility, multi-dimensional realizations of this process are called for, where especially the transition zone between open and closed field lines should be investigated in highly resolved MHD simulations. From the linear MHD spectroscopic viewpoint, these regions bring all ingredients for rich dynamics together: current sheets (tearing), shear flow (Kelvin-Helmholtz), changing radiative loss effects, thermodynamic and heating prescriptions, \ldots

In that context, the 2D axisymmetric study by \cite{Schlenker2021} already realizes TNE cycles and associated rain in a streamer plus solar wind setup, where the (turbulent) consequences of the TI/TC condensation formation within the closed helmet structure can also be traced beyond its tip. The simulation adopts a total heating prescription that splits into two exponential functions that serve as a (weaker) background and as a footpoint heating source, respectively. Numerical challenges are relaxed by artificially breaking the symmetry in the equatorially located streamer, and by forcing a high temperature ($10^5$ K) chromosphere boundary to act as a mass reservoir for wind and streamer zone. Increasing the resolution ($1024^2$ on a radially stretched 30 $R_\odot$ domain) showed a clear tendency for changing from bursty rain formation to a more continuous, smaller-scale rain manifestation. Synthetic Doppler maps looking down on the streamer show speed variations that reflect the falling rain as well as the chromospheric evaporations, and clearly locate at the streamer periphery, in accord with observations.
Further work should clarify the role of condensation formation, the role of interchange reconnection, and the possibility to have combined tearing and TI/TC instabilities (as in \cite{Sen2022,Sen2023,Jordi2025}) at these open/closed field boundaries, on the acceleration and internal variability of the (slow) solar wind.

That interesting thermodynamic evolutions can appear in more realistic, complex topologies was observationally established by \cite{Mason2019}, where rain was witnessed in the typical spine-fan topology of a parasitic dipole within a unipolar region. This involves a coronal null point, and in three separate active regions characterized by this topology, rain appeared not only on the closed loops inside the fan, but also near the separatrices (spines and fan, and near the null). It was suggested that both interchange reconnection and thermal non-equilibrium (with the basic underlying TI process) should be at play in such settings. That reconnection itself can act as a trigger for TI condensations was observed by \cite{Kohutova2019}. It was shown that reconnection between a (prominence-carrying) flux rope and surrounding coronal field led to impulsive heating in the higher coronal loop regions. This observation supports the emerging view that TI can develop anywhere in the corona, provided some trigger mechanism brings the region into local runaway conditions. Hence, exclusively footpoint-located heating is not a necessary ingredient, although it is known to aid in condensation formation.

\begin{figure}[ht]
\centering
\includegraphics[width=0.95\textwidth,height=12cm]{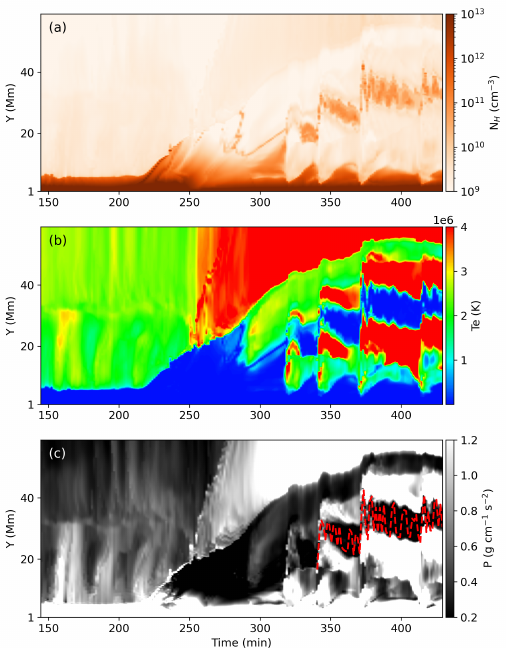}
\caption{Height-time diagrams of density (top), temperature (middle) and pressure (bottom), taken along the center of a gradually rising flux rope and its underlying reconnection current sheet. Notice how repeated reconnection injects prominence matter (dense/cool) in the bottom of the flux rope, which rises each time an injection occurs. The filament oscillates (red curve, tracing the density structure) in response. From~\cite{Li2025}.}\label{f-li25}
\end{figure} 

A first numerical demonstration of this reconnection-induced rain pathway was recently presented in \cite{Beatrice2025}, where a two-fluid (plasma-neutral) coronal-only model of a spine-fan topology was found to develop rain. Although gravity was not incorporated in the study, the reconnection enforced by a bottom boundary driving was sufficient to trigger higher density seeds that became TI unstable. The two-fluid effects included ionization and recombination processes, with local velocity decoupling effects at play. Interestingly, compared to the earlier MHD simulations of rain as triggered by evaporation-condensation, the actual rain formation was acting much faster in this reconnection-triggered pathway. It would be valuable to perform these 3D spine-fan studies in two complementary ways: (1) in the frozen-field settings of Section~\ref{s-ffhd}, where by construction the interchange/reconnection processes are suppressed; and (2) in more realistically stratified chromosphere to coronal settings, with full multi-dimensional MHD thermodynamic effects.

As an example of the latter, \cite{Li2025} studied flux-emergence in a 2.5D chromosphere to coronal simulation, varying the main parameters of the emerging field region. This can lead to plasmoid-forming current sheets bounding the emergence region, where multithermal jets and filaments could form and evolve in self-consistent evolutions. Depending on the orientation of the emerged field, a particular emergence-driven prominence formation scenario was identified where Rayleigh-Taylor and Kelvin-Helmholtz related fine structure can result. Indeed, by locally emerging field that is oriented orthogonal to the original dipolar arcade, one can form a filament channel, and it was shown that repeated reconnection below the forming flux rope can bring in cool matter in bursts, that collects as a rising prominence. This is shown in the time-distance views of Fig.~\ref{f-li25}, where the vertical height variation is oriented along the slanted current sheet underneath the flux rope. This flux-emergence driven prominence scenario should again be revisited in full 3D settings, where the repeated injection of chromospheric matter due to reconnection is expected to lead to interesting thread-like morphologies.

\section{Postflare rain}\label{s-pfrain}

\begin{figure}[ht]
\centering
\includegraphics[width=\textwidth]{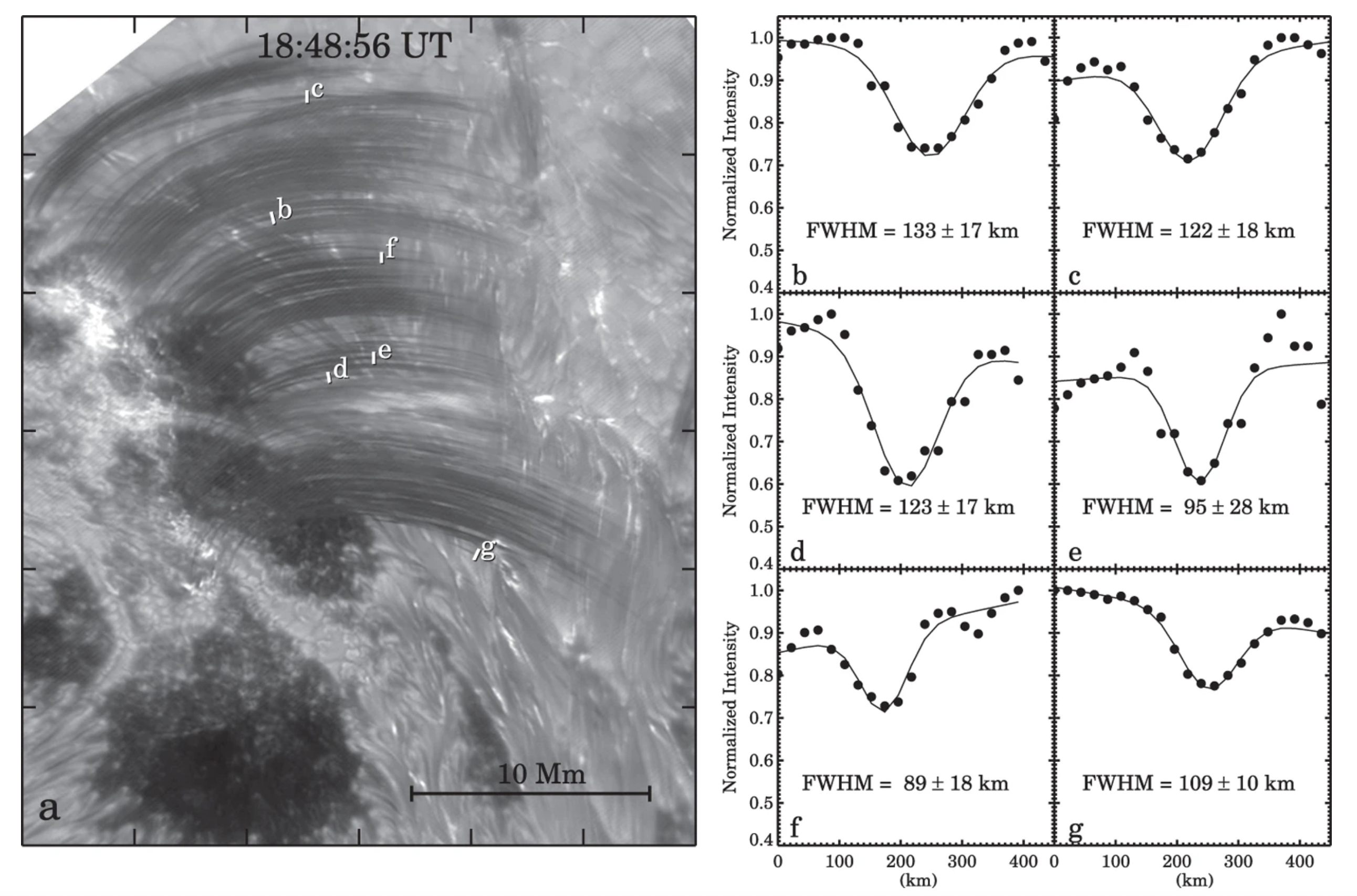}
\caption{H$\alpha$ image (left panel) of a flare arcade, showing condensations following arched flare loops. The six panels (b,c,d,e,f,g) on the right quantify intensity variations across indicated slits on the arched loops, allowing to quantify rain strand widths around 100 km. From~\cite{Jing2016}.}\label{f-jing}
\end{figure}

As mentioned in our introductory Section~\ref{sec-intro}, a rather spectacular demonstration of multiphase dynamics with spontaneous condensations forming is witnessed in association with solar flares. \cite{Mason2022} analysed 241 flares, and thereby discovered a positive correlation between GOES flare class (i.e. energy released) and the duration of postflare rain events. In accord with \cite{Svestka2007}, we note that reconnected lower-lying loops are an integral part of the entire eruptive flare loop system, and should not be termed `postflare' loops, but we adhere to  calling rain that may develop after the initial phase as postflare. \cite{Jing2016} used the adaptive optics at the Goode Solar Telescope at Big Bear to showcase unprecedented fine structure during an M-class flare, where the high cadence (28 s) H$\alpha$ views presented intricate details of flare ribbons, reconnected flare loops refilled with chromospheric material and postflare coronal rain. Brightenings associated with rain clumps impacting the chromosphere could be identified, and individual rain strands with cross-sections of about 100 km (see Fig.~\ref{f-jing}) were found throughout.

By analysing IRIS-spectra and EUV images, \cite{Seray2024} investigated the mass and energy cycle in a C2.1 flare, where phases of chromospheric evaporation and coronal rain could be clearly identified, throughout the evolving flare. Rain quantity and intensity was found to increase by factors from three to six, when comparing preflare to gradual flare phases. 

This postflare rain poses context-specific challenges to modelers, where reconnection-based energy release causes multi-million degree plasma to ultimately transform into chromospheric condensations and sometimes postflare rain is already observed from the peak of the impulsive phase onwards. In the many 1D hydro models discussed in Section~\ref{s-1d}, footpoint located heating acts favorably to induce TI driven condensations, but significant flare loop thermodynamics in the standard flare model results from energy transport down from the reconnection site to the chromosphere due to non-thermal particle beams. The lighter electrons are usually invoked in this process, but analytic models for both electrons and protons that collisionally interact with a cold hydrogen target (the chromosphere) exist \citep{Emslie1978}, and they quantify the energy deposition rate as a function of column depth. The parameters involved are then setting the injected electron beam flux spectrum (with a cut-off and a spectral index). These beams are accelerated at the reconnection site by kinetic processes beyond the usual MHD models, then guided by the magnetic fields, making the beams penetrate and heat the chromosphere. This causes chromospheric evaporation which rapidly refills flare loops with denser matter, as these beams act for few minutes at most. 

In \cite{Reep2020}, 1D hydrodynamic models of beam-injected flare loops that incorporate radiative losses with non-equilibrium ionization effects, thermal conduction and account for area expansion of the loop, provided a rather negative answer: it was impossible to get rain condensation under typical parameters for flare loop systems. The model parameters changed both the loop geometry as well as the electron beam parameters (such as duration and energy cut-off). When a secondary weak footpoint heating (besides the beam-driven effects) was added, rain condensations could occur. Again focusing on short, hot loops impacted by impulsive electron beam heating, \cite{Reep2022} varied the area expansion factors, and studied changes from semi-circular to elliptic loop shapes. The 1D assumption still implies that the loop shape only enters as a projected gravity as explained in Section~\ref{s-1d}, but a puzzling finding was the lack of postflare coronal rain formation throughout all models explored. 

A novel ingredient in the 1D radiative hydro models of impulsively heated (due to electron beams) flare loops is the incorporation of spatio-temporally varying elemental abundances. Chromospheric evaporation shifts the local abundances, and changes the local heat-loss effects. It was shown in~\cite{Reep2025} that this ingredient suffices to cause rain condensations in impulsively heated or flare loops. By enriching the 1D hydro model with a continuity (advection) equation for the spatiotemporal abundance of the low first-ionization potential (low-FIP, where low is below 10 eV, like Fe, Si, Mg) elements, and incorporating this factor in the cooling curve prescription, one could trigger rain where earlier models failed. \cite{Reep2025B} followed this up with further 1D simulations, suggesting that the occurence of postflare coronal rain and the initial location of assumed low-FIP enhancements that get advected into the loop may well be correlated. We note that these works concentrated on the normal FIP effect under solar flare conditions, while observations also identified a reverse FIP effect in flares \citep{Laming2021}.

\begin{figure}[ht]
\centering
\includegraphics[width=\textwidth]{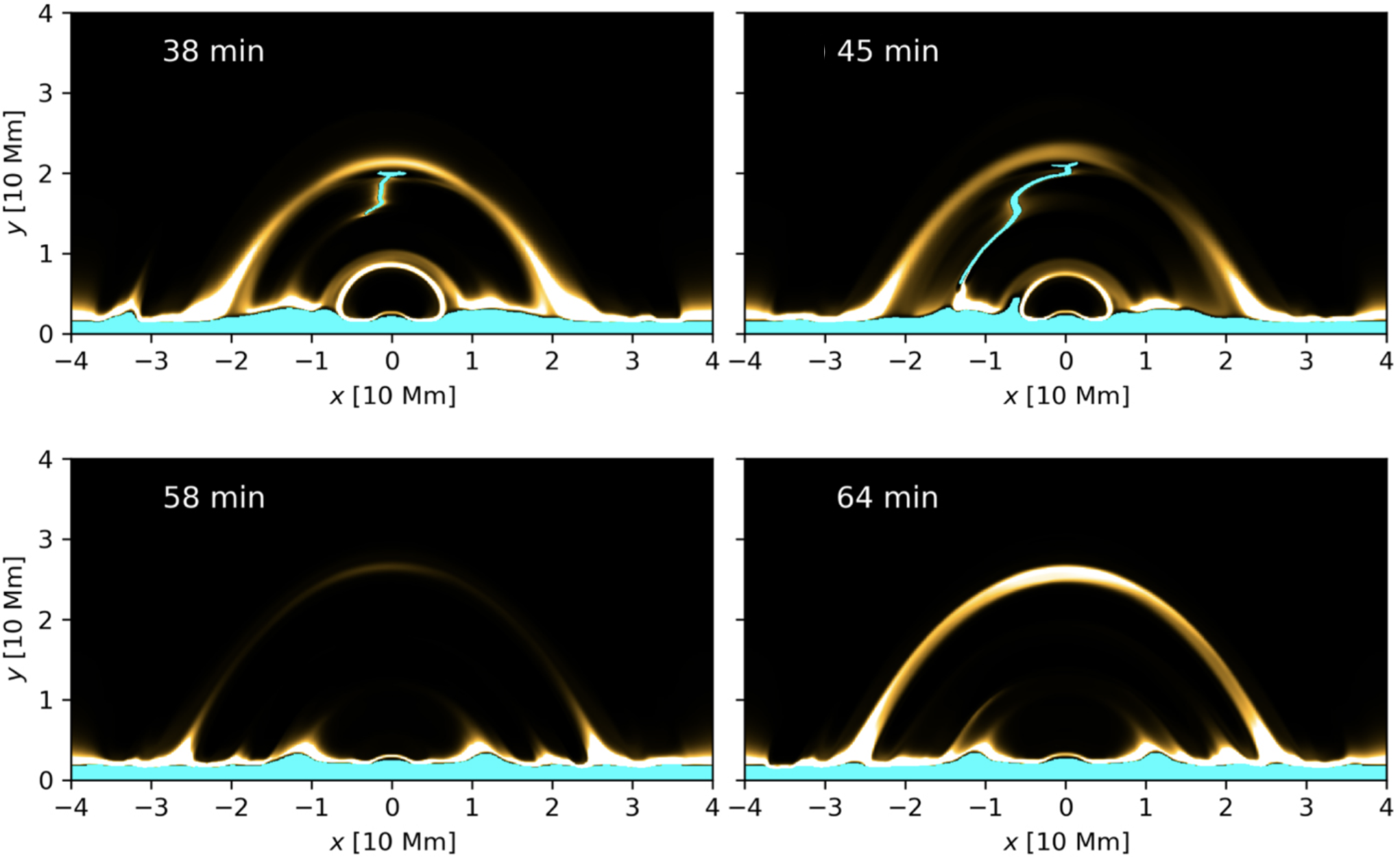}
\caption{Synthetic EUV views on postflare rain arcades demonstrate how hot loops may temporarily disappear from view, due to rain triggered on neighboring arcade fieldlines. The rain (and the chromospheric) matter is indicated in blue. From~\cite{Ruan2021}.}\label{f-loopdim}
\end{figure}

While these 1D models focus on more advanced physical ingredients accompanying chromosphere to coronal conditions in flare loops, various multi-dimensional MHD models emerged that do produce postflare rain due to thermal instability. By following a standard 2.5D flare model (involving an initial vertical current sheet) from the preflare, through the impulsive and into the gradual phase, \cite{Ruan2021} showed postflare rain appearing in two consecutive episodes that both lasted about 15-20 minutes. The rain formed about half an hour after the impulsive phase, and these MHD models did not include any non-thermal beams that act to transport energy rapidly and non-locally. An interesting multi-dimensional finding was that coronal rain formation on neighboring loops could cause hot, bright loops (as visible in EUV) to temporarily disappear from view, until these loops in turn refilled. This is illustrated in Fig.~\ref{f-loopdim}. The flare loop refilling in these models is due to wave and plasma reflections on the transition region, as well as due to thermal conduction. Also Lorentz forces enter, as loops contract or expand in an overall non-force-free evolution. Future models must combine the insights of 1D beam models into such multi-dimensional flare evolutions, as already achieved for the purely impulsive flare phases in \cite{Ruan2020,Malcolm23,Malcolm24}. We note that the non-local feedback of parametrized non-thermal electron beams in these self-consistent multi-dimensional models \citep{Ruan2020,Malcolm23,Malcolm24} follows the same prescriptions used in 1D flare loop evolutions (such as those in \cite{Reep2020}), namely variants of an analytic cold target model from \cite{Emslie1978}. However, the multi-dimensional integrated beam+MHD models  go way beyond the 1D findings and also can address the role of trapping of nonthermal electrons in favorable magnetic bottle configurations, explaining loop top hard X-ray sources.

Another 2.5D realization of postflare rain was presented in \cite{Samrat2024}, where an initially non force-balanced configuration developed multiple (homologous) erupting magnetic flux ropes, to eventually settle to a post-eruptive configuration. Only the coronal region was simulated, and about half an hour after the last eruption, TI driven condensations were demonstrated. While the rain blob dynamics clearly matched observational behavior in terms of their arched pathways and speeds, the actual condensation timings and the ad-hoc initial conditions of a non-equilibrium state are aspects to improve upon in further model efforts. It is interesting to note that this purely coronal model produced postflare rain, without any role for chromospheric evaporation.

\begin{figure}[ht]
\centering
\includegraphics[width=0.49\textwidth,height=5cm]{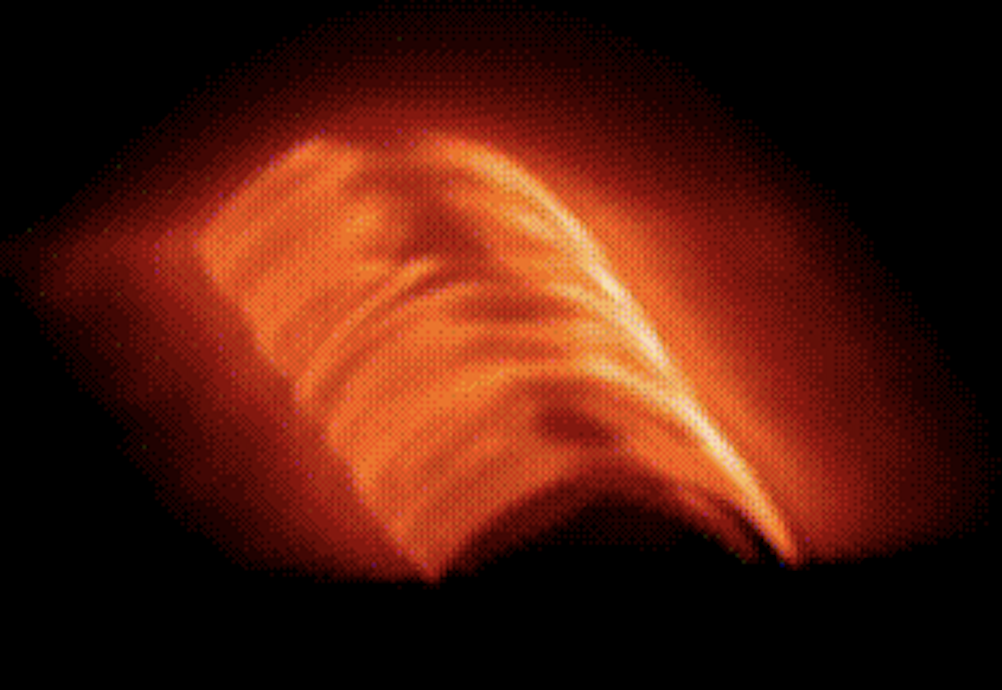}\includegraphics[width=0.49\textwidth,height=5cm]{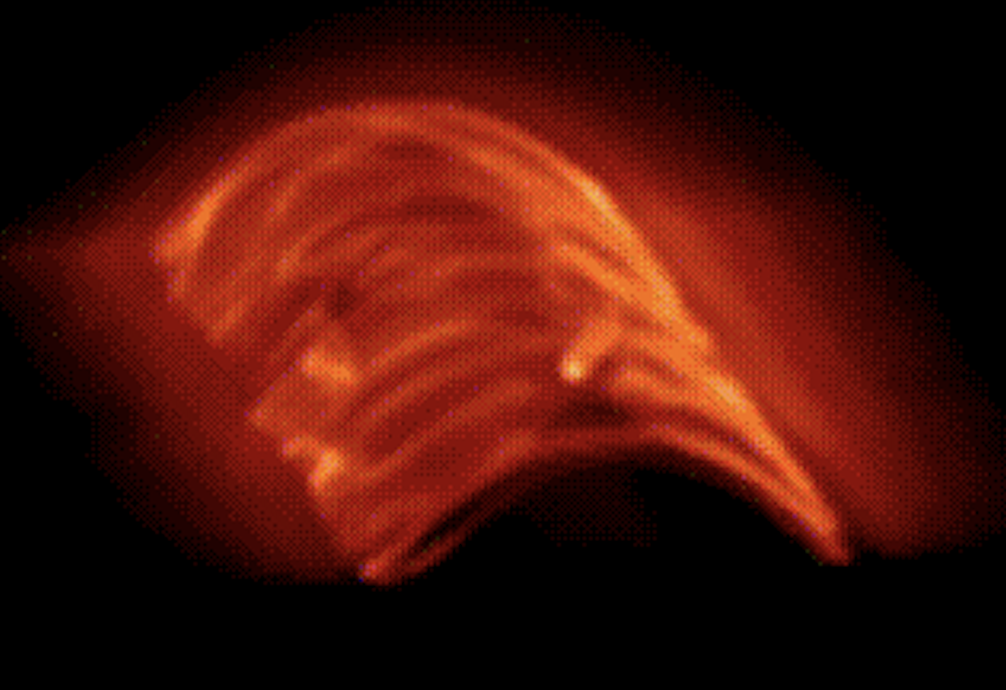}
\includegraphics[width=0.49\textwidth,height=3cm]{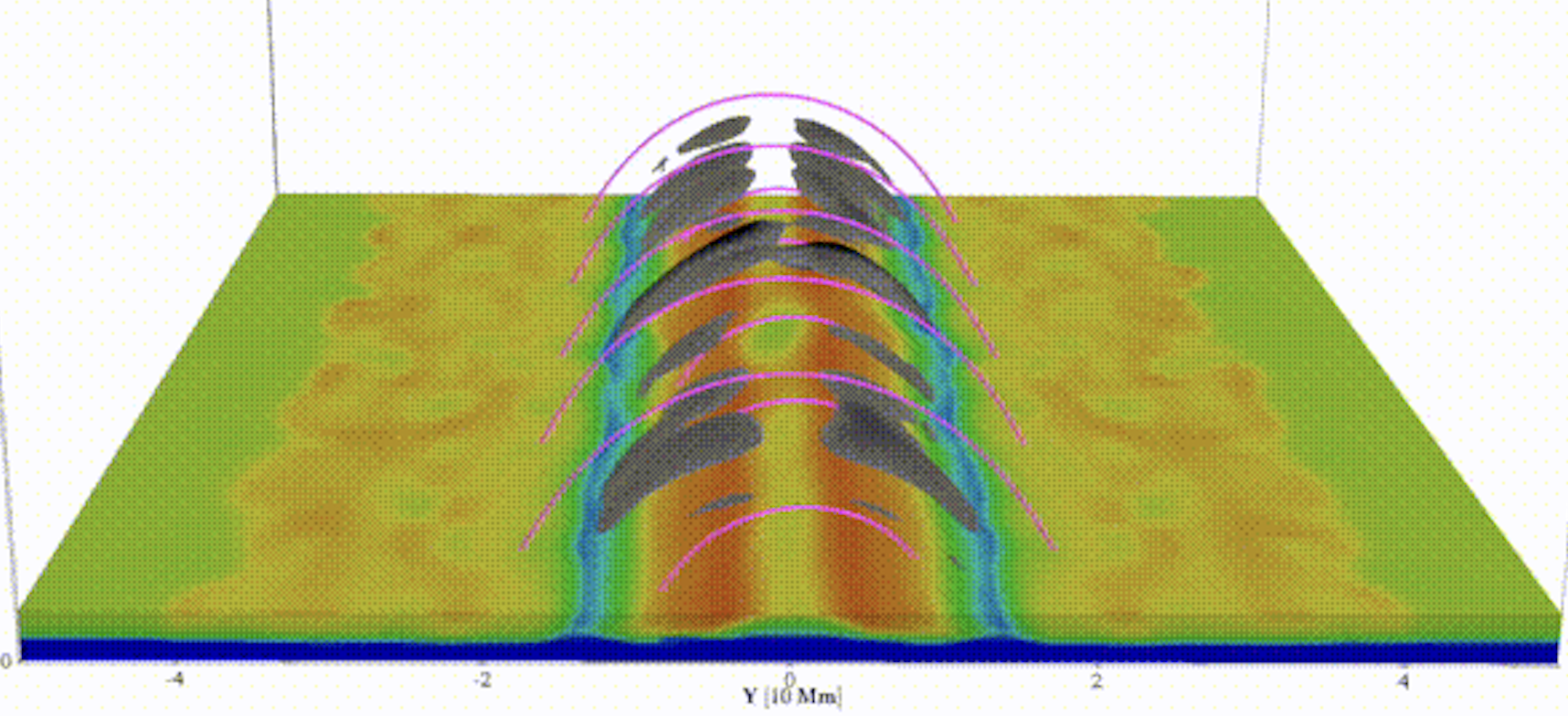}\includegraphics[width=0.49\textwidth,height=3cm]{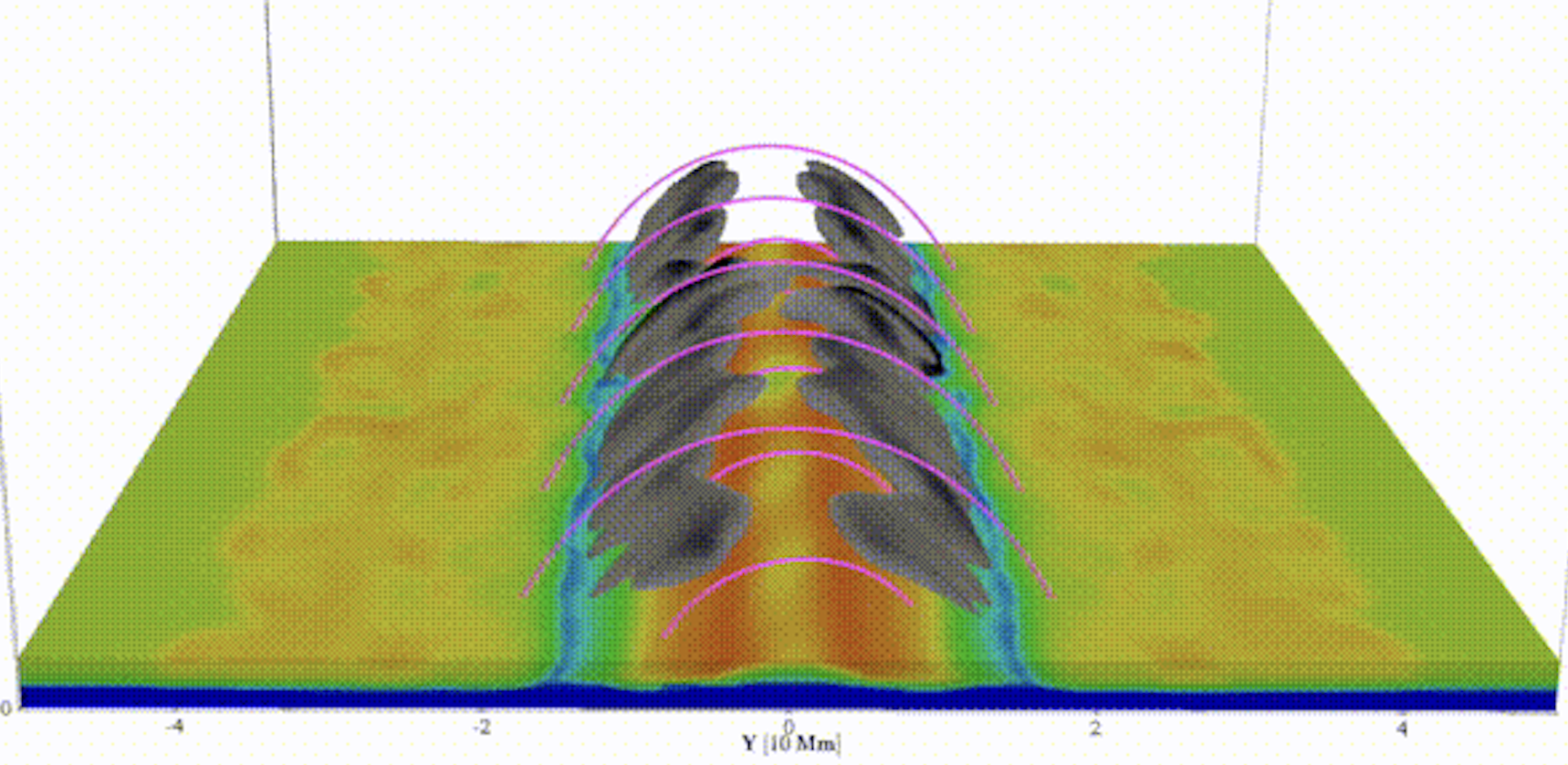}
\caption{Synthetic EUV views on postflare rain arcades (top), taken several minutes apart. Later rain events happen on higher loops, and cool dense matter follows the loop shapes closely. Bottom panels show selected field lines, the temperature variation (in color) and temperature isocontours (in grey) that highlight the rain draining down. From~\cite{Ruan2024}.}\label{f-pfrain}
\end{figure}

Recently, the modeling for postflare rain progressed to a full 3D MHD model by \cite{Ruan2024}, who followed the standard flare from onset to decay, producing postflare rain that closely matches several aspects of observed counterparts. In standard flare settings, a vertical current sheet is combined with an anomalous resistivity prescription, that triggers reconnection and realizes the flare loop formation. With the model atmosphere spanning from chromosphere to corona, the impact of reconnection downflows on the flare arcade induces Kelvin-Helmholtz turbulence in the loop tops during the impulsive phase \citep{Ruan2023}, that can be followed by Richtmeyer-Meshkov instabilities near and below the termination shock \citep{Shen2022}. \cite{Ruan2023} demonstrated that the (impulsive phase) loop top turbulence in the model is in perfect agreement with nonthermal velocity distributions inferred from Hinode EUV-Imaging Spectroscopy. The Richtmeyer-Meshkov deformations relate to supra-arcade downflows. The flare loop system is found to realize a non-force-free magnetic topology in the entire gradual phase, where the vertical plasma pressure variation combines to an overall force-balanced state \citep{Ruan2024}. In that non-force-free setting, Rayleigh-Taylor turbulence develops throughout the loop top and current sheet regions. All these processes ultimately create the right conditions for TI runaway condensations in the refilled flare loops, again relatively late in the gradual phase. Fig.~\ref{f-pfrain} shows synthetic EUV views, and temperature isocontour views, to show field-guided condensations raining down the loops at observed speeds and morphologies. Note how rain is seen in isocontours of temperature to follow the arched loop shapes, and how later rain events manifest on higher lying loops. When compared to Fig.~\ref{f-jing}, there is finer detail with more strands visible in the observational H$\alpha$ views. \cite{Seray2024} made a detailed observational analysis of the role of chromospheric evaporation and postflare rain to the energy cycle of a C-class flare, and used total pressure equilibrium to deduce the flare loop magnetic field strengths during impulsive and gradual phases, finding a decrease from 24 to 11 Gauss, rather similar to the 3D \cite{Ruan2024} model.

In summary, while we currently do have both 1D and multi-D numerical manifestations of postflare rain, further insights are expected when combining their key ingredients: the role of non-local energy depositions by beams must still be included in multi-D postflare rain simulations, while the sophisticated 1D models now point to abundance effects as key to get postflare rain in impulsively heated loops. The multi-dimensional models so far only develop rain with a clear delay beyond the impulsive phase, and poses challenges to observed events with rain observed a few minutes beyond the peak. This should be investigated across flare classes, in simulations that handle both beams and evolving abundances in 2D and 3D setups.

\section{Beyond solar condensations}\label{s-beyond}

The basic mechanism of thermal instability was recognized from the start to be universally applicable \citep{Parker1953,Field1965,Balbus1986,Waters2019,Falle2020,KeppensTC2025}, capable of explaining condensation formation without invoking self-gravity. The gas dynamic HD and ideal MHD equations are also fully scale-invariant \citep{goed2019}, and hence can be applied to many different astrophysical contexts. Here, we highlight some examples of condensation formation models that closely relate to the solar case, where very similar numerical HD or MHD approaches were used.

Exoplanet survey missions like Kepler and the Transiting Exoplanet Survey Satellite (TESS) also boosted research on stellar flares, where optical stellar flare lightcurves show strong similarities (usually an impulsive phase with gradual decay) with the solar case.  Superflares on M dwarf stars have inferred energies above $10^{32}$ ergs, and a good fraction of those has a highly impulsive peak followed by a second, more Gaussian peak. In \cite{Yang2023}, 1D hydro simulations of flare loop dynamics in fixed-shape semi-circular loops were performed, where flare energy injections act as localized, time-dependent heating sources at loop apex and footpoints. Postflare loop condensations formed due to thermal instability. The gradual phase was analyzed by translating the obtained thermodynamic evolution to an optical light curve, assuming  optically thin (free-free and free-bound) continuum emission from fully ionized plasma from the coronal part of an off-limb flare loop. The postflare coronal rain condensations were shown to lead to a pronounced secondary peak, consistent with the late phase stellar flares observed by TESS. In \cite{Wollmann2023}, H$\alpha$ spectral line asymmetries in a stellar flare for the dMe star AD Leo were explained by  non-LTE radiative transfer computations for rain clouds in free fall along a fixed arcade, and synthetic line profiles are consistent with the observed 50 $\mathrm{km}\,\mathrm{s}^{-1}$ red wing enhancements, signaling downflows. 

\cite{Peng2017} presented a 2.5D MHD simulation of the formation of a galactic prominence, which closely follows the levitation-condensation model from \cite{Kaneko2015B}. What differs is the scale of the setup (a domain of order $400 \times 400$  pc$^2$), the time scale of the evolution (following the evolution for 120 My), as well as the typical thermodynamic conditions. The latter also imply a different cooling curve that recognizes that matter in the galactic central region can be in two stable states: a cold neutral medium (order 10-100 K) or a warm neutral medium (around 10$^4$ K), with a thermally unstable regime in between. In all other respects, the simulation is performed like the solar case, where an initial arcade gets deformed by footpoint motions to a flux rope, that rises and forms a prominence by thermal instability. After the levitation-condensation process, the simulated galactic prominence was a monolithic structure of about 60 pc vertically, and 7 pc in width, making it of order $7 \times 10^4 \, M_\odot$ in mass. This structure could be related to observations of molecular loops in the central galactic region. A detailed study of the observational properties of these galactic loops is found e.g. in \cite{Torii2010}.

Motivated by cool star surveys studying H$\alpha$ asymmetries that indicate the presence of moving, cool dense matter in their surroundings, \cite{DaleyYates2023} produced a model of coronal rain on a young sun. The solar rotation rate (and the typical coronal magnetic field) was originally much higher than its present value, since solar wind mediated angular momentum loss occurred throughout its main sequence lifetime. Adopting a solar rotation period of 1 day, and a polar field strength of 100 G, one encounters new numerical challenges to model the solar atmosphere and wind under such conditions. This is best done in the co-rotating frame (introducing centrifugal and Coriolis forces), and by solving equivalent nonlinear equations that split off the strong dipole magnetic field. With the necessary ingredients of heat-loss in the energy equation, it was found that coronal rain could again develop within the closed field regions, with downflow speeds that agree with the ones in spectral H$\alpha$ line asymmetries. Quasi-periodic behavior (rain cycles) could be identified, with periods of 37 hours, somewhat longer than those known for the present solar case.

\begin{figure}[ht]
\centering
\includegraphics[width=0.9\textwidth]{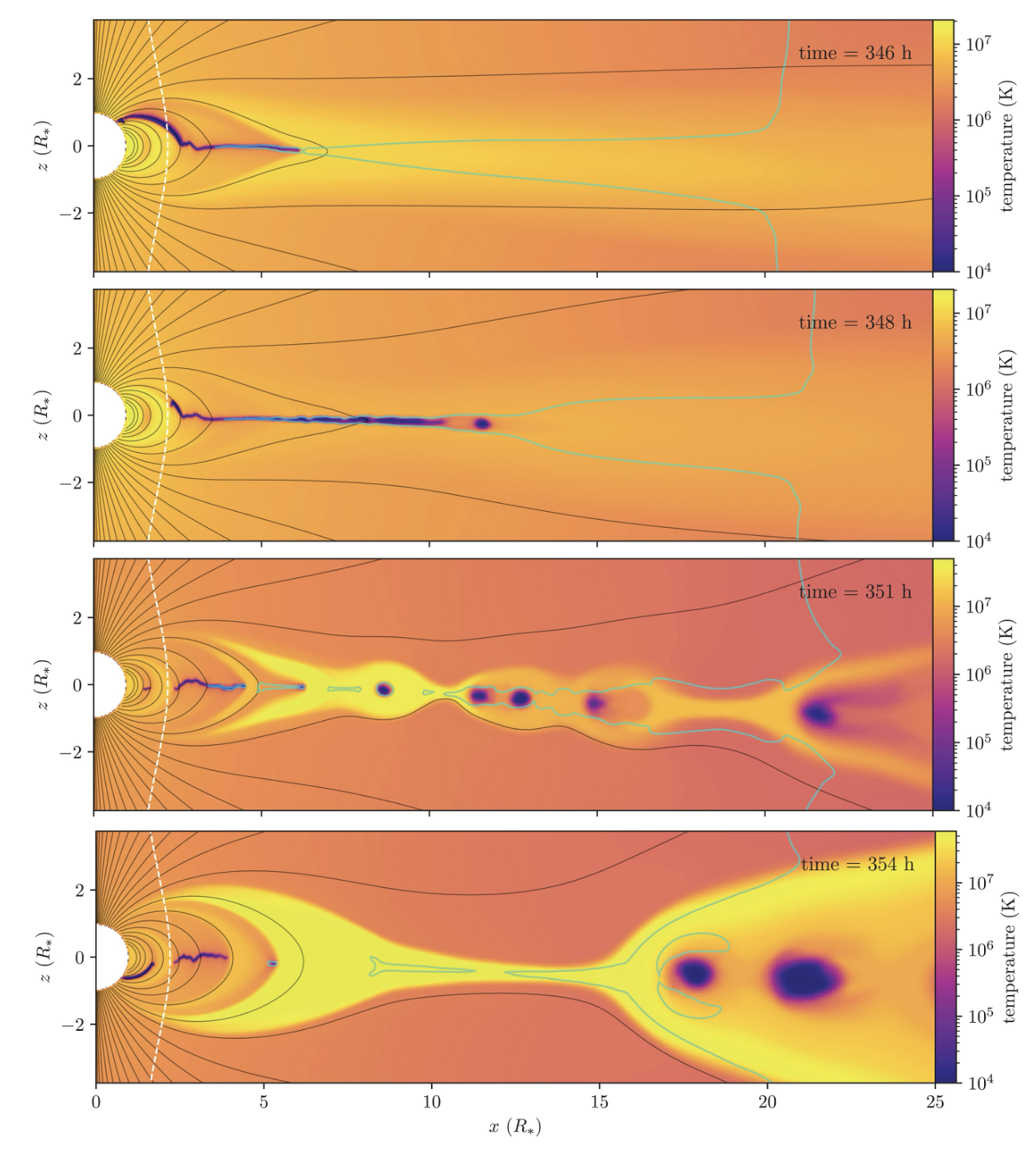}
\caption{Temperature views on the axisymmetric stellar magnetosphere evolution of a rapidly rotating, strongly magnetized solar-type star. The corotation radius is indicated in white dashes, the magnetic field lines are shown, and a single centrifugal ejection event is shown at different epochs. From~\cite{DaleyYates2023}.}\label{f-simon}
\end{figure}

The same H$\alpha$ profiles from stellar coronae also revealed the possibility of so-called slingshot prominences, which could be located near the co-rotation radius where a balance between centrifugal forces and gravity occurs, and then get ejected. \cite{DaleyYates2024} simulated what happens in rotating, highly magnetized stellar atmospheres and winds, this time confronting a case with a 0.38 day versus a 3.8 day rotation period. They correspondingly adjusted the magnetic field strength and the overall heating and radiative loss prescriptions. As shown in Fig.~\ref{f-simon}, the faster rotating, centrifugal magnetosphere shows condensation formation near the co-rotation radius (at roughly 2 stellar radii) where the part forming beyond this radius gets centrifugally ejected. The equatorial current sheet where also the condensation resides then tears and plasmoids get ejected and merge. These ejections could be followed up to the domain size of 50 stellar radii (double that shown in Fig.~\ref{f-simon}). During the simulated episode of nearly 400 hours, the simulation demonstrated about 18 of these slingshot prominences, going out at a typical 813 km/s. Compared with the lower rotation rate (longer period) cases, which formed more rain-like events without breakouts, these centrifugal regimes with slingshot prominence ejections are important when quantifying stellar mass loss rates and may impact stellar rotational evolution through modified angular momentum loss rates. Both regimes (slow and fast rotating solar-type stars) rely on thermal instability to trigger the in-situ condensations.

Finally, we note that many studies exist in the context of multiphase matter formation and evolution in the interstellar (ISM) to intergalactic medium (IGM), or even in the intercluster medium (ICM). These studies routinely invoke TI to explain how cool, dense matter can form in-situ \citep[e.g.][]{Sharma2012} and usually feature background flows representing outflow (winds) or accretion configurations. 

\section{Open problems and challenges}\label{s-open}

We end this review by listing a number of challenges that persist when modeling condensation formation in the solar atmosphere (or in related astrophysical) settings. It thereby remains imperative for modelers to confront observations of coronal rain and prominences in an even more quantitative fashion, e.g. by means of synthetic spectra in multiple wavebands, in direct correspondence with existing instrument specifications. 
\begin{itemize}
\item While high-resolution numerical models do reveal aspects of the (observed) fine-structure in prominences that form inside flux ropes, it is  unexplored how much of this fine-structure is encoded in its spectrum of linear normal (eigen)modes, which can host global as well as ultra-localized thermal instabilities from the thermal continuum \citep{KeppensTC2025}. The link between spectral theory where eigenmodes can be computed exactly, versus detailed linear as well as nonlinear evolutions from given initial conditions, is particularly relevant for prominence and coronal rain studies, since they share the common physical origin of radiatively driven runaway. How such eigenmode-based insights connect to the findings in 1D hydro `loop' settings, where varying field line geometry, parametrized heating and radiative loss curve selection revealed cyclic behavior with and without condensation formation (thermal non-equilibrium) is part of ongoing research. There, we can benefit from model efforts on multiphase behavior in interstellar, circumgalactic or intercluster medium, where TI is known to drive cold filaments forming out of the hot phase \citep{Sharma2012}. The major difference in a solar setting is the intricate thermodynamic coupling between chromosphere and corona across a pre-existing transition region, and the energy and mass cycling between these regions. These solar-specific couplings have been explored extensively in 1D loop settings. The intrinsic nonlinearity in the governing hydrodynamic equations allows for limit cycle behaviour, which must be linked to the (evolution of) the normal mode spectrum: non-adiabatically modified p-modes and thermal continuum modes can be computed for any moving (even force-unbalanced) background variation. In actual multi-dimensional MHD settings, the role of finite perpendicular thermal conduction is known to introduce fine-structured magnetothermal  eigenmodes which can be overstable \citep{vdl1991B,vdlslab1991}, and their consequences on nonlinear evolutions have never been  explored to date. This may well be linked to the important aspect of what dictates coronal rain blob widths, which are near (or below) current resolution limits. Observations indicate that the thickness of a rain clump is constant as it falls (Fig.~10 in \cite{Seray2023}), with lengths an order of magnitude larger, although especially the widths are known to be overestimated.
 \item  Many findings on filaments or prominences quantify Doppler shifts, with recent examples like \cite{Karki2025} providing evidence for counterstreaming velocities in quiescent filament strands at the arc-second scale. One of the outstanding problems to filament formation models is to understand how such multi-strand structure can form, persist and evolve dynamically. These counterstreaming mass flows can show up as a Doppler bullseye pattern when seen at the limb as a prominence. \cite{Zhou2025} provide a 3D MHD model (with an ad-hoc inserted filament mass) which nicely reproduces the effect of localized coronal-jet-related heating: counterstreaming flow patterns develop naturally in the flux rope and prominence body. It is still an open question whether one can have both counterstreaming as well as rotational dynamics established during the in-situ formation. Rotational motions have been identified in (lower dimensional) formation models \citep{Valeriia2023}. Further spectroscopic quantifications must augment synthetic images. These should include multi-dimensional, non-LTE physics, at varying viewing angles, that incorporate the prominence- or coronal-rain-specific situation of irradiated structures suspended in the solar corona, as initiated in \cite{Osborne2025}.
\item In a somewhat related context, the internal structure and existence of `prominence tornadoes' is also heavily debated \citep{Gunar2023}, where perceived helical motions that suggest rotation may be argued away through projection effects and oscillatory or counterstreaming motions. There is as yet no model that actually realizes a tornado-like, strongly rotating magnetic structure which forms and  carries prominence condensations. That an overall force-balanced, stationary and axisymmetric structure can exist has been shown in \cite{Luna2018}, where the added centrifugal forces enrich the force balance of our Eq.~(\ref{mombal}). In that sense, there is clearly a theoretical possibility to form counterparts of the well-known weather phenomenon, enriched by condensations due to thermal instability. As the thermal continuum in helical, rotating structures is a robust ingredient of the (unstable) MHD spectrum \citep{Hermans2024}, follow-up nonlinear studies could explore what kind of structures can develop in the stratified settings of our solar corona.
\item The role of partial ionization on condensation onset and evolution is only studied so far in simplified settings: the first 1D plasma-neutral model of prominence formation on a fixed loop \citep{Veronika2025} ignores all multi-dimensional aspects, while the 3D two-fluid setup to trigger coronal rain of \cite{Beatrice2025} did not include gravity, but did include important ionization and recombination aspects. Multi-dimensional setups are needed to explore how partial ionization, recombination, as well as realistic stratification all impact condensation formation. Decoupling effects for coronal rain blobs have been studied for initially inserted density structures in a simple stratified medium, such as in 1D settings by \cite{Oliver2016} or in 2D by \cite{David2022}. Collisional interactions play a role in secondary structure formation when RTI development is followed from initial density inversions \citep{Beatrice2021}, but these effects are localized at thin edges of RTI plumes and are likely beyond observational limits. Nevertheless, the PCTR variation implies important changes in collisional frequencies, that surely impact the physical evolutions.
\item Multi-dimensional MHD models must still improve on the way in which important net radiative cooling effects impact condensations and their internal structuring, as identified by 1D non-LTE radiative transfer treatments of idealized (isobaric and isothermal) prominence slabs. Recently, \cite{Gunar2025} tabulated net radiative cooling rates (or NRCRs), as well as electron density and ionization degrees, for voxels that are aware of their location with respect to illumination sources, in particular differentiating between vertical, horizontal bottom or horizontal top bounding surfaces. These tables provide a significant update to early efforts by \cite{Kuin1991}. Knowing the distance into the condensation from the bounding surface, as well as the local pressure and temperature, these NRCRs quantify the full optically thin to thick radiative transfer effects in a consistent fashion, and could be used as local source terms in numerical simulations which spontaneously develop and evolve radiatively driven condensations. These NRCRs can modify the internal temperature structure of each individual rain blob or larger-scale prominence, but it will be non-trivial to make clear choices between the given three bounding surface orientations at each numerical grid cell, as e.g. Rayleigh-Taylor deformations lead to strongly curved boundaries, and shadowing effects when neighboring blobs appear are not easily incorporated. Nevertheless, non-LTE net heating-cooling is known to affect all cool plasma embedded within the corona, especially for temperatures below 30000 K, such as present in spicules, cool jets or coronal loops and in prominence and rain settings. A recent review and discussion is provided in \cite{Heinzel2025}. Improving the details of non-LTE processes may lead to more physically motivated internal (low) temperature values in condensations. Temperatures as low as 2000 K for coronal rain blobs were inferred observationally in \cite{Antolin2015}, to be contrasted with even lower quiet sun chromosphere locations identified in 2D radiative MHD simulations \citep{Leenaarts2011}.
\item We do not yet fully understand how the heating-cooling balance dictates the detailed thermodynamical evolutions in loops or open field line structures. It is certain that the presence of coronal rain and the morphology of filaments and prominences somehow encodes information on this balance, and the models thus far suggest that we can trigger condensation formation in almost any magnetic topology. The 1D hydro models do narrow down the search for the most important physical ingredients, with e.g. the role of footpoint heating as agreed in most models, or of elemental abundances \citep{Reep2025} seemingly crucial for postflare rain condensations. Stepping to multi-dimensional settings, with actual MHD dynamics both along and across the loop diameter, can drastically modify findings based on 1D loop models alone, as e.g. postflare rain was succesfully modeled in 2D and 3D flare loop setups \citep{Ruan2021,Ruan2024}, while 1D models fail unless abundance variations are incorporated. That coronal rain, in quiescent or flare settings, is so similar to filaments and prominences, agrees with the basic thermal instability mechanism acting universally to explain in-situ condensations, even in other astrophysical settings. How it triggers rather different fine-structure according to the magnetic topology or heating prescription is as yet an open question. In that context, we need model efforts like \cite{Mok2016}, where detailed active region evolutions that showed coronal rain were translated to synthetic EUV observations. These efforts already showed encouraging correspondence with actual observations showing clear non-equilibrium (i.e. non-steady) thermal evolutions. 
\item Most models we discussed here do not include sub-photospheric regions, which will obviously impact the chromospheric to coronal responses. A preview on a comprehensive simulation of a solar prominence that forms by a combination of injection and in-situ condensation is shown in Fig.~\ref{f-lisa}, where a bipolar arcade is evolved from subphotosphere to corona. A seed cool plasma parcel is injected in a pre-existing magnetic dip, 
due to underlying convective processes, and then feeds further condensation formation. Such models, and the role of more sophisticated radiative processes from optically thick to thin (occurring across the unit optical depth surface $\tau_{500}=1$ for wavelengths at 500 nm as indicated in Fig.~\ref{f-lisa}) are urgently needed.

\begin{figure}[ht]
\centering
\includegraphics[width=0.9\textwidth]{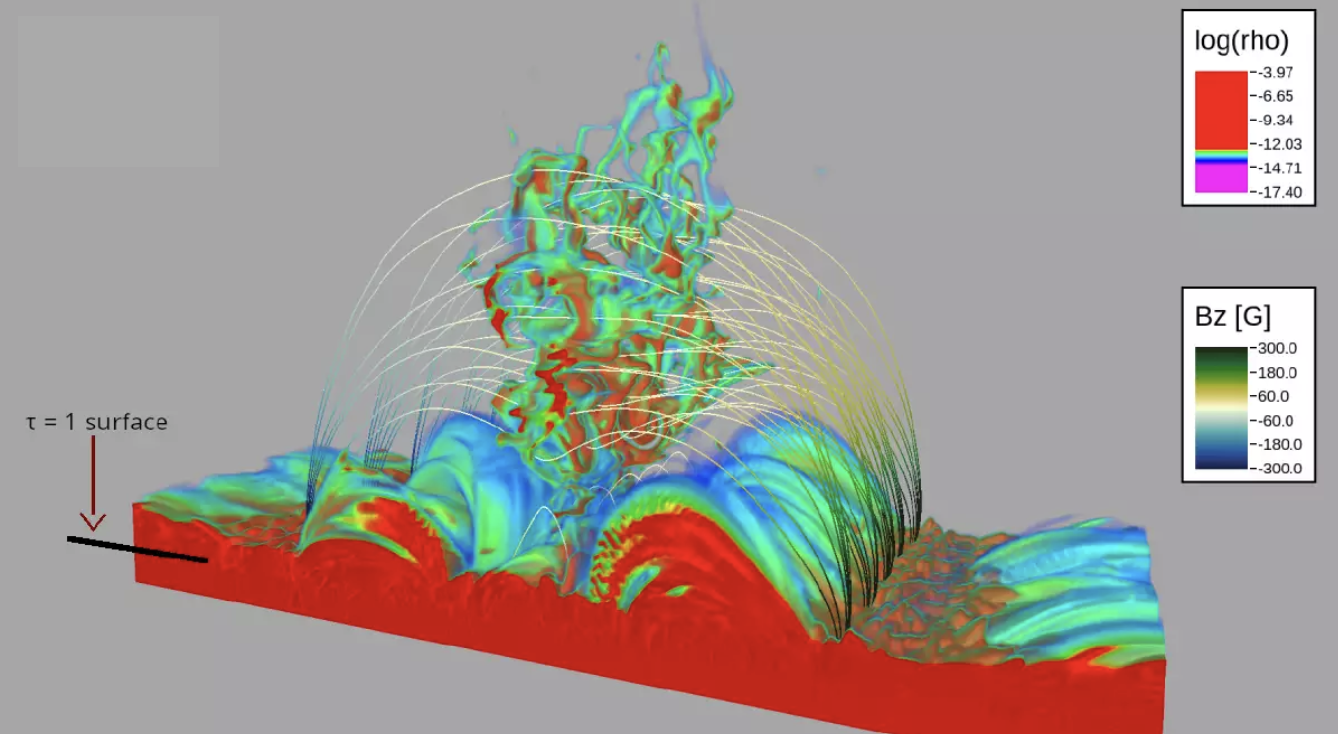}
\caption{A prominence that formed by a combination of injection and in-situ condensation, within a dipped arcade that is inserted in a sub-photosphere to coronal box. Work under review \citep{Zessner2025}, figure kindly provided by Lisa-Marie Zessner.}\label{f-lisa}
\end{figure}

\item Further model efforts should try and follow the entire life cycle of filaments, including formation, internal dynamics, and eventual disappearance. This is perhaps best achieved in a data-driven approach, based on detailed observations. This may use observed magnetogram evolutions as bottom boundary information, much like the one simulated in Fig.~\ref{f-chunfig}, but then including all thermodynamic processes. Efforts have been done for global solar models using magnetofrictional evolutions \citep{Mackay2022}, and one may perhaps at first combine such global magnetic field quantifications with follow-up frozen-field hydro simulations (from Section~\ref{s-ffhd}), to get filament matter included. Also in simulations where prominences get ejected, we are as yet missing important details on the thermodynamic fine-structure, with observations showing many falling filament fragments that impact chromospheric layers. Whether prominence drainage differs from dynamics observed in coronal rain settings is an open question. It is also worth noting that current space weather forecasting frameworks (such as \cite{icarus2025}), which use simplified models to represent the ejecta, usually ignore the presence of cool matter alltogether, and incorporating the details of prominences and rain in them may well be needed for more accurate predictions on their geo-effectiveness. Surely, inclusion of multithermal matter could play a role in the initial trigger to unstable and erupting evolutions. Observational findings on coronal rain as a trigger to prominence instability and resulting CMEs are collected in \cite{Vasha2022}.
\end{itemize}

In the end, our numerical resolutions should at least challenge or exceed observational ones, in order to make further modeling progress or even predict novel dynamical features. In that respect, the recent breakthrough in the use of adaptive optics to make observations near or at the solar limb \citep{Schmidt2025} is quite astonishing, revealing finer scale details in coronal rain or in prominence dynamics than ever reported (better than 70 km or 0.1 arcsec). \cite{Schmidt2025} show postflare loop strands at the diffraction limit (of 64 km, implying that native strand widths could have been down to 10 km or less), plasmoid features that strongly evolved within minutes, and prominence and coronal rain details that are missed in current state-of-the-art models. \cite{dkist2025} shows that coordinated observations between Daniel K. Inouye Solar Telescope (DKIST) and Solar Orbiter offer a wealth of observational data, including detailed coronal rain dynamics. \cite{Tamburri2025} identified post-reconnection flare loops in DKIST H$\alpha$ images with widths down to 21 km, further lowering the coronal rain strand widths inferred as in \cite{Jing2016} (see also Fig.~\ref{f-jing}). This poses severe computational challenges to full 3D flare models. The rich possibilities offered by interacting MHD waves and instabilities in gravitationally stratified, radiating, magnetically structured solar and stellar atmospheres are undoubtedly used by nature, and continue to form the inspiration for theoretical and numerical researchers.

\bmhead{Acknowledgements}
RK acknowledges funding from the KU Leuven C1 project C16/24/010 UnderRadioSun and the Research Foundation Flanders FWO project G0B9923N Helioskill. We thank both referees for extensive, constructive comments and suggestions.


\bibliography{sn-bibliography}%


\end{document}